\documentclass[twocolumn]{openjournal}

\usepackage{xcolor}
\usepackage{textgreek}
\usepackage[utf8]{inputenc}
\usepackage[english]{babel}
\usepackage[T1]{fontenc}
\usepackage{verbatim}
\usepackage[normalem]{ulem}
\usepackage{orcidlink}
\usepackage{soul}
\usepackage{hyperref}
\usepackage{color,colortbl}
\usepackage{tensind}
\usepackage{amsmath}
\usepackage{amssymb}

\definecolor{linkcolor}{rgb}{0.0,0.3,0.5}
\hypersetup{
    unicode=true,
    colorlinks=true,
    linkcolor=linkcolor,
    citecolor=linkcolor,
    filecolor=linkcolor,
    urlcolor=linkcolor,
}

\tensordelimiter{?}
\DeclareGraphicsExtensions{.bmp,.png,.jpg,.pdf}

\graphicspath{{./figs/}}

\makeatletter

\def\frontmatter@affiliationfont{%
  \normalfont\footnotesize
  \setlength{\baselineskip}{10.2pt}%
}

\makeatother

\begin{document}

\title{Quality assessment of spectroscopic data reduction pipelines using artificial intelligence: scrutinizing the Data Release 2 from the DESI survey}

\author{
V.~Torres-Gomez$^{1,*}$\orcidlink{0009-0004-9975-7093},
J.~Su\'arez-P\'erez$^{2}$\orcidlink{0000-0002-0896-8134},
J.~E.~Forero-Romero$^{1,3}$\orcidlink{0000-0002-2890-3725},
S.~Bailey$^{4}$\orcidlink{0000-0003-4162-6619},
A.~Kremin$^{4}$\orcidlink{0000-0001-6356-7424},
B.~Dey$^{5,6}$\orcidlink{0000-0002-5665-7912},
R.~P.~Nathan$^{7}$,
S.~Panda$^{8,\dagger}$\orcidlink{0000-0002-5854-7426},
J.~Aguilar$^{4}$,
S.~Ahlen$^{9}$\orcidlink{0000-0001-6098-7247},
D.~Bianchi$^{10,11}$\orcidlink{0000-0001-9712-0006},
D.~Brooks$^{7}$,
T.~Claybaugh$^{4}$,
A.~de la Macorra$^{12}$\orcidlink{0000-0002-1769-1640},
P.~Doel$^{7}$,
S.~Ferraro$^{4,13}$\orcidlink{0000-0003-4992-7854},
A.~Font-Ribera$^{14,15}$\orcidlink{0000-0002-3033-7312},
E.~Gazta\~naga$^{16,17,18}$\orcidlink{0000-0001-9632-0815},
S.~{Gontcho A Gontcho}$^{19}$\orcidlink{0000-0003-3142-233X},
G.~Gutierrez$^{20}$,
K.~Honscheid$^{21,22,23}$\orcidlink{0000-0002-6550-2023},
C.~Howlett$^{24}$\orcidlink{0000-0002-1081-9410},
R.~Joyce$^{8}$\orcidlink{0000-0003-0201-5241},
S.~Juneau$^{8}$\orcidlink{0000-0002-0000-2394},
D.~Kirkby$^{25}$\orcidlink{0000-0002-8828-5463},
O.~Lahav$^{7}$\orcidlink{0000-0002-1134-9035},
M.~Landriau$^{4}$\orcidlink{0000-0003-1838-8528},
L.~Le~Guillou$^{26}$\orcidlink{0000-0001-7178-8868},
M.~E.~Levi$^{4}$\orcidlink{0000-0003-1887-1018},
M.~Manera$^{27,15}$\orcidlink{0000-0003-4962-8934},
A.~Meisner$^{8}$\orcidlink{0000-0002-1125-7384},
R.~Miquel$^{14,15}$,
J.~Moustakas$^{28}$\orcidlink{0000-0002-2733-4559},
S.~Nadathur$^{17}$\orcidlink{0000-0001-9070-3102},
N.~Palanque-Delabrouille$^{29,4}$\orcidlink{0000-0003-3188-784X},
W.~J.~Percival$^{30,31,32}$\orcidlink{0000-0002-0644-5727},
F.~Prada$^{33}$\orcidlink{0000-0001-7145-8674},
I.~P\'erez-R\`afols$^{34}$\orcidlink{0000-0001-6979-0125},
G.~Rossi$^{35}$,
L.~Samushia$^{36,37,38}$\orcidlink{0000-0002-1609-5687},
E.~Sanchez$^{39}$\orcidlink{0000-0002-9646-8198},
D.~Schlegel$^{4}$,
H.~Seo$^{40}$\orcidlink{0000-0002-6588-3508},
R.~Sharples$^{41,42}$\orcidlink{0000-0003-3449-8583},
J.~Silber$^{4}$\orcidlink{0000-0002-3461-0320},
G.~Tarl\'{e}$^{43}$\orcidlink{0000-0003-1704-0781},
B.~A.~Weaver$^{8}$,
and H.~Zou$^{44}$\orcidlink{0000-0002-6684-3997}
}

\email{* v.torresg23@uniandes.edu.co}
\altaffiltext{$\dagger$}{Gemini Science Fellow}

\affiliation{
$^{1}$ Departamento de F\'isica, Universidad de los Andes, Cra. 1 No. 18A-10, Edificio Ip, CP 111711, Bogot\'a, Colombia\\
$^{2}$ Tecnol\'ogico de Monterrey, Escuela de Ingenier\'ia y Ciencias, CP 45010, Zapopan, M\'exico\\
$^{3}$ Observatorio Astron\'omico, Universidad de los Andes, Cra. 1 No. 18A-10, Edificio H, CP 111711 Bogot\'a, Colombia\\
$^{4}$ Lawrence Berkeley National Laboratory, 1 Cyclotron Road, Berkeley, CA 94720, USA\\
$^{5}$ Department of Astronomy \& Astrophysics, University of Toronto, Toronto, ON M5S 3H4, Canada\\
$^{6}$ Department of Physics \& Astronomy and Pittsburgh Particle Physics, Astrophysics, and Cosmology Center (PITT PACC), University of Pittsburgh, 3941 O'Hara Street, Pittsburgh, PA 15260, USA\\
$^{7}$ Department of Physics \& Astronomy, University College London, Gower Street, London, WC1E 6BT, UK\\
$^{8}$ NSF NOIRLab, 950 N. Cherry Ave., Tucson, AZ 85719, USA\\
$^{9}$ Department of Physics, Boston University, 590 Commonwealth Avenue, Boston, MA 02215, USA\\
$^{10}$ Dipartimento di Fisica ``Aldo Pontremoli'', Universit\`a degli Studi di Milano, Via Celoria 16, I-20133 Milano, Italy\\
$^{11}$ INAF-Osservatorio Astronomico di Brera, Via Brera 28, 20122 Milano, Italy\\
$^{12}$ Instituto de F\'{\i}sica, Universidad Nacional Aut\'onoma de M\'exico, Circuito de la Investigaci\'on Cient\'{\i}fica, Ciudad Universitaria, Cd. de M\'exico C.~P.~04510, M\'exico\\
$^{13}$ University of California, Berkeley, 110 Sproul Hall \#5800, Berkeley, CA 94720, USA\\
$^{14}$ Instituci\'o Catalana de Recerca i Estudis Avan\c{c}ats, Passeig de Llu\'{\i}s Companys, 23, 08010 Barcelona, Spain\\
$^{15}$ Institut de F\'{\i}sica d'Altes Energies (IFAE), The Barcelona Institute of Science and Technology, Edifici Cn, Campus UAB, 08193, Bellaterra (Barcelona), Spain\\
$^{16}$ Institut d'Estudis Espacials de Catalunya (IEEC), c/ Esteve Terradas 1, Edifici RDIT, Campus PMT-UPC, 08860 Castelldefels, Spain\\
$^{17}$ Institute of Cosmology and Gravitation, University of Portsmouth, Dennis Sciama Building, Portsmouth, PO1 3FX, UK\\
$^{18}$ Institute of Space Sciences, ICE-CSIC, Campus UAB, Carrer de Can Magrans s/n, 08913 Bellaterra, Barcelona, Spain\\
$^{19}$ University of Virginia, Department of Astronomy, Charlottesville, VA 22904, USA\\
$^{20}$ Fermi National Accelerator Laboratory, PO Box 500, Batavia, IL 60510, USA\\
$^{21}$ Center for Cosmology and AstroParticle Physics, The Ohio State University, 191 West Woodruff Avenue, Columbus, OH 43210, USA\\
$^{22}$ Department of Physics, The Ohio State University, 191 West Woodruff Avenue, Columbus, OH 43210, USA\\
$^{23}$ The Ohio State University, Columbus, 43210 OH, USA\\
$^{24}$ School of Mathematics and Physics, University of Queensland, Brisbane, QLD 4072, Australia\\
$^{25}$ Department of Physics and Astronomy, University of California, Irvine, 92697, USA\\
$^{26}$ Sorbonne Universit\'e, CNRS/IN2P3, Laboratoire de Physique Nucl\'eaire et de Hautes Energies (LPNHE), FR-75005 Paris, France\\
$^{27}$ Departament de F\'{\i}sica, Serra H\'unter, Universitat Aut\`onoma de Barcelona, 08193 Bellaterra (Barcelona), Spain\\
$^{28}$ Department of Physics and Astronomy, Siena University, 515 Loudon Road, Loudonville, NY 12211, USA\\
$^{29}$ IRFU, CEA, Universit\'e Paris-Saclay, F-91191 Gif-sur-Yvette, France\\
$^{30}$ Department of Physics and Astronomy, University of Waterloo, 200 University Ave W, Waterloo, ON N2L 3G1, Canada\\
$^{31}$ Perimeter Institute for Theoretical Physics, 31 Caroline St. North, Waterloo, ON N2L 2Y5, Canada\\
$^{32}$ Waterloo Centre for Astrophysics, University of Waterloo, 200 University Ave W, Waterloo, ON N2L 3G1, Canada\\
$^{33}$ Instituto de Astrof\'{\i}sica de Andaluc\'{\i}a (CSIC), Glorieta de la Astronom\'{\i}a, s/n, E-18008 Granada, Spain\\
$^{34}$ Departament de F\'isica, EEBE, Universitat Polit\`ecnica de Catalunya, c/Eduard Maristany 10, 08930 Barcelona, Spain\\
$^{35}$ Department of Physics and Astronomy, Sejong University, 209 Neungdong-ro, Gwangjin-gu, Seoul 05006, Republic of Korea\\
$^{36}$ Abastumani Astrophysical Observatory, Tbilisi, GE-0179, Georgia\\
$^{37}$ Department of Physics, Kansas State University, 116 Cardwell Hall, Manhattan, KS 66506, USA\\
$^{38}$ Faculty of Natural Sciences and Medicine, Ilia State University, 0194 Tbilisi, Georgia\\
$^{39}$ CIEMAT, Avenida Complutense 40, E-28040 Madrid, Spain\\
$^{40}$ Department of Physics \& Astronomy, Ohio University, 139 University Terrace, Athens, OH 45701, USA\\
$^{41}$ Centre for Advanced Instrumentation, Department of Physics, Durham University, South Road, Durham DH1 3LE, UK\\
$^{42}$ Institute for Computational Cosmology, Department of Physics, Durham University, South Road, Durham DH1 3LE, UK\\
$^{43}$ University of Michigan, 500 S. State Street, Ann Arbor, MI 48109, USA\\
$^{44}$ National Astronomical Observatories, Chinese Academy of Sciences, A20 Datun Road, Chaoyang District, Beijing, 100101, P.~R.~China
}


\begin{abstract}
Large spectroscopic surveys now collect data at a scale that makes traditional visual inspection impractical.
We present an unsupervised pipeline for spectroscopic quality assessment that requires no labeled training data: it combines Uniform Manifold Approximation and Projection (UMAP) for dimensionality reduction with Friends-of-Friends (FoF) clustering to isolate anomalous spectra for targeted review.
We apply this pipeline to 58{,}291{,}334 spectra across 14{,}199 tiles from DESI Data Release 2, the largest application of this type of approach to date, processing each tile independently to produce a tile-level outlier catalog.
In each tile, the pipeline separates a dense core of typical spectra from small, isolated components and singletons, yielding a total of 1{,}095{,}816 outlier candidates; the mean tile-level outlier fraction is $\sim$1.96\% overall, with values of 0.76\% and 2.36\% for the dark and bright main-survey programs, respectively.
From the visual inspection of 391 outlier candidates from the dark and bright programs of the main survey, we find that $66.8^{+4.6}_{-5.0}$\% exhibit identifiable spectral anomalies consistent with known reduction and calibration effects.
By contrast, only $4.1^{+2.5}_{-1.7}$\% carry a non-zero quality flag from the standard reduction pipeline, demonstrating that the method provides a complementary quality-assessment layer to existing pipeline diagnostics and recovers a substantial population of problematic spectra that standard diagnostics miss.
Extrapolating to the main-survey catalog, we estimate that approximately $218{,}000^{+32{,}000}_{-31{,}000}$ candidate outliers are free of identifiable reduction artifacts and may correspond to genuine atypical spectra in the context of DESI.
The pipeline is scalable, reproducible, and directly comparable across successive data releases, making it a practical quality-assurance monitor for DESI and future multi-object spectroscopic surveys.
\end{abstract}


\section{Introduction}
\label{sec:intro}

Large spectroscopic surveys now collect data at a scale that was unimaginable a decade ago.
The Dark Energy Spectroscopic Instrument (DESI) \citep{Levi_2013, DESI_Collaboration_2022} is a clear example: it gathers spectra from more than one million targets every month \citep{desiops}, and its first data release alone delivered high-confidence redshifts for 18.7 million unique targets \citep{desi2026dr1}.
At this scale, traditional quality assurance (QA) based on visual inspection becomes impractical.
New automated methods are needed to identify problematic spectra quickly and reliably.

Data quality is critical, as systematic effects can propagate into biases in redshift estimation, target classification, and large-scale structure measurements \citep{Ross_2016, Bault_2025, Krolewski_2025, Adame_2025}.
At the same time, spectral anomalies are scientifically interesting in their own right: detecting them can reveal unexpected instrumental effects or genuine astrophysical variability \citep{way_2012}.
Ensuring the integrity of survey data is therefore both an operational necessity and a scientific opportunity.

Previous efforts in automated spectroscopic QA have relied on a range of machine-learning techniques \citep{Nun_2014, Baron_2016, Muthukrishna_2022, liang2023outlierdetectiondesibright, Cook_2024}. Among these, unsupervised methods are especially useful, as they require no labeled training data. Instead, they learn what a ``normal'' spectrum looks like directly from the data, enabling us to flag objects that deviate from that norm. Earlier work in DESI and similar surveys has shown that projecting spectra into low-dimensional embeddings makes outliers and instrument failures visible as isolated structures \citep[see, e.g.,][]{Sanchez_Saez_2021, Rosito_2023, Lan_2023, Alexander_2023, Nicolaou_2026}.

Our approach relies on the standard manifold assumption underlying dimensionality reduction: spectra are high-dimensional vectors, but their wavelength bins are strongly correlated by continua, spectral features, redshift, and calibration
effects, so they occupy a structured lower-dimensional subset of flux space \citep{Yip_2004a, Sharbaf_2023}. This premise has been exploited in SDSS spectroscopy using PCA/Karhunen--Lo\`eve methods \citep{Yip_2004a, Yip_2004b, McGurk_2010} and more recently, with nonlinear latent-variable models such as variational autoencoders \citep[see, e.g.,][]{Portillo_2020, Ortiz_2025}.

We build on the method first introduced by \citet{Suarez_Art_2023} and extended to DESI quality assessment by \citet{umap-proceedings}. The pipeline combines Uniform Manifold Approximation and Projection (UMAP; \citealt{Mcinnes_2020}) for dimensionality reduction with Friends-of-Friends (FoF;  \citealt{HuchraGeller_1982}) clustering to identify outliers. It requires no labels, operates on the standard pipeline outputs, and preserves the DESI identifiers needed for direct retrieval and audit of flagged spectra (see Sections~\ref{subsec:desiproc} and~\ref{sec:dataproc}). This design makes it a practical complement to the existing visual-inspection workflow: it concentrates human effort on the spectra most likely to be problematic, rather than replacing human judgment altogether.

We apply this framework to 58{,}291{,}334 coadded spectra from DESI Data Release 2 (DR2), using a tile-by-tile processing scheme that makes survey-scale analysis feasible. This is, to our knowledge, the largest application of this type of method to a spectroscopic dataset.

The remainder of this paper is organized as follows.
Section~\ref{sec:desi} reviews the DESI spectroscopic pipeline, covering instrument and observations (Section~\ref{subsec:instrument}), target selection (Section~\ref{subsec:targetselection}), and the data-reduction workflow (Section~\ref{subsec:desiproc}). Section~\ref{sec:dataproc} describes the processing of DESI DR2 coadded spectra and the structure used for the analysis. Section~\ref{sec:methods} details the unsupervised anomaly-detection methodology based on UMAP embeddings and FoF clustering.
Section~\ref{sec:results} presents survey-scale results. Section~\ref{sec:visinspection} discusses characteristic spectral anomalies derived by targeted visual inspection. Sections~\ref{sec:disc} and~\ref{sec:concl} summarize our conclusions and outline directions for integration into DESI operations and future data releases.

\section{DESI spectroscopic observations and data processing}
\label{sec:desi}

DESI is a 5{,}000-fiber multi-object spectrograph mounted on the 4-m Mayall telescope at Kitt Peak National Observatory, built to conduct a wide-field spectroscopic survey of galaxies, quasars, and stars \citep{DESI_Collaboration_2022, desiops}. By collecting millions of spectra over $\sim14{,}000~\mathrm{deg}^2$, DESI maps the large-scale structure of the Universe and delivers massive redshift samples that improve constraints on cosmic expansion and structure formation \citep{Ruiz-Macias_2021}. The survey's primary target classes provide complementary tracers that probe distinct epochs and environments \citep{Myers_2023}, yielding a unified, statistically powerful dataset for both extragalactic and Galactic science.

DESI divides observations into three programs selected in real time based on sky conditions \citep{desiops}. The \emph{dark program} is observed when conditions are best and targets faint extragalactic objects; it accounts for approximately 59\% of total observing time. The \emph{bright program} runs under moderate conditions and targets brighter galaxies and Milky Way stars ($\sim$35\% of time). A \emph{backup program} of bright stars is observed only in the poorest conditions ($\sim$6\% of time) \citep{desiops}. The assignment of specific target classes to each program is described in Section~\ref{subsec:targetselection}.

Main-survey operations were preceded by a Survey Validation (SV0--SV3) phase in which observing strategies and the spectroscopic reduction pipeline were iteratively refined \citep[for a comprehensive description, see][]{Lan_2023}. Reductions across SV and the main survey were produced with a sequence of tagged pipeline releases, and subsequent data releases incorporate these updates.

DESI data are organized into successive public \emph{data releases} (DRs), each corresponding to a fixed spectroscopic processing run applied to a well-defined set of observations. These releases serve as the reproducible units against which QA tools, such as the one presented here, can be benchmarked and compared across time.

The first data release, DR1 \citep{desi2026dr1}, covers the initial 13 months of main-survey operations (May 2021 through June 2022), together with a uniform reprocessing of the Survey Validation data. The primary spectroscopic production for DR1 is named \textit{Iron}. 
DR1 delivers high-confidence redshifts for 18.7 million unique targets
\citep{desi2026dr1}, enabling measurements of two-point clustering \citep{DESI_2024_II},
BAO from galaxies and quasars \citep{DESI_2024_III}, BAO from the Lyman-$\alpha$ forest
\citep{DESI_2024_IV}, full-shape clustering \citep{DESI_2024_V}, and cosmological
constraints \citep{DESI_2024_VI}.

The second data release, DR2, extends the survey baseline to three years of main-survey observations (through April 2024). The underlying spectroscopic production is named \textit{Loa}, an updated reprocessing that corrected an error in the coaddition of spectra from separate exposures present in the preceding \textit{Kibo} run \citep{desi2025dr2bao}. DR2 extends the first-year DR1 sample to the first three years of DESI observations \citep{desi2025lya}.

The DR2 BAO analysis uses more than 14 million galaxies and quasars selected for BAO measurements alone \citep{desi2025dr2bao}. The Lyman-$\alpha$ forest sample also roughly doubles that of DR1, with forest measurements in over 820{,}000 quasar spectra \citep{desi2025lya}. The spectra analyzed in the present work (58{,}291{,}334 coadded spectra across 14{,}199 tiles) are drawn from this DR2 processing run.

The step from DR1 to DR2 also provides a concrete motivation for automated quality assessment. 
The pipeline fix between \textit{Kibo} and \textit{Loa} affected approximately 0.1\% of measured redshifts \citep{desi2025dr2bao}, illustrating how subtle processing changes can propagate into downstream science products relevant for clustering-based cosmological analyses \citep{Krolewski_2025}. 
A scalable, data-driven monitoring framework is therefore useful not only for identifying problematic spectra within a single release, but also for tracking changes in data quality across successive pipeline releases.

\subsection{Instrument and Observations}
\label{subsec:instrument}

\begin{figure}[t]
  \centering
  \includegraphics[width=\linewidth]{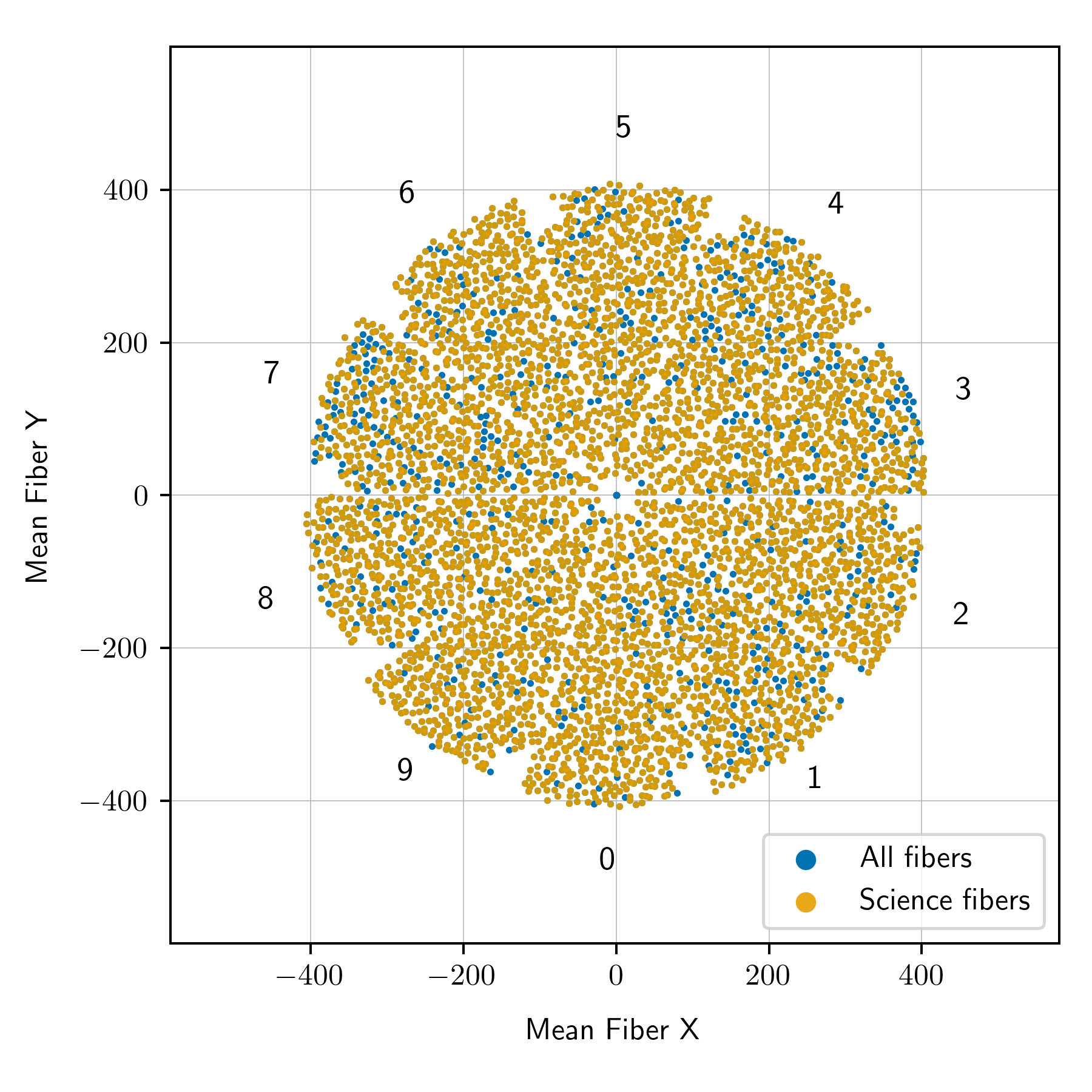}
  \caption{DESI focal plane. Mean fiber positions are aggregated over the tiles analyzed in this work. Blue points show all 5{,}000 spectroscopic fiber positions, while orange points show the subset of fiber positions that were assigned to science targets in at least one of those tiles. Petal identifiers (0--9) are annotated; each petal has 500 fibers feeding one bench spectrograph.}
  \label{fig:focalplane}
\end{figure}

The DESI instrument collects light from astronomical sources through a new wide-field corrector and focuses it onto a $3.2^{\circ}$ diameter focal plane \citep{desicollaboration2016desiexperimentiiinstrument, miller2023opticalcorrectordarkenergy, desiops}.
The focal plane comprises 5{,}000 robotic fiber positioners arranged in ten wedge-shaped modules called ``petals'' (Fig.~\ref{fig:focalplane}) \citep{Silber_2022, DESI_Collaboration_2022, desiops}. 
Each petal hosts 500 positioners, connects to one dedicated bench spectrograph, and contains a Guide-Focus Array (GFA) imaging camera \citep{desiops, Poppett_2024, edr_paper}. 
Four GFAs deliver out-of-focus images used to determine the telescope focus, while the other six provide in-focus images used for guiding, point-spread-function measurements, and throughput monitoring \citep{desiops}. The ten petals are designed to work independently, so a problem in one petal does not affect the others \citep{desiops}.

Each spectrograph splits the collected light into three wavelength channels, or ``arms'': blue (B), red (R), and near-infrared (Z), covering together the range $3600$--$9800$\,\AA\ \citep{desiops}. Observations are executed at fixed sky pointings called ``tiles,'' each matching the $3.2^{\circ}$ field of view. The survey covers approximately $14{,}000$\,deg$^2$ of high-Galactic-latitude sky, divided into 9{,}929 dark-time tiles and 5{,}676 bright-time tiles \citep{desiops}. Each region of the sky is covered on average by 5.2 dark-program passes and 3.2 bright-program passes \citep{desiops}.

A dedicated subset of fibers on every petal is assigned to blank-sky positions and spectrophotometric standard stars, used for sky-background estimation and absolute flux calibration \citep{edr_paper}. As a result, the effective number of fibers available for science targets per exposure is below 5{,}000. Fields whose on-sky target density exceeds this allocation are scheduled for multiple passes to maintain completeness and to reduce the impact of close-pair fiber collisions \citep{desicollaboration2016desiexperimentisciencetargeting}.

Each night, DESI typically observes roughly twenty tiles containing $\sim$100{,}000 sources. By the following morning, the offline pipeline automatically calibrates the exposures, extracts the spectra, subtracts the sky background, and estimates redshifts \citep{Guy_2023, desiops}.

Exposure depth is controlled in real time by the Exposure Time Calculator (ETC), which uses measurements of seeing, sky brightness, and atmospheric transparency from the GFAs and sky monitor to decide when a tile has reached its target signal-to-noise ratio \citep{desiops}. Dark-program observations aim for 1{,}000 seconds of effective time, while bright-program observations aim for 180 seconds \citep{desiops}. 
This system allows DESI to produce spectra of relatively uniform quality even when observing conditions change during the night.

It is also possible that a tile need be observed on multiple exposures, possible on different nights. 
This often happens when observing conditions dynamically evolve requiring a long exposure (the maximum time for a single exposure is 1800 seconds), which forces the observer to finish the current ongoing exposure without the required effective time to be accumulated for that tile.
If the required effective time is not reached in a single exposure, additional exposures may be taken later during the same observing night or on a different night. As a result, a single tile can be completed through multiple exposures.

For each exposure, the fiber assignment algorithm computes the positioner-target mapping for the tile and writes this assignment to the raw data as the \texttt{FIBERMAP} HDU \citep{Guy_2023}. 
The table records, for each positioner, the assigned category (science/sky/standard), unique identifiers (e.g., \texttt{TARGETID}), on-sky coordinates, and configuration metadata (e.g., \texttt{TILEID}, \texttt{PETAL\_LOC}, \texttt{FIBER}). 
We use these fields throughout this work to index spectra by tile and petal and to localize anomalies at the petal/fiber level.

\subsection{Target Selection}
\label{subsec:targetselection}

Spectroscopic targets are defined by DESI's target-selection package \texttt{desitarget}, which implements the survey's photometric selection algorithms \citep{Myers_2023}. The input imaging comes from the DESI Legacy Imaging Surveys (DECaLS, BASS, and MzLS), which provide photometry in the $g$, $r$, and $z$ optical bands, augmented with infrared measurements from \emph{WISE} in the $W1$ and $W2$ bands \citep{Zou_2017, Dey_2019, Moustakas_2023}. From this combined dataset, \texttt{desitarget} applies uniform imaging-quality masks, star--galaxy separation criteria, and color--magnitude--morphology cuts to build the science target samples. The resulting selections are encoded as per-object targeting bitmasks (e.g., \texttt{DESI\_TARGET}, \texttt{BGS\_TARGET}, \texttt{MWS\_TARGET}) that record which program each object belongs to. 
These bitmasks propagate through fiber assignment and spectroscopic reduction, ensuring that the target selection is reproducible across the full survey footprint \citep{Myers_2023, Ross_2025}.

The primary target samples are partitioned across three observing programs. In the \emph{dark program}, DESI observes three extragalactic tracers: luminous red galaxies (LRGs) at $0.4 < z < 1.0$ \citep{Zhou_2023}, emission-line galaxies (ELGs) at $0.6 < z < 1.6$ \citep{Raichoor_2020, Raichoor_2023}, and quasars (QSOs) at $0.9 < z < 3.5$ \citep{Myers_2023, Chaussidon_2023}. In the \emph{bright program}, DESI observes the magnitude-limited Bright Galaxy Survey (BGS) at $z < 0.4$ \citep{Ruiz-Macias_2021, Hahn_2023, Juneau_2025} and stellar targets from the Milky Way Survey (MWS) \citep{Cooper_2023}. Finally, the \emph{backup program} is activated only when observing conditions are too poor for the bright program, and it targets bright Milky Way stars that can be observed even under bad seeing or high sky background \citep{desiops}. Backup targets include halo giant candidates (\texttt{BACKUP\_GIANT}) and general faint stars (\texttt{BACKUP\_FAINT}), and they carry the lowest target priorities in the survey---below even the filler secondary classes \citep{desiops}. Because completeness and spatial homogeneity are not primary requirements for the backup program, its tiles are not optimized for airmass and are observed at zero hour angle \citep{desiops}.

In addition to these primary samples, DESI also observes \emph{secondary} and \emph{tertiary} targets, designed to facilitate bespoke science programs beyond the primary survey goals \citep{desi2026dr1}. Secondary targets are interleaved with regular targets on normal survey tiles and can be selected from any source of imaging, although most have counterparts in the Legacy Surveys \citep{Dey_2019, desi2026dr1}. Tertiary targets are similar in nature, but they are observed on dedicated \emph{special tiles} that are kept separate from the main survey and its Main Target List (MTL) strategy \citep{desi2026dr1}. In DR1, approximately 2\% of observing time was spent on such special tiles, covering a range of scientific and technical use cases \citep{desi2026dr1}. These special tiles carry \texttt{SURVEY=special} in the data model and \texttt{PROGRAM=other} for non-standard programs, making them easy to identify in the data \citep{desi2026dr1}.

\subsection{DESI Data Processing}
\label{subsec:desiproc}

Raw DESI spectroscopic exposures are reduced by an automated pipeline that delivers wavelength- and flux-calibrated one-dimensional spectra, together with redshift and classification estimates \citep{Guy_2023}. The pipeline runs independently for each of the ten spectrographs, processing the three spectral arms (B, R, and Z) described in Section~\ref{subsec:instrument} \citep{Guy_2023}.

Initial processing removes instrumental signatures and artifacts, including electronic offsets, flat-field variations, bad pixels, and cosmic rays, yielding cleaned detector frames \citep{Guy_2023}. 
Wavelength solutions, spectral trace positions, and a two-dimensional point-spread function (PSF) are derived from arc-lamp exposures and nightly calibrations. 
Science exposures are then extracted via PSF forward modeling to produce, for each fiber, a flux vector, an inverse-variance vector, and a pixel-level mask \citep{Guy_2023}. 
Relative throughput variations across fibers are corrected using fiber flats; the sky background is modeled from the dedicated sky fibers on each petal and subtracted; and absolute flux calibration is obtained from spectrophotometric standard stars. 

When a target has been observed in multiple exposures, these are inverse-variance coadded to improve signal-to-noise while preserving spectral resolution \citep{Guy_2023}. 
The coaddition status of each fiber is recorded in the \texttt{COADD\_FIBERSTATUS} bitmask, which we use in Section~\ref{sec:dataproc} to retain only fibers with no reduction-level flags.
This is true for targets observed on different tiles, but more relevant for this work is the case when the same tile is observed on multiple exposures. 
In our algorithm we use the fully coadded data for the tile over all available exposures for that tile.

Redshifts and spectroscopic classifications are obtained with \texttt{redrock}, DESI's automated template-fitting code \citep{edr_paper}. 
\texttt{Redrock} fits a grid of galaxy, quasar, and stellar templates over a range of redshifts and returns three main outputs: a best-fit redshift \texttt{Z}, a spectroscopic type \texttt{SPECTYPE} (one of \texttt{STAR}, \texttt{GALAXY}, or \texttt{QSO}), and a quality bitmask \texttt{ZWARN} that flags potentially unreliable redshift solutions \citep{edr_paper, Brodzeller_2023}.

\section{Our Processing of DESI Spectra}
\label{sec:dataproc}

This section describes how we prepare the DESI DR2 data for the anomaly-detection pipeline. We start from the standard pipeline outputs described in Section~\ref{subsec:desiproc}, apply a minimal quality filter, and organize the data in a format suitable for tile-by-tile processing.

\subsection{Inputs}

The inputs to this analysis are the calibrated, coadded one-dimensional spectra produced by the DESI reduction pipeline \citep{Guy_2023, edr_paper}. For each spectrum, we use the flux, the inverse variance, and the pixel-level mask that marks problematic bins. We also include the standard \texttt{redrock} outputs for each target: the best-fit redshift \texttt{Z}, the spectroscopic classification \texttt{SPECTYPE} and the quality bitmask \texttt{ZWARN} \citep{edr_paper}. 

Importantly, we do not impose cuts on \texttt{ZWARN} prior to the UMAP embedding, so that spectra with problematic redshift fits are not excluded and can be recovered as candidate outliers by the pipeline. For later stratification of results, each spectrum is labeled by its observing program, target class, and \texttt{redrock} spectroscopic type.

\subsection{Fiber-Level Preselection}

Before running our pipeline, we apply a minimal preselection to remove spectra that are not suitable for analysis. We keep only targets with \texttt{COADD\_FIBERSTATUS}~$= 0$, meaning the coaddition step completed without any recorded reduction flags (see Section~\ref{subsec:desiproc}). We also require \texttt{OBJTYPE = TGT}, which excludes sky fibers.
After this preselection, the working sample comprises 58{,}291{,}334 spectra across 14{,}199 tiles.

\subsection{Tile-by-Tile Processing}

Processing is carried out independently for each DESI tile, following the modular layout of the focal plane (Section~\ref{subsec:instrument}). For each independent run, we fit a separate UMAP using only the spectra observed in that field, and then apply FoF clustering to the resulting two-dimensional embedding. Therefore, the candidates identified by this pipeline are local outliers within a given observation, rather than outliers with respect to a single global embedding of the full DESI data set.

For each selected spectrum, we read the flux and inverse-variance arrays in the three DESI spectrograph arms (B, R, and Z), together with the wavelength grid of each arm. The wavelength arrays are not used as input features; they are used only to define the column ordering of the spectral matrix. Since the DESI coadded spectra are provided on a common wavelength grid for each arm, the B, R, and Z flux arrays are concatenated, producing one flux vector per spectrum with length $N_\lambda \approx 7{,}900$ wavelength bins.

The UMAP embedding is computed from the resulting flux matrix only. The inverse-variance matrix is retained with the same shape for possible downstream quality checks, but it
is not used as an input feature to the model. We also attach the \texttt{redrock} products (\texttt{Z}, \texttt{SPECTYPE}, \texttt{ZWARN}) and the survey metadata needed for traceability: \texttt{TILEID}, \texttt{PETAL\_LOC}, \texttt{FIBER}, and the \texttt{FIBERMAP} association between fibers and targets \citep{Guy_2023}.

\subsection{No Spectral Preprocessing}

We do not apply any denoising, continuum normalization, or band-specific filtering before the embedding step. This choice is deliberate: our goal is to detect spectra that look unusual compared to the rest of the spectra on the tile. 
Reduction artifacts, such as flux discontinuities at arm boundaries, abnormal noise patterns, or calibration residuals, are therefore part of the signals we want to identify. 
Applying any normalization or filtering could suppress exactly these features and reduce the sensitivity of the method.

\subsection{Storage}

Processed data are stored in an HDF5 file structure organized by tile. Each spectrum is saved as a concatenated flux vector together with its inverse-variance and mask arrays, preserving the original per-arm information for full traceability. The metadata fields described above are stored alongside each spectrum. This layout makes it straightforward to retrieve any flagged spectrum and inspect it using standard DESI tools, as described in Sections~\ref{subsec:umap} and~\ref{sec:visinspection}.

\section{Outlier Detection Method}
\label{sec:methods}

The data prepared in Section~\ref{sec:dataproc} enter an unsupervised two-stage pipeline applied independently to each tile. First, UMAP projects each spectrum from a high-dimensional flux space into a two-dimensional embedding that preserves local neighborhood structure. Second, FoF clustering partitions that embedding into connected components. Spectra that form small, isolated groups, or that appear as singletons, are flagged as candidate outliers for further inspection.

UMAP provides a practical compromise between the simplicity of PCA and the flexibility of more expensive nonlinear methods. PCA is fast, deterministic, and widely used for SDSS spectra \citep[e.g.,][]{Yip_2004a}, but its linear projection can miss nonlinear spectral structure driven by continua, spectral features, redshift, and calibration effects. t-SNE is also effective for visualizing local neighborhoods \citep{vanderMaaten_2008}, but UMAP was designed as a scalable manifold-learning algorithm with visualization quality competitive with t-SNE and superior runtime performance \citep{Mcinnes_2020}.

Latent-variable methods, such as variational autoencoders, can reconstruct spectra and capture nonlinear variation in compact latent spaces \citep{Portillo_2020}, but this level of modeling is beyond our operational goal. We therefore use UMAP as a fast, nonlinear, neighborhood-preserving embedding for tile-level QA and candidate prioritization.

This approach, introduced for DESI quality assessment by \citet{Suarez_Art_2023}, has two practical advantages. It requires no labeled training data: the method learns what is ``normal'' directly from the distribution of spectra within each tile. It also produces interpretable outputs: because each flagged spectrum retains its \texttt{TILEID}, \texttt{PETAL\_LOC}, and \texttt{FIBER} identifiers, candidates can be retrieved and reviewed using standard DESI tools (Section~\ref{sec:visinspection}).

This pipeline produces two primary outputs: (i) a compact set of candidate outliers for targeted visual inspection, and (ii) class-stratified outlier fractions per tile and fiber that can be tracked across data releases. These outputs are described in Sections~\ref{sec:results} and~\ref{sec:visinspection}.

\subsection{UMAP}
\label{subsec:umap}

UMAP is a nonlinear dimensionality reduction method that places high-dimensional data points into a low-dimensional space while trying to keep nearby points close together. Given an input metric $M_e$, it builds a weighted $k$-nearest-neighbor graph over the data and learns coordinates in the embedding space by minimizing a cross-entropy objective between the high- and low-dimensional graph structures.

Let $X \in \mathbb{R}^{N \times N_\lambda}$ be the data matrix, where each row $x_i \in \mathbb{R}^{N_\lambda}$ is the concatenated flux vector of spectrum $i$ described in Section~\ref{sec:dataproc}. UMAP learns a map $\phi: \mathbb{R}^{N_\lambda} \to \mathbb{R}^2$ and returns two-dimensional coordinates
\begin{equation}
    y_i = \phi(x_i) \in \mathbb{R}^2, \qquad i = 1, \ldots, N,
\end{equation}
which we stack into the matrix $Y \in \mathbb{R}^{N \times 2}$. Neighborhoods in the embedding are inherited from the $k$-nearest-neighbor graph in $X$: spectra that appear isolated or in low-density regions of $Y$ are therefore those that are atypical with respect to their local neighborhood in the original flux space.

We train UMAP independently for each tile, rather than on the full DR2 data set. This choice is motivated by both computational and operational considerations. First, the full DR2 sample analyzed in this work contains approximately 58 million spectra, making a single global UMAP embedding computationally impractical. Second, a tile-based implementation is naturally aligned with the DESI observing and reduction workflow: spectra are acquired and processed on a tile-by-tile and night-by-night basis, so the method can be run continuously as part of a near-real-time quality-assurance system rather than only after a full survey or data release has been assembled.

Furthermore, training within a single tile maximizes the homogeneity of observational conditions, since spectra co-observed on the same tile share the same night, airmass, seeing, sky background, and instrumental configurations. 
Embedding a heterogeneous mixture of conditions in a single UMAP would conflate astrophysical diversity with observational variance, potentially masking or mimicking genuine outliers.

Finally, processing individual tiles allows candidate outliers to be directly connected to focal-plane position, fibers, and petals, which is essential for identifying recurring instrumental or reduction issues associated with specific hardware elements.

Four main hyperparameters control the embedding: the number of neighbors $N_n$ (locality scale), the embedding dimension $D = 2$ (fixed for visualization and downstream grouping), the similarity metric $M_e$, and the minimum distance $M_d$ (how tightly points may pack in the embedding). We adopt the configuration of \citet{Suarez_Art_2023}, fix a random seed for reproducibility, and keep all hyperparameters constant across tiles (Table~\ref{table:params}). The coordinate matrix $Y$ then serves as input to the FoF stage. Lastly, we keep the UMAP set-operation mix ratio, $S_{\rm mix}$, fixed at its default value of 1.0, so that the fuzzy simplicial set is constructed using the standard UMAP union operation.

\subsection{Friends-of-Friends}
\label{subsec:fof}

FoF defines groups as connected components of a graph whose edges join pairs of points separated by less than a fixed linking length $\ell$. Since UMAP preserves local neighborhood connectivity through its underlying nearest-neighbor graph, we use FoF as a simple connectivity-based criterion to identify groups that are separated from the main spectral distribution in the embedding. In our setting, the points are the UMAP coordinates $y_i \in \mathbb{R}^2$. We connect nodes $i$ and $j$ whenever their Euclidean distance satisfies
\begin{equation}
    \| y_i - y_j \|_2 \leq \ell.
\end{equation}
The resulting undirected graph $G = (V, E)$ is decomposed into connected components, each of which we call a FoF group; components of size one are singletons.

Two hyperparameters control this step. The linking length $\ell$ is the main sensitivity knob: smaller values fragment the graph into many small groups, while larger values merge groups together. The minimum size $N_{\min}$ defines the boundary between a group (which represents a characteristic spectral morphology shared by several spectra) and a candidate outlier. For each connected component $C$, we flag as candidates all singletons and all components with $|C| < N_{\min}$; larger components are treated as part of the normal population. We adopt the same $(\ell, N_{\min})$ values as \citet{Suarez_Art_2023} (Table~\ref{table:params}).

The result is a clean partition of each tile's embedding into a dense background of locally typical spectra and a small set of isolated points and groups, which are the tile-level candidate outliers examined in Section~\ref{sec:visinspection}.

Under this tile-based strategy, the term ``outlier'' should be understood locally: candidates are spectra that are atypical with respect to the population observed in the same tile, rather than global outliers with respect to the entire DR2 sample. This local definition is appropriate for the QA goal of the pipeline, since many problematic cases are expected to be tied to observing conditions, tile-level reductions, or focal-plane/fiber-dependent effects.

\begin{table}[t!]
  \centering
  \renewcommand{\arraystretch}{1.25}
  \setlength{\tabcolsep}{10pt}
  \begin{tabular}{lc}
    \hline\hline
    Hyperparameter & Value \\
    \hline
    $N_n$ (UMAP neighbors)       & 45     \\
    $M_d$ (minimum distance)     & 1.0    \\
    $M_e$ (metric)               & cosine \\
    $S_{\rm mix}$ (set-operation mix ratio) & 1.0 \\ 
    $\ell$ (FoF linking length)  & 0.15    \\
    $N_{\min}$ (minimum group size) & 5   \\
    \hline
  \end{tabular}
  \caption{UMAP and FoF hyperparameters used throughout this work, following \citet{Suarez_Art_2023}. All values are held constant across tiles.}
  \label{table:params}
\end{table}

\subsection{Hyperparameter Selection}
\label{subsec:hyperparams}

The UMAP and FoF hyperparameters adopted in this work follow the configuration established by \citet{Suarez_Art_2023}, which was derived from a systematic exploration of the parameter space on early DESI data. To identify the optimal configuration, a grid of values was evaluated for the three free UMAP parameters: $N_n \in \{5, 15, 25, 35, 45\}$, $M_d \in \{0.0, 0.25, 0.5, 0.75, 1.0\}$, and $M_e \in \{\text{Euclidean, Bray--Curtis, cosine}\}$, with the embedding dimension fixed at $D = 2$ throughout. 
The resulting projections were assessed visually on a representative observation night (20200314, SV0), with the selection criterion being the degree to which known object classes (stars, galaxies, and QSOs) cluster together while anomalous spectra are pushed into small, isolated components.

The combination $N_n = 45$, $M_d = 1.0$, and $M_e = \text{cosine}$ was found to produce the most interpretable embedding: a large $N_n$ encodes more global spectral structure into the neighborhood graph, $M_d = 1.0$ spreads the embedding and prevents the main population from collapsing into an indistinct core, and cosine distance, which is sensitive to the shape of the flux vector rather than its overall normalization, yields the clearest separation between spectral classes. For the FoF stage, a linking length of $\ell = 0.15$ and a minimum group size of $N_{\min} = 5$ were adopted; these values balance sensitivity to genuine outliers against the inclusion of small chance groupings.

We keep both the UMAP and FoF hyperparameters fixed across all tiles to make the output comparable and operationally reproducible. Although each tile can differ in observing conditions and number of valid spectra, tuning the embedding separately for every tile or observing night would introduce an additional layer of subjectivity and would be impractical for a daily QA workflow. Using a single parameter set instead provides a stable reference configuration: changes in the number or spatial distribution of candidates can then be interpreted as changes in the data or observing conditions, rather than as changes in the anomaly-detection setup.

\section{Results}
\label{sec:results}

We apply the UMAP+FoF pipeline described in Section~\ref{sec:methods} to all 14{,}199 DR2 tiles. 
We first illustrate the structure of a single tile embedding, then examine how candidate outlier fractions are distributed across the full survey at the tile, fiber, and petal levels. 
We also report the processing times to characterize the computational scaling of the method.

\subsection{Single-tile illustration}
\label{subsec:single_tile}

\begin{figure}[t!]
  \centering
  \includegraphics[width=\linewidth]{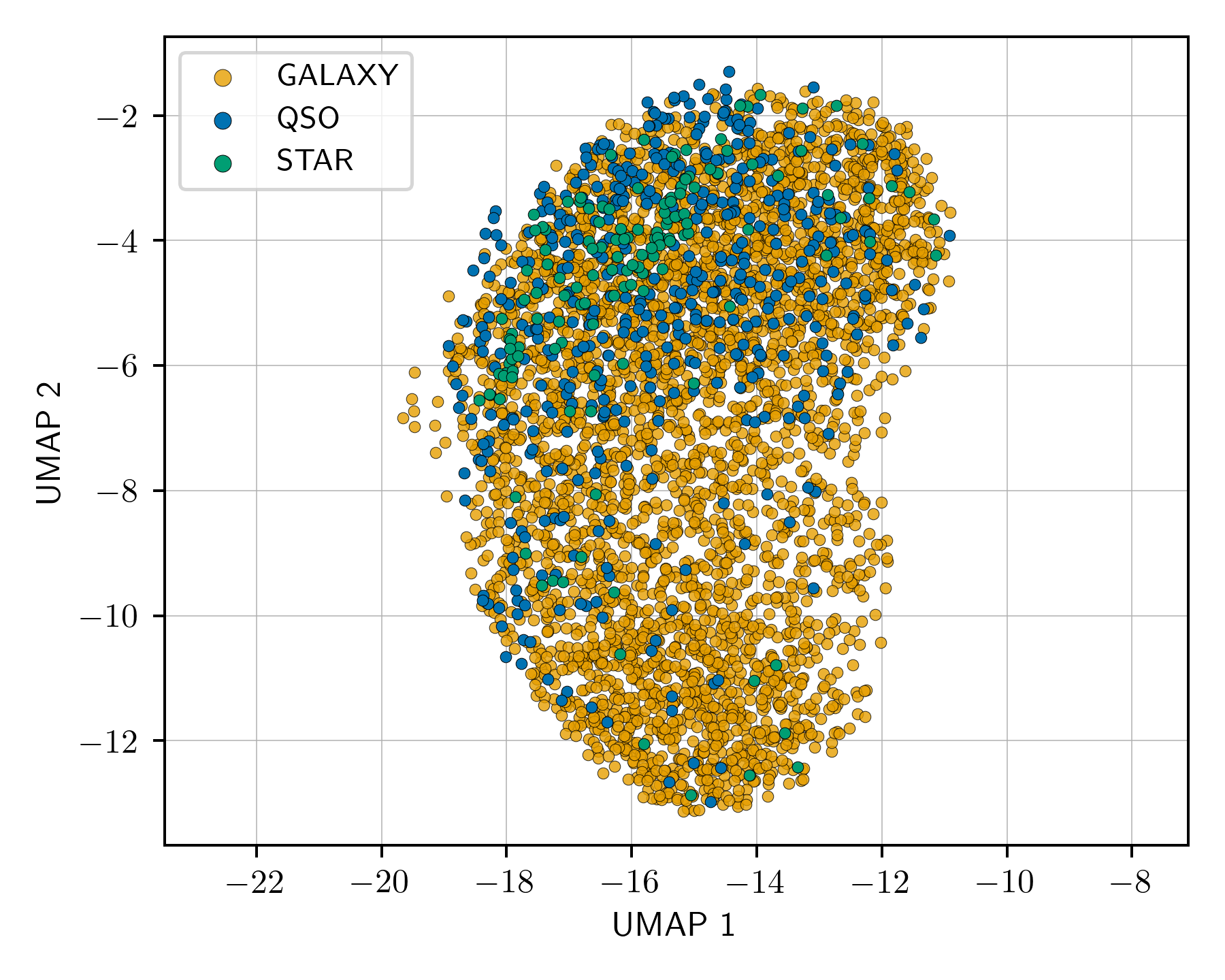}
  \caption{UMAP embedding for \texttt{TILEID}\,8643, containing 4{,}105 spectra after quality and target-type selection, and colored by \texttt{redrock} spectroscopic class (STAR, GALAXY, QSO). Galaxies dominate the main embedding, while stars and quasars appear with substantial overlap with the galaxy distribution.}
  \label{fig:umap_spectype}
\end{figure}

Figure~\ref{fig:umap_spectype} shows the two-dimensional UMAP embedding for \texttt{TILEID}\,8643, which contains 4{,}105 spectra after the selection described in Section~\ref{sec:dataproc}, and corresponds to a tile with an above-median number of candidate outliers that we use throughout as a running example. 
Spectra are color-coded by \texttt{redrock} spectroscopic class. Galaxies dominate the main population, while stars and quasars appear as partially clustered subpopulations that overlap with the galaxy distribution rather than forming fully separated regions. This behavior is expected, as the embedding is built from the full concatenated flux vector and is not trained using the \texttt{redrock} classes as labels.

\begin{figure}[t!]
  \centering
  \includegraphics[width=\linewidth]{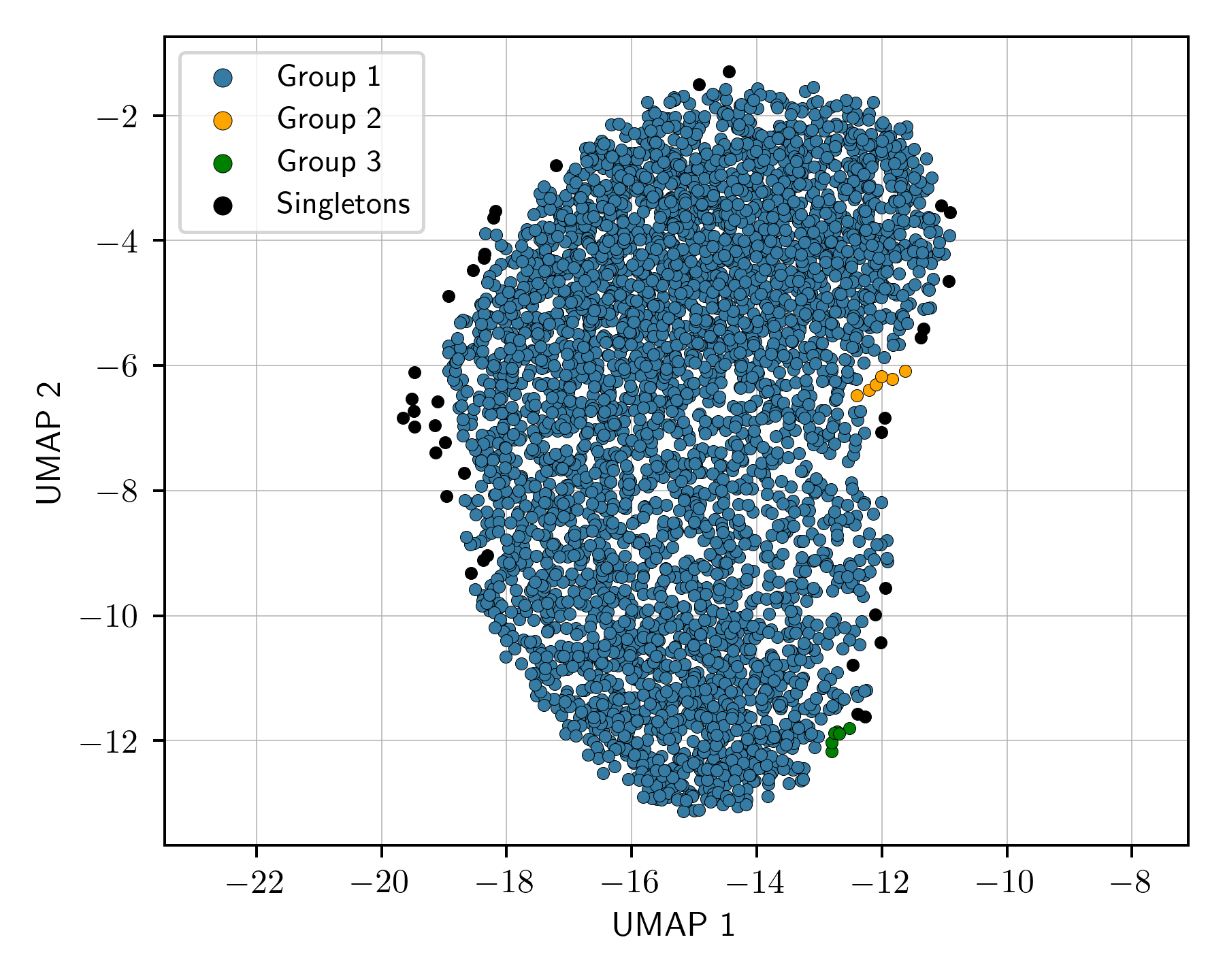}
  \caption{Same embedding as Fig.~\ref{fig:umap_spectype}, now colored by FoF group. The central core corresponds to the main population of typical spectra; small groups and singletons in the periphery are the candidate outliers (singletons and groups with $|C| < N_{\min}$; see Section~\ref{subsec:fof}).}
  \label{fig:umap_fof}
\end{figure}

Applying FoF to this embedding (Fig.~\ref{fig:umap_fof}) reveals a single high-density core containing the vast majority of spectra, together with a small number of compact groups and isolated singletons in the low-density borders. 
These peripheral points are the candidate outliers.

\begin{figure}[t!]
  \centering
  \includegraphics[width=\linewidth]{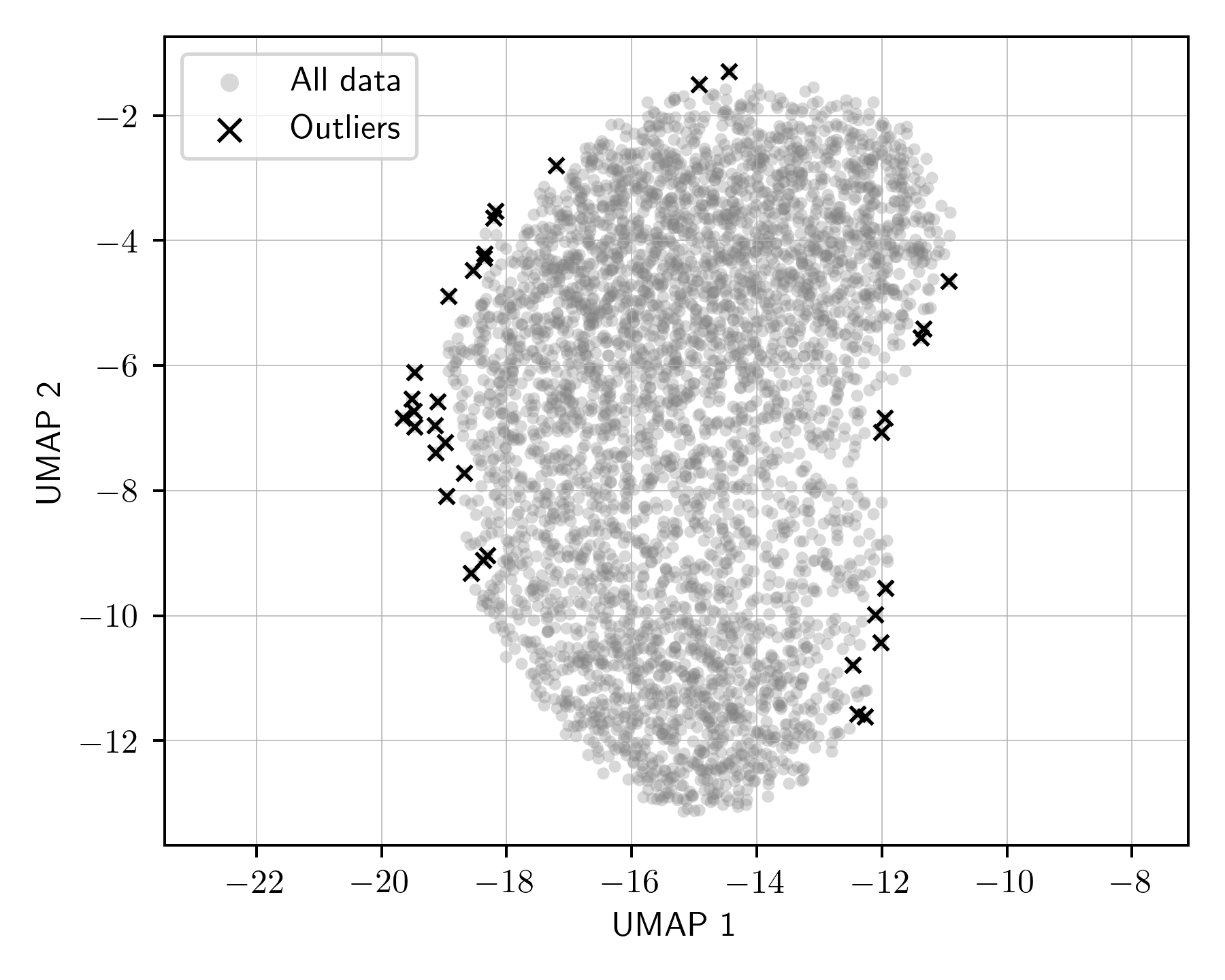}
  \caption{UMAP embedding for \texttt{TILEID}\,8643. Gray points show all spectra; black crosses mark candidates (singletons and groups with $|C| < N_{\min}$).}
  \label{fig:outliers}
\end{figure}

Figure~\ref{fig:outliers} shows the same embedding with all spectra in gray and candidates marked with black crosses. For this tile, the pipeline identifies $34$ candidates, corresponding to $0.83\%$ of the spectra.
Candidates lie almost exclusively in sparse regions, consistent with the FoF linking length $\ell$ and minimum-size threshold $N_{\min}$ (Table~\ref{table:params}). 
This view makes clear that the method reduces the set requiring human review to a small fraction of the tile.

\subsection{Tile-level distribution}
\label{subsec:out_analysis}

\begin{figure*}[t!]
  \centering
  \includegraphics[width=\linewidth]{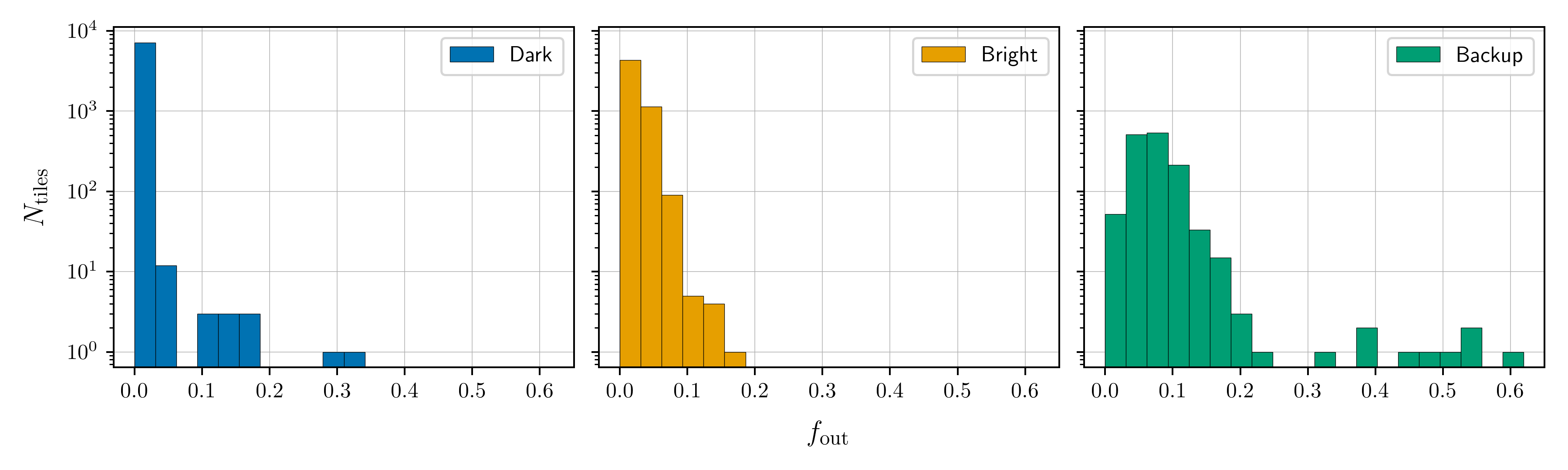}
  \caption{Outlier fraction $f_{\rm out}$ per tile across DR2, separated by observing program (Dark, Bright, and Backup), shown with fixed-width bins and a logarithmic count axis. Normalizing by the number of spectra reduces sample-size effects. Most tiles have candidate fractions of a few percent or less; tiles with $f_{\rm out}>0.2$ are rare and occur primarily in the Backup program.}
  \label{fig:fracs}
\end{figure*}

To account for differences in the number of spectra processed per tile, we examine the candidate fraction $f = N_{\rm out}/N_{\rm spec}$, where $N_{\rm out}$ is the number of candidate outliers and $N_{\rm spec}$ is the number of spectra processed for a given tile. 
Figure~\ref{fig:fracs} shows this fraction separated by observing program. The mean candidate fraction is lowest in the Dark program, $\langle f_{\rm out}\rangle_{\rm Dark}=0.76\%$, increases in the Bright program, $\langle f_{\rm out}\rangle_{\rm Bright}=2.36\%$, and reaches its highest value in the Backup program, $\langle f_{\rm out}\rangle_{\rm Backup}=7.31\%$. These values correspond to a survey-wide mean of $\sim1.96\%$. 

Most tiles have $f_{\rm out}$ at the percent level, while very high candidate fractions are rare: tiles with $f_{\rm out}>0.2$ represent only 0.03\% of Dark tiles, 0.00\% of Bright tiles, and 0.73\% of backup tiles. This indicates that the highest candidate rates are confined to a small subset of tiles and are not only driven by larger numbers of processed spectra.

\subsection{Focal plane and spectrograph dependence}
\label{subsec:fplane}

\begin{figure*}[t!]
  \centering
  \includegraphics[width=\linewidth]{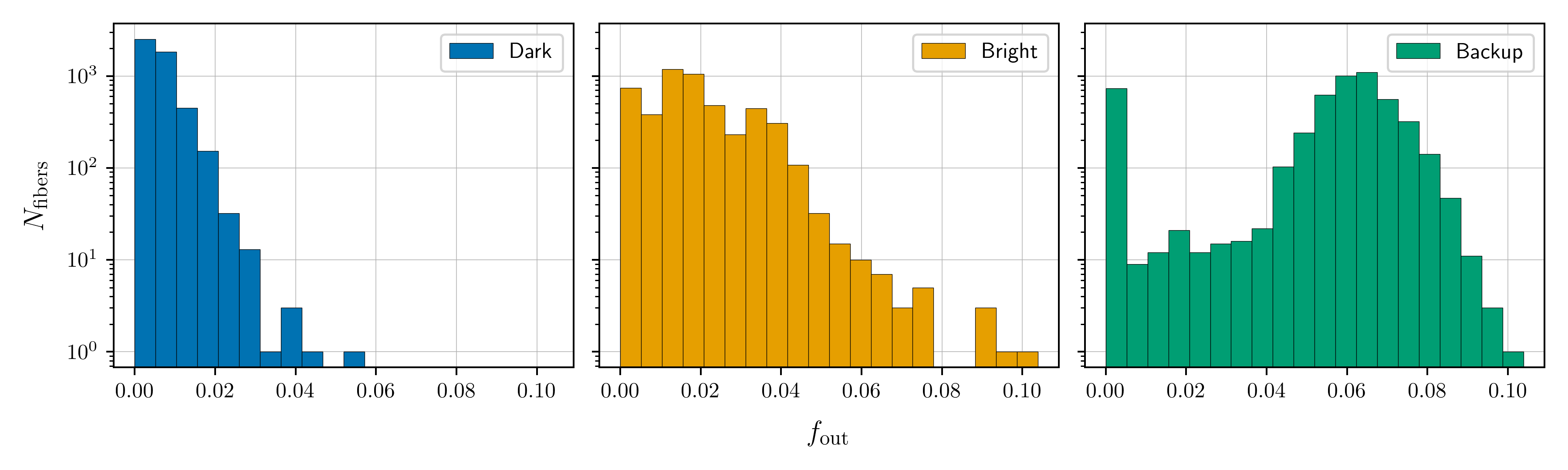}
  \caption{Distribution of candidate outlier fraction per science fiber across DR2, separated by observing program. Most fibers show fractions at the percent level, and a small high-fraction tail is present.}

  \label{fig:fiber_dist}
\end{figure*}

We next examine the distribution of candidate outlier fractions across individual science fibers. 
Figure~\ref{fig:fiber_dist} shows this distribution separated by observing program. 
For the dark and bright programs, the distributions are similarly shaped, with most fibers clustered at low fractions and a small high-fraction tail: a minority of fibers reach values above 0.05 (approximately 0.02\% of dark fibers and 1.14\% of bright fibers). 
The backup program shows a notably broader distribution, with a larger fraction of fibers at elevated outlier rates, consistent with the more heterogeneous and less optimized observing conditions of that program. 
Across all three programs, the bulk of fibers remain at the percent level or below, confirming  that high outlier rates are confined to a small subset of fibers rather than being a survey-wide feature.

\begin{figure}[t!]
  \centering
  \includegraphics[width=\linewidth]{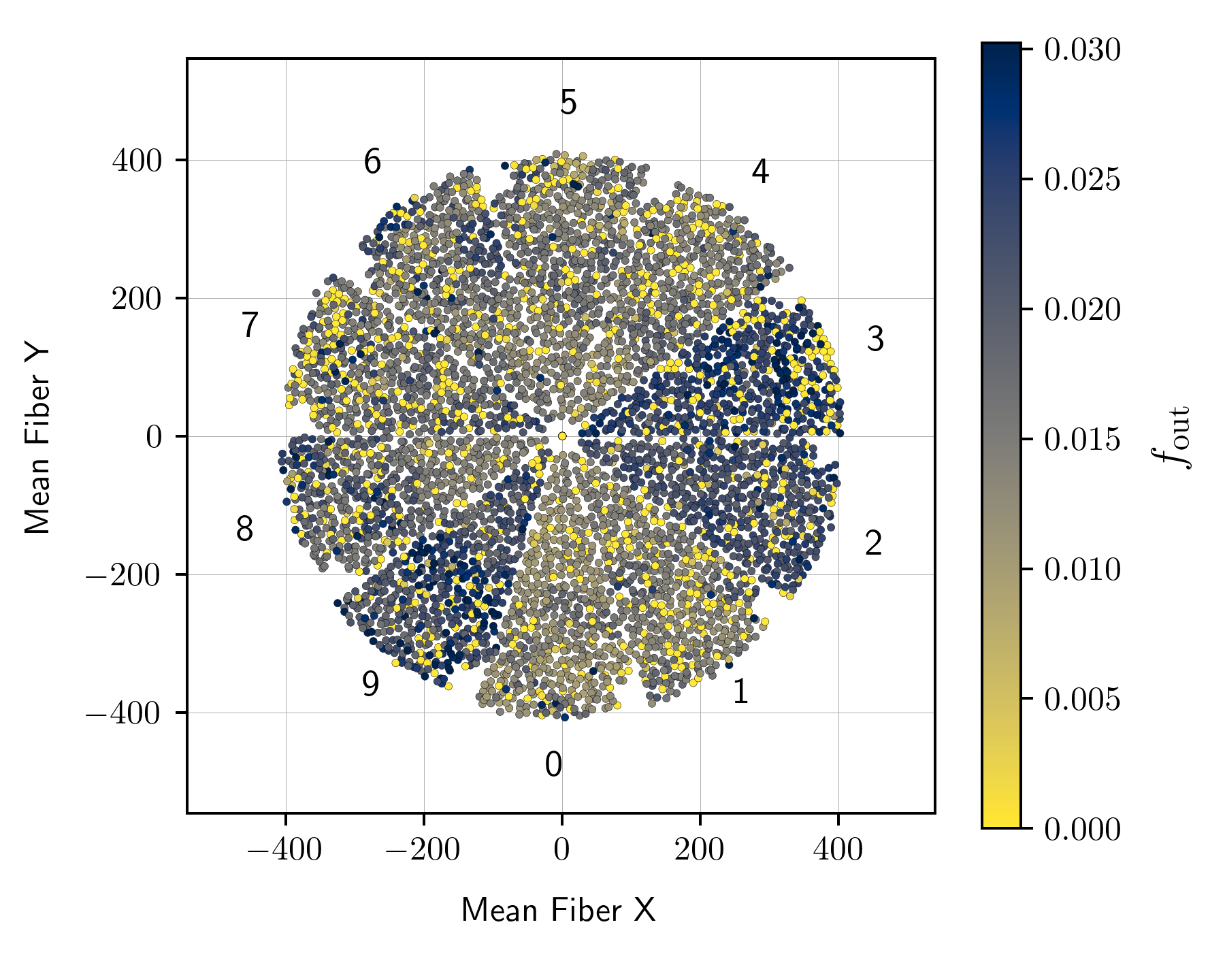}
  \caption{Focal plane map of outlier fraction per fiber. Points mark science fibers at their mean focal plane coordinates; color encodes outlier candidate fraction, and petal identifiers (0--9) are annotated. The distribution is largely uniform, with localized regions of elevated fractions.}
  \label{fig:fp_map}
\end{figure}

Projecting these fractions onto the focal plane (Fig.~\ref{fig:fp_map}) reveals that the  distribution is largely uniform across the $3.2^\circ$ field of view, with most fibers 
showing outlier fractions below $\sim$2\%. 
However, there seem to be petals with higher outlier fractions, hinting at spectrograph-level  differences rather than a smooth focal-plane gradient. These localized enhancements motivate  a closer examination of petal- and fiber-level trends.

\begin{figure*}[t!]
  \centering
  \includegraphics[width=\linewidth]{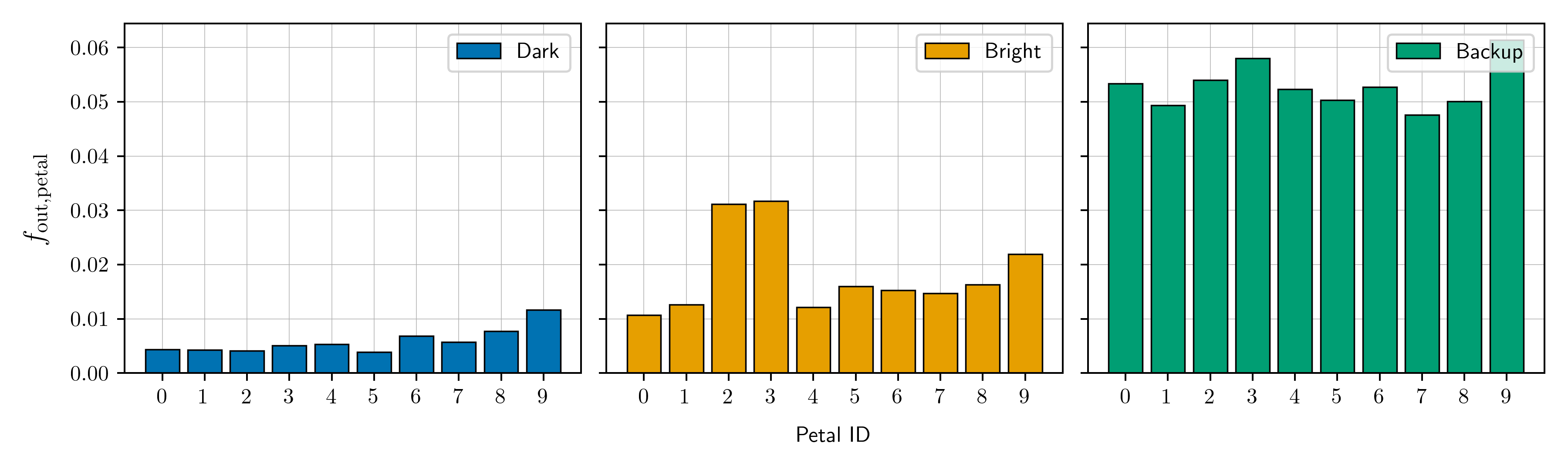}
  \caption{Candidate outlier fraction per \texttt{PETALID} across DR2, separated by observing program. This normalization accounts for differences in petal contribution across tiles and observing programs. All petals contribute non-zero fractions, with substantial petal-to-petal variation.}
  \label{fig:out_petal}
\end{figure*}

Aggregating by \texttt{PETALID} shows a clear program-dependent trend after normalizing  by the number of spectra processed in each petal (Fig.~\ref{fig:out_petal}). 
We define  the petal-level candidate fraction as $f_{\rm out,petal} = N_{\rm out,petal} / N_{\rm  spec,petal}$, where $N_{\rm out,petal}$ is the number of candidate outliers associated  with a given petal and $N_{\rm spec,petal}$ is the corresponding number of valid spectra  processed in it. 
This normalization accounts for differences in the number of spectra 
contributed by each petal across observing programs and tiles. 
In the dark program,  fractions are low and with a small gradient of increasing fractions with increasing \texttt{PETALID}. 
The bright program shows  higher fractions overall, with petals 2 and 3 standing out as clear outliers above the rest. 
The backup program exhibits the highest fractions overall.
Since each focal-plane module feeds a dedicated bench spectrograph 
(Section~\ref{subsec:instrument}), this variation suggests that differences between spectrographs contribute to the candidate rates, with the effect being most pronounced under the less controlled observing conditions of the bright and backup programs.

\begin{figure*}[t!]
  \centering
  \includegraphics[width=\linewidth]{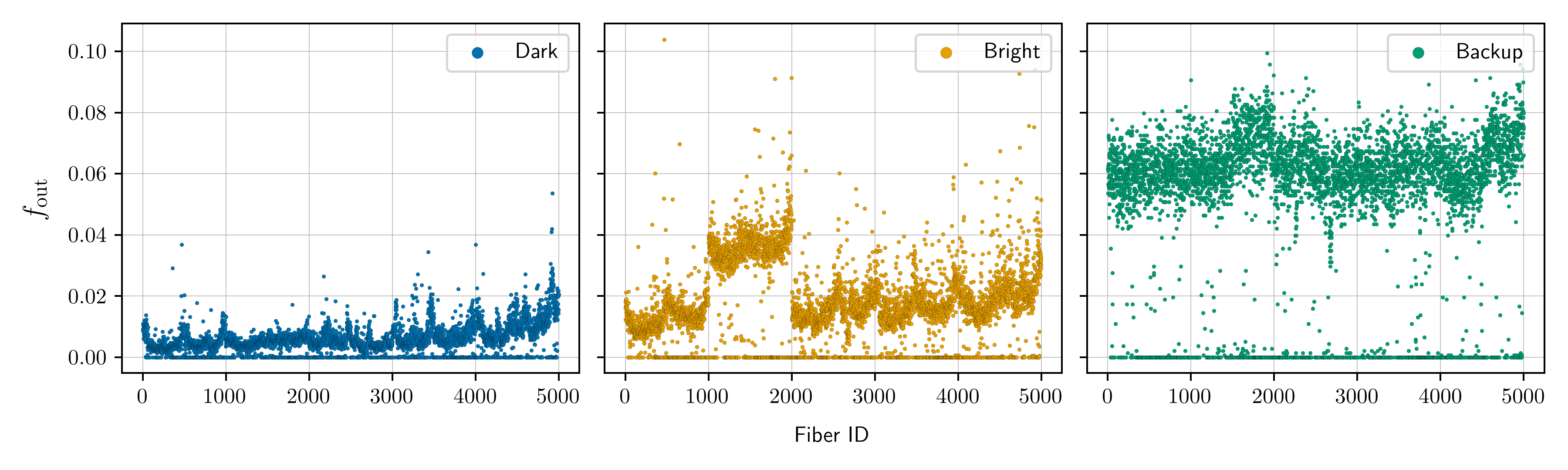}
  \caption{Candidate outlier fractions per fiber as a function of \texttt{FIBERID}, separated by observing program. The $\sim$500-fiber periodicity reflects the petal/spectrograph segmentation of the focal plane. Candidate fractions increase toward the edges of each 500-fiber spectrograph block, consistent with small edge-dependent differences in spectral resolution, focus, throughput, or noise properties.}
  \label{fig:outliers_vs_fiberid}
\end{figure*}

Ordering fibers by \texttt{FIBERID} makes the instrumental pattern more explicit  (Fig.~\ref{fig:outliers_vs_fiberid}). 
A quasi-periodic modulation with a period of  $\sim$500 is visible across all three programs, reflecting the ten-petal structure of 
the focal plane described in Section~\ref{subsec:instrument}. 

The baseline level of the  candidate fraction increases from the dark program to the bright and backup programs,  consistent with the program-level trends reported above. 
In all  programs, the scatter and upper envelope of the candidate fraction increase near  \texttt{FIBERID} $\approx 500\,k$, corresponding to the edges of each 500-fiber  spectrograph block, though the effect is most pronounced in the bright and backup programs. 
This indicates that fibers close to the edges of the spectrograph footprint are more frequently associated with candidate outliers, regardless of observing  program. This pattern traces edge-dependent structure in the fiber direction on the detector, complementing the wavelength-direction edge effects seen in arm-join discontinuities during visual inspection (Section~\ref{sec:visinspection}).

The UMAP selection is sensitive to spectra that differ from the local population in the tile. Fig.~\ref{fig:fracs} shows that high candidate fractions are concentrated in a minority of tiles, while Fig.~\ref{fig:outliers_vs_fiberid} shows that many high-fraction fibers lie near spectrograph-channel boundaries. Such differences can arise from small but repeatable instrumental variations, including changes in throughput, CCD noise properties, spectral resolution near the edges of each spectrograph channel, or focus variations associated with fiber placement in the slithead {\citep[see, e.g.,][]{Krolewski_2025}}. 
These effects might make spectra more likely to appear as isolated points in the embedding even when the redshift fit remains acceptable. 
In this sense, the candidates trace both obvious reduction artifacts and subtler instrumental differences that are useful for QA but are not necessarily failures of the spectroscopic pipeline.

\begin{figure}[t!]
  \centering
  \includegraphics[width=\linewidth]{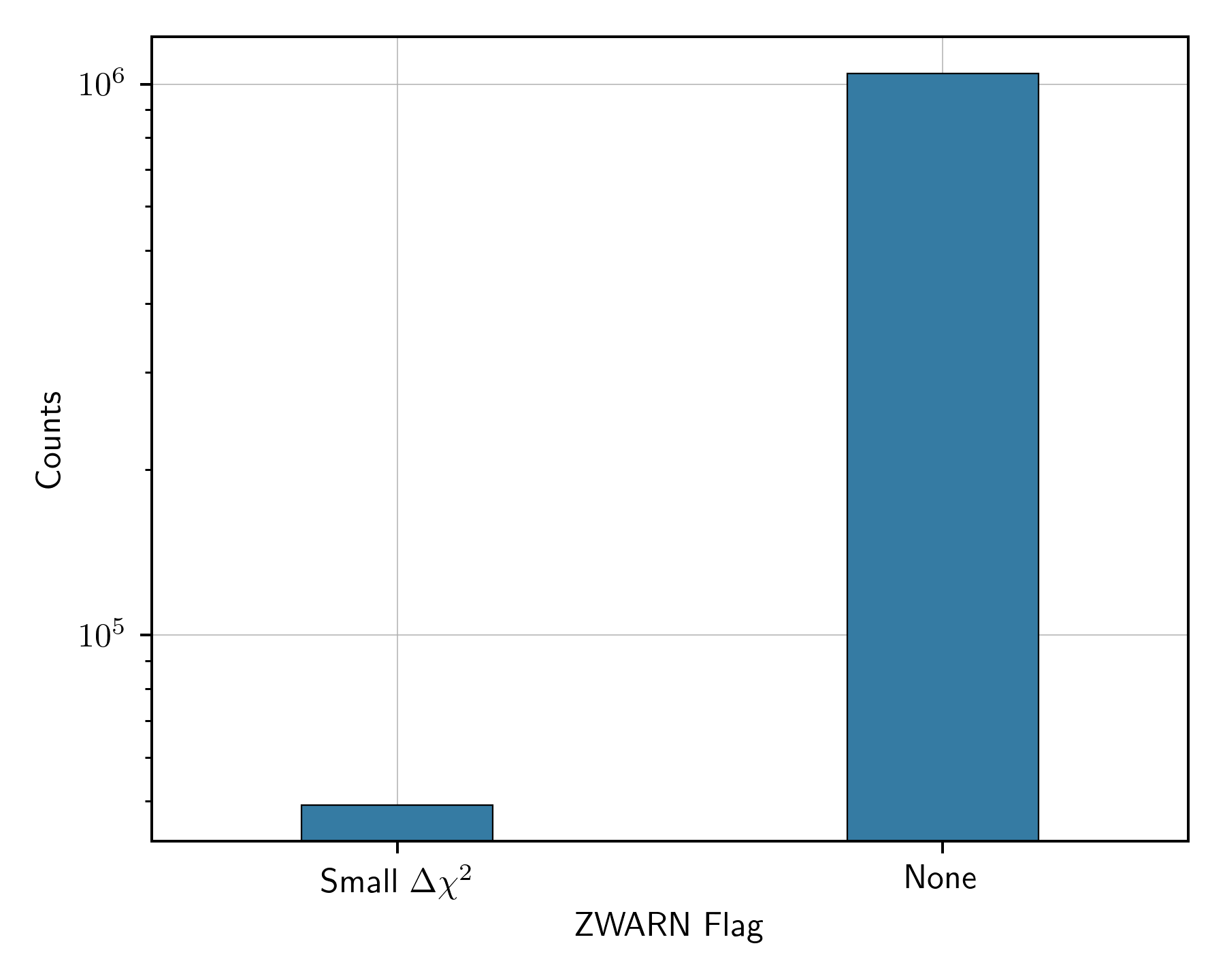}
  \caption{Distribution of \texttt{ZWARN} flag categories for the candidate set on a logarithmic scale. The majority of candidates have \texttt{ZWARN}~$= 0$, indicating they were not flagged by the standard pipeline quality bitmask. This shows that the UMAP+FoF selection identifies a population missed by standard redshift-quality flags.}
  \label{fig:outliers_zflags}
\end{figure}

Figure~\ref{fig:outliers_zflags} shows the distribution of \texttt{ZWARN} flags within the candidate set. 
The majority of candidates carry \texttt{ZWARN}~$= 0$, meaning the standard redshift-quality diagnostics did not flag them. 
This indicates that our embedding-based selection identifies a complementary set of problematic spectra compared to the pipeline's own quality bitmasks, and motivates the no-\texttt{ZWARN}-cut strategy described in Section~\ref{sec:dataproc}.

\begin{figure}[t!]
  \centering
  \includegraphics[width=\linewidth]{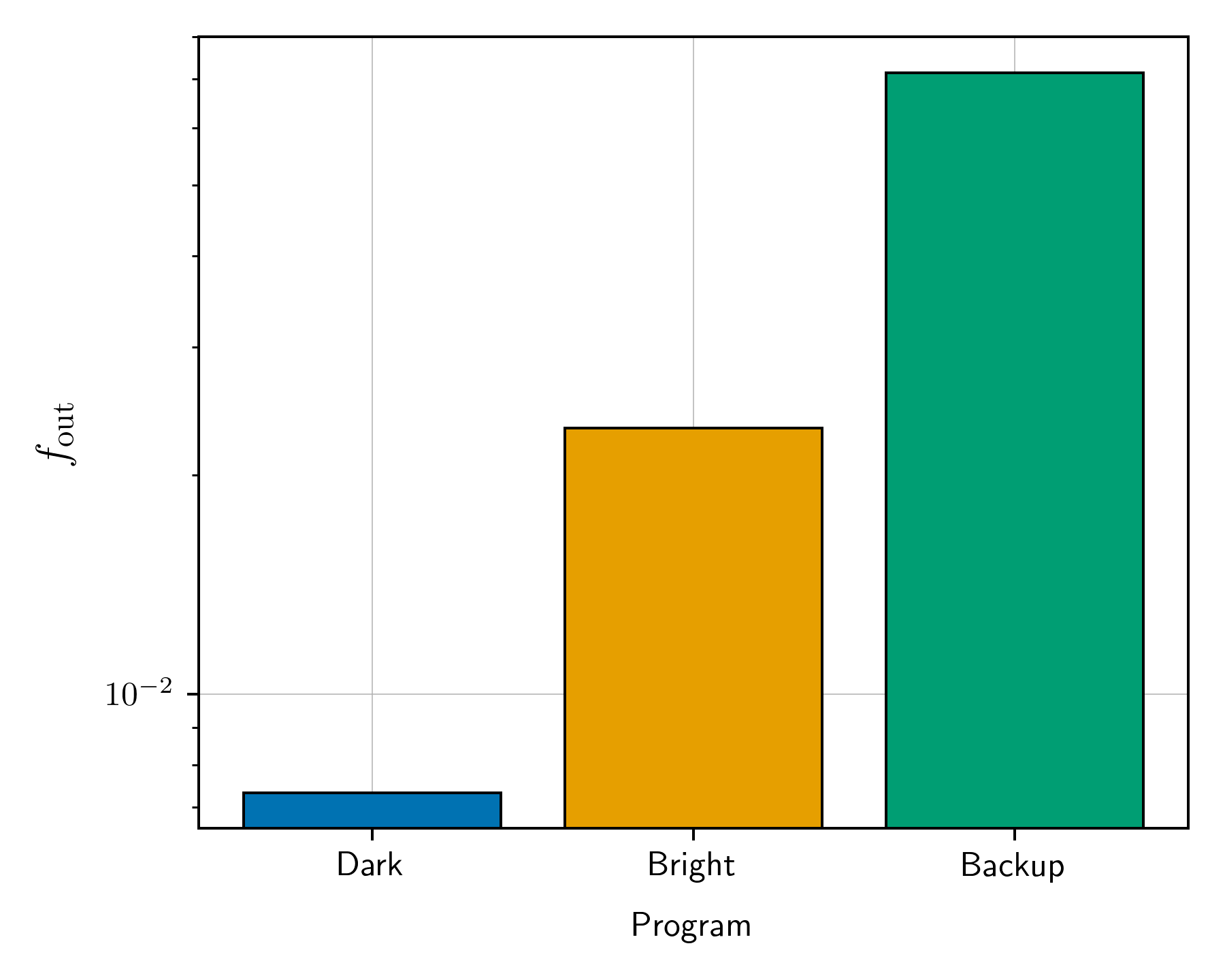}
  \caption{Candidate fraction by observing program on a logarithmic scale. All programs contribute, spanning a range of candidate fractions. The variation indicates that the outlier rate depends on observing conditions and target populations.}
  \label{fig:outliers_program}
\end{figure}

Finally, Fig.~\ref{fig:outliers_program} shows that candidates are drawn from all three DESI observing programs (Dark, Bright, and Backup). The candidate fraction varies by observing program, with higher values in Bright and Backup programs than in the Dark program. This distribution reflects the observing-program composition of the analyzed spectra, rather than a sensitivity limited to a single target or program class.

\subsection{Performance}
\label{subsec:performance}

\begin{figure}[t!]
  \centering
  \includegraphics[width=\linewidth]{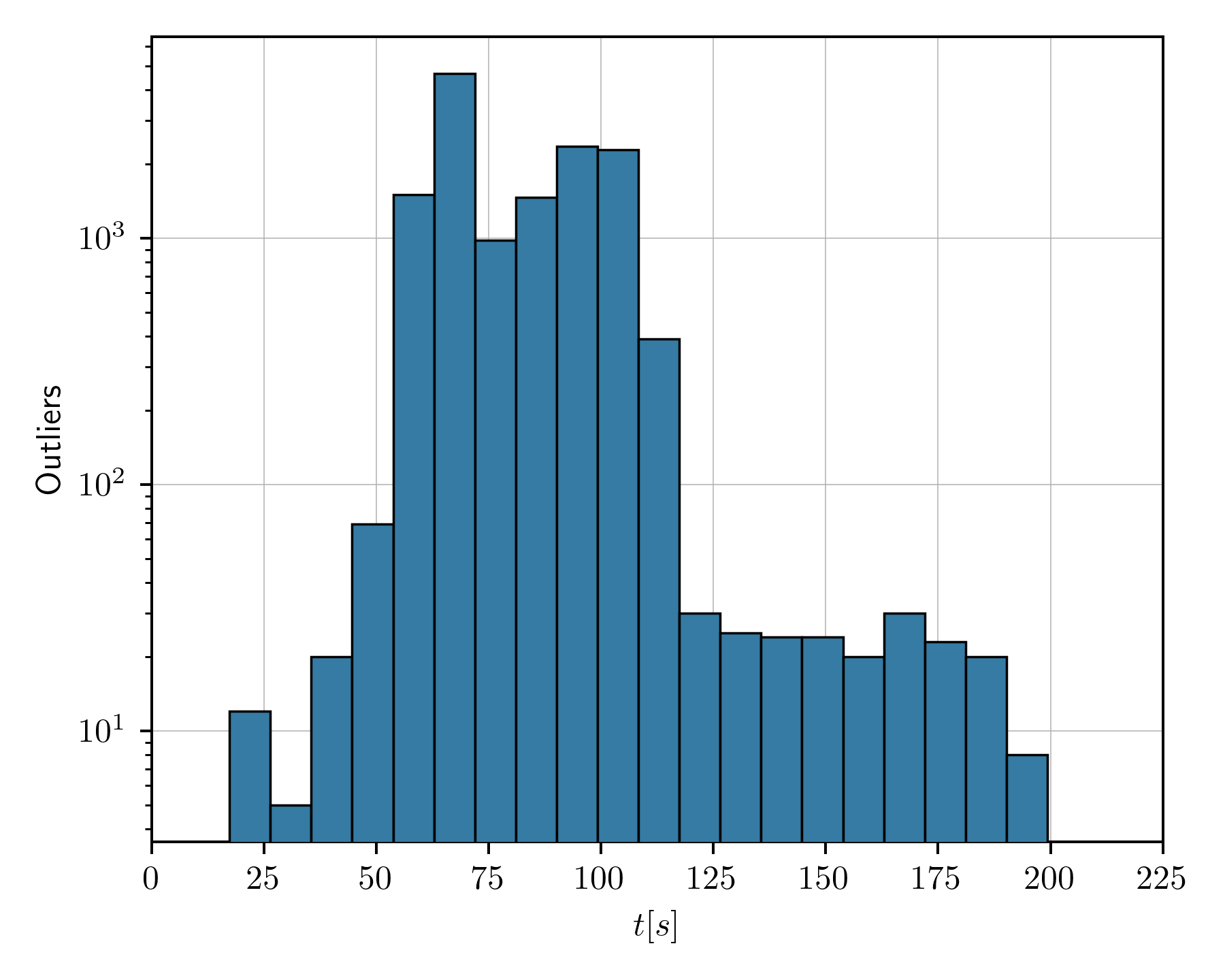}
  \caption{Histogram of wall-clock time per tile, $t_{\rm tile}$, in seconds across DR2. Runtimes concentrate around a typical value, with a minority of outlying tiles. The narrow runtime distribution indicates that the implementation has a predictable computational cost for routine tile-level processing.}
  \label{fig:perf_times}
\end{figure}

Figure~\ref{fig:perf_times} shows the distribution of wall-clock processing time per tile, denoted $t_{\rm tile}$. Runtimes cluster around a common value for most tiles, with a minority of noticeably slower cases. The scaling is approximately linear with the number of spectra processed in each tile, denoted $N_{\mathrm{spec}}$, which we summarize as
\begin{equation}
  t_{\rm tile} \approx \alpha\,N_{\mathrm{spec}} + \beta,
  \label{eq:runtime}
\end{equation}
where $\alpha$ is the time per spectrum and $\beta$ is a fixed overhead that depends on hardware and implementation. This linear behavior is expected for UMAP at fixed hyperparameters and confirms that the tile-by-tile strategy described in Section~\ref{sec:dataproc} scales to the full DR2 footprint without bottlenecks.

\begin{figure}[t!]
  \centering
  \includegraphics[width=\linewidth]{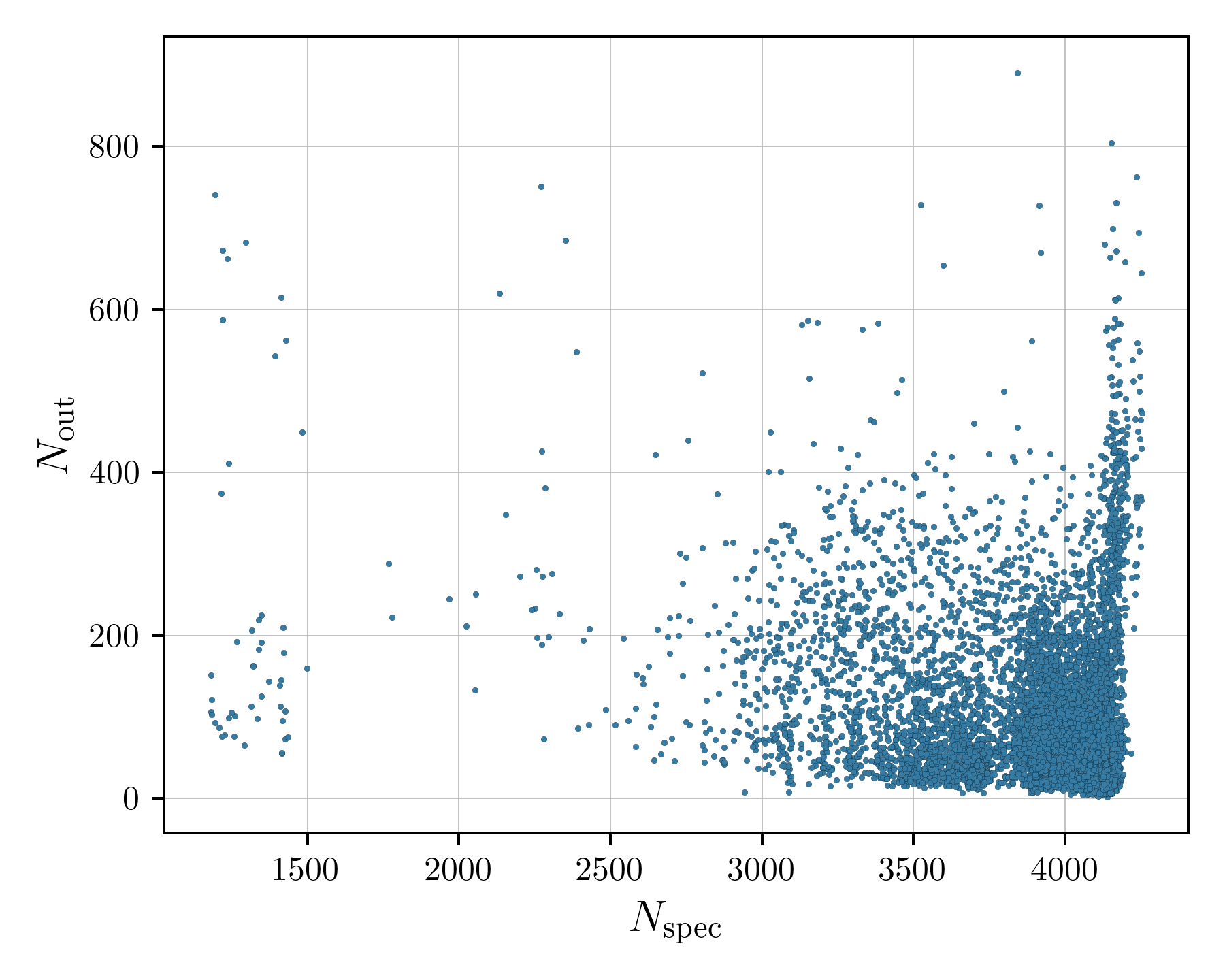}
  \caption{Candidate outlier count $N_{\mathrm{out}}$ versus number of spectra $N_{\mathrm{spec}}$ per tile. A broad positive association is present, with substantial scatter consistent with additional tile-level factors. The scatter at fixed $N_{\rm spec}$ shows that candidate counts are not determined by sample size alone.}
  \label{fig:out_vs_spec}
\end{figure}

Figure~\ref{fig:out_vs_spec} shows $N_{\mathrm{out}}$ versus $N_{\mathrm{spec}}$ per tile. A broad positive trend is present, as tiles with more spectra tend to produce more candidates, though the scatter is large. 
The discrete vertical bands in $N_{\mathrm{spec}}$ reflect common tile configurations in the survey, and the wide spread in $N_{\mathrm{out}}$ at fixed $N_{\mathrm{spec}}$ indicates that tile-level factors beyond sample size, such as observing conditions or local calibration quality, also affect the fractions.

\section{Visual Inspection of Candidate Outliers}
\label{sec:visinspection}

The Loa outlier catalog contains approximately 1.1 million entries, each identified by a \texttt{TARGETID}, \texttt{TILEID}, \texttt{NIGHT}, and \texttt{FIBER} number.
To perform the visual inspection, we restrict the sample to main-survey tiles (\texttt{SURVEY = main}), excluding commissioning (CMX) and survey-validation (SV) observations.

The resulting tiles are split by observing program: the dark program contains 6{,}671 tiles with a mean of $28.7 \pm 14.6$ outliers per tile (range 2--155), and the bright program contains 5{,}165 tiles with a mean of $89.7 \pm 48.4$ outliers per tile (range 10--380).
To obtain a representative but unbiased sample, we randomly select three tiles from each program; the selected tiles and their outlier counts are listed in Table~\ref{tab:vi_tiles_loa}.

\begin{table}[h]
\centering
\renewcommand{\arraystretch}{1.25}
  \setlength{\tabcolsep}{10pt}
\caption{Tiles selected for visual inspection from the Loa main survey.}
\label{tab:vi_tiles_loa}
\begin{tabular}{llrrr}
\hline
Program & TILEID & $N_{\rm out}$ & $N_{\rm anomaly}$ & ZWARN$\neq$0 \\
\hline
dark   & 1406  & 10  &  5  &  1 \\
dark   & 3473  & 30  & 23  &  2 \\
dark   & 3833  & 39  & 35  & 10 \\
\hline
bright & 25300 &  68 & 28  &  1 \\
bright & 22175 & 101 & 45  &  1 \\
bright & 22097 & 143 & 125 &  1 \\
\hline
\end{tabular}
\end{table}

The selected dark tiles (10--39 outliers) bracket the dark-program mean of 28.7, and the selected bright tiles (68--143 outliers) bracket the bright-program mean of 89.7.
The goal of this stage is not to reclassify every flagged spectrum, but to answer two questions: when anomalies are present, are they consistent with known reduction or calibration effects, and what fraction of the candidates show recognizable spectral anomalies. We address the first question in Section~\ref{subsec:pipeline} through representative examples, and quantify the second in Section~\ref{subsec:inspection}.

\subsection{Recurrent anomaly types}
\label{subsec:pipeline}

Visual inspection reveals three recurrent morphologies among the candidates, each pointing to a specific reduction or calibration effect.

\begin{figure*}[t!]
  \centering
  \includegraphics[width=\linewidth]{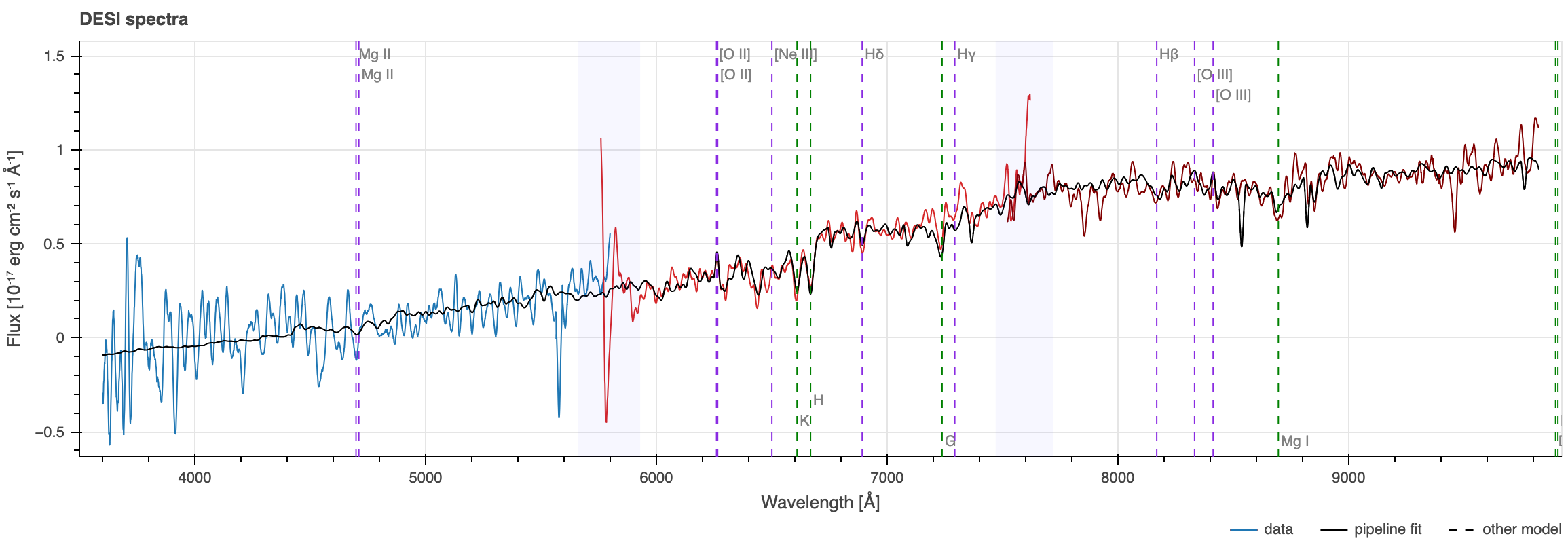}
  \caption{Galaxy targeted as an LRG and classified by \texttt{redrock} as
\texttt{SPECTYPE=GALAXY}, showing arm-join flux mismatches near
5{,}800\,\AA\ (B--R boundary) and 7{,}600\,\AA\ (R--Z boundary). The continuum shifts abruptly at each arm boundary while the \texttt{redrock} fit (black) remains smooth (\texttt{TARGETID}~39628374682901440, \texttt{TILEID}~3473,
  \texttt{FIBERID}~470).}
  \label{fig:jump_example}
\end{figure*}

\textbf{Arm-join discontinuity:}
The most common pattern is an abrupt step in flux level at the transition between spectrograph arms (B$\to$R or R$\to$Z), while the \texttt{redrock} model remains smooth across the boundary. 
This effect is consistent with imperfect relative flux calibration between arms. 
Figure~\ref{fig:jump_example} shows a galaxy (\texttt{TILEID}~3473,
\texttt{FIBERID}~470) where the continuum shifts abruptly at both arm boundaries, producing a visible step.

\begin{figure*}[t!]
  \centering
  \includegraphics[width=\linewidth]{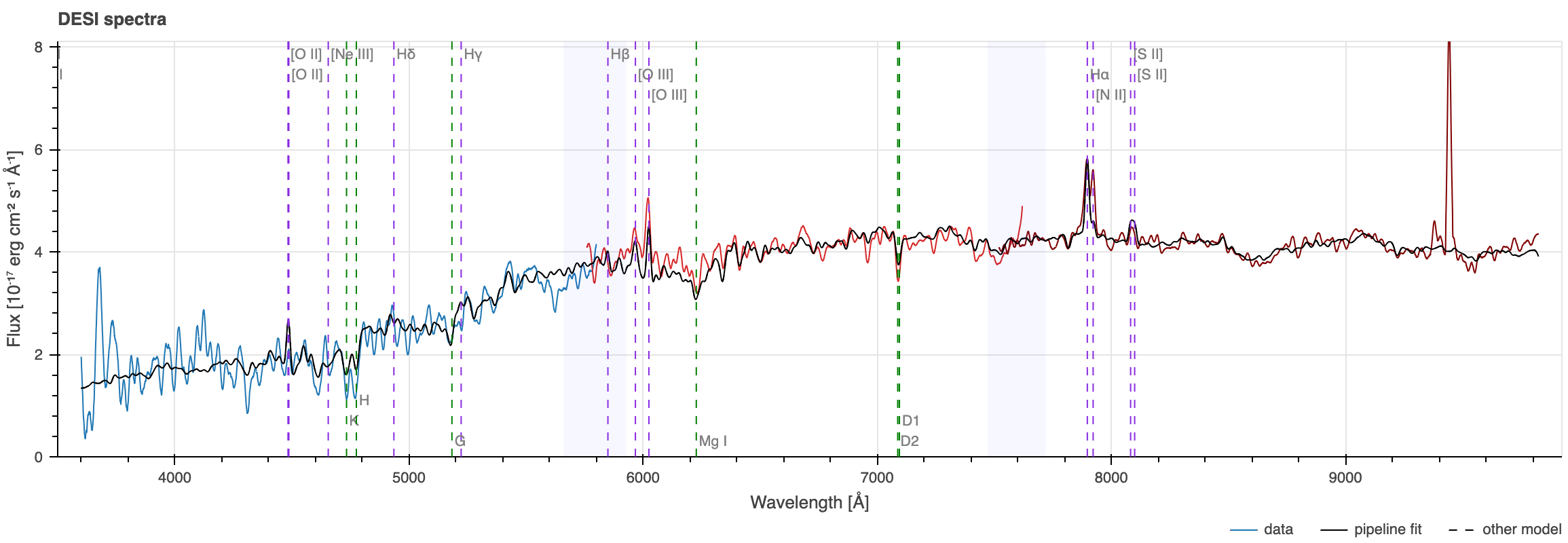}
  \caption{Galaxy targeted as a BGS object and classified by \texttt{redrock} as
\texttt{SPECTYPE=GALAXY}, showing spurious Z-arm emission. A strong, narrow peak appears in the Z arm with no associated \texttt{redrock} emission line at the fitted redshift, while the B and R arms follow  the pipeline fit (black) closely. The feature is consistent with sky-subtraction   residuals (\texttt{TARGETID}~39633232366404175, \texttt{TILEID}~22175,  \texttt{FIBERID}~3881).}
  \label{fig:emission_example}
\end{figure*}

\textbf{Spurious Z-arm emission:}
A second pattern shows narrow emission features confined to the Z arm that are not reproduced by the \texttt{redrock} best-fit model and have no counterpart emission line at the measured redshift.
These features are likely residuals from imperfect sky subtraction rather than intrinsic galaxy emission. Figure~\ref{fig:emission_example} shows a representative case (\texttt{TILEID}~22175, \texttt{FIBERID}~3881) where a sharp peak appears in the Z arm while the B and R arms are well described by the pipeline fit.

\begin{figure*}[t!]
  \centering
  \includegraphics[width=\linewidth]{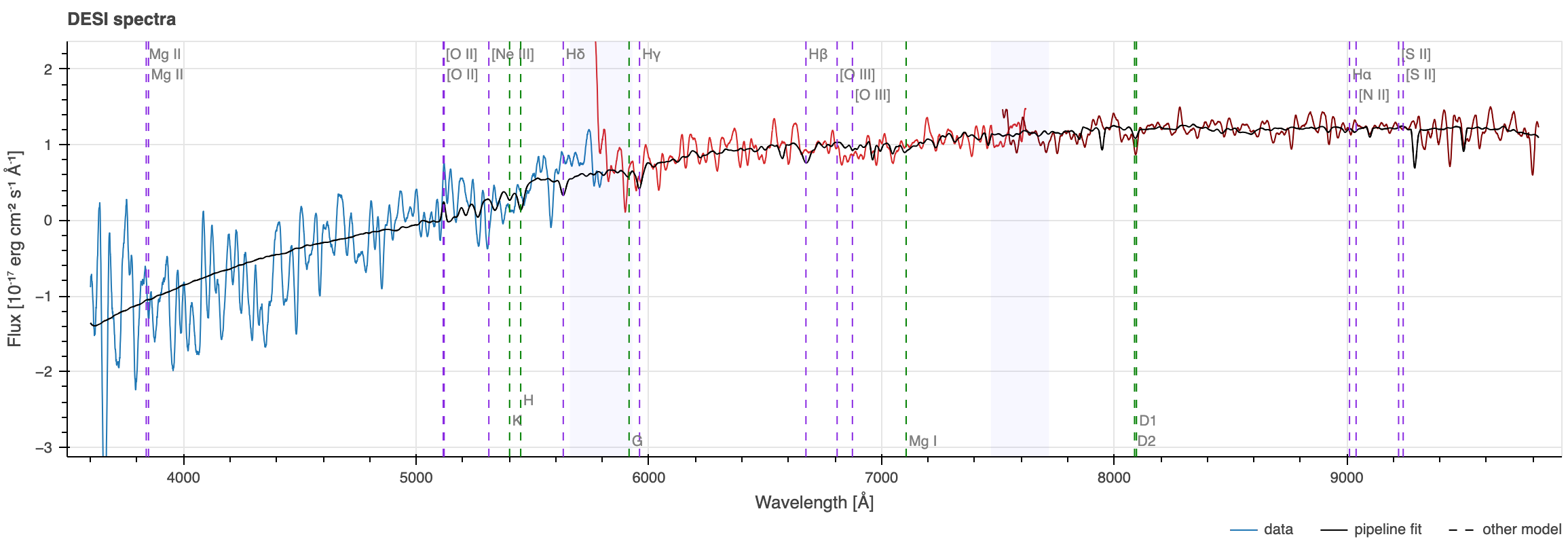}
  \caption{Galaxy targeted as an LRG and classified by \texttt{redrock} as
\texttt{SPECTYPE=GALAXY}, showing a depressed blue continuum. The B arm (blue) is
  depressed below zero over a broad wavelength interval, while the R and Z arms (red)
  remain positive. The black curve is the \texttt{redrock} best-fit model
  (\texttt{TARGETID}~39633251953804852, \texttt{TILEID}~25300,
  \texttt{FIBERID}~4393).}
  \label{fig:negblue_example}
\end{figure*}

\textbf{Negative blue continuum:}
A third class shows a systematically depressed, often negative, flux in the B arm over broad wavelength intervals, while the R and Z arms remain unaffected. Sky subtraction can over-subtract if the sky model is slightly inaccurate, pushing the residual continuum below zero.
Figure~\ref{fig:negblue_example} shows a representative case: a galaxy
(\texttt{TARGETID}~39633251953804852, \texttt{TILEID}~25300, \texttt{FIBERID}~4393) where the B-arm continuum is below zero while the redder arms remain positive.

These morphologies are naturally interpreted in the context of the petal and spectrograph architecture described in Section~\ref{subsec:instrument}, and are spatially consistent with the elevated candidate fractions near spectrograph-channel boundaries reported in Section~\ref{subsec:fplane}. This pattern indicates that the embedding is sensitive to small, repeatable differences in spectral properties across the focal plane. 
However, these differences should not be read as synonymous with ``wrong'': some outlier candidates may still correspond to spectra with correct redshifts but slightly different features relative to the local tile population.

\subsection{Quantitative assessment}
\label{subsec:inspection}

To measure the purity of the candidate selection, we visually assessed all 391 outlier spectra across the six tiles listed in Table~\ref{tab:vi_tiles_loa}.
Each spectrum was classified as ``Anomaly'' if it showed a recognizable spectroscopic problem or reduction artifact (arm-dependent flux gaps, spurious Z-arm emission or negative continuum), and as ``Average'' otherwise.

Of the 391 inspected spectra, 261 ($66.8^{+4.6}_{-5.0}$\%) show a visually identifiable  spectral anomaly, indicating substantial selection purity for the \texttt{UMAP}+FoF  pipeline. 
The confirmed anomaly fraction is notably higher in dark tiles 
($79.7^{+8.3}_{-10.5}$\%) than in bright tiles ($63.5^{+5.3}_{-5.6}$\%). 
This difference  likely reflects the distinct target compositions of the two programs: dark tiles are  dominated by faint extragalactic targets (LRGs, ELGs, and QSOs) whose spectra have lower  signal-to-noise ratios and are therefore more susceptible to arm-join discontinuities and  sky-subtraction residuals, while bright tiles include a larger fraction of high  signal-to-noise stellar and BGS spectra where reduction artifacts may be diluted by the  stronger continuum signal. 
Observing condition differences between the two programs 
(Section~\ref{subsec:instrument}) may also contribute to this trend.
\footnote{All uncertainties are 95\% Clopper--Pearson intervals~\citep{Clopper1934} ($\alpha = 0.05$), computed as the $\alpha/2 = 0.025$ and $1-\alpha/2 = 0.975$ quantiles of the Beta distribution, $B(0.025;\,k,\,n-k+1)$ and $B(0.975;\,k+1,\,n-k)$, where $k$ is the number of spectra classified as Bad and $n$ is the total number of candidates inspected in each program.}.

In contrast, only $4.1^{+2.5}_{-1.7}$\% of the inspected spectra carry a non-zero \texttt{ZWARN} flag from the reduction pipeline, consistent with the survey-wide rate of $5.94$\% computed over all dark and bright main-survey outliers, which confirms that our six-tile sample is representative.

This large gap between the visually confirmed anomaly rate and the \texttt{ZWARN} rate confirms that the embedding-based selection provides a complementary diagnostic to the pipeline's own quality bitmasks and recovers a substantial population of problematic spectra that standard diagnostics miss.

Applying the program-specific bad fractions to the full Loa main-survey catalog (191{,}411 dark and 463{,}543 bright outlier spectra, totaling $\sim$655{,}000 entries), we estimate that approximately $218{,}000^{+32{,}000}_{-31{,}000}$ outliers show no identifiable spectral anomaly and may therefore correspond to genuine atypical spectra in the context of DESI.

\section{Discussion}
\label{sec:disc}

The results in Sections~\ref{sec:results} and~\ref{sec:visinspection} show that the UMAP+FoF pipeline produces a stable and informative partition of Loa spectra. Here we summarize what these results mean in practice, discuss the limitations of the approach, and outline directions for future work.

\subsection{What the results tell us}

At the tile level, candidate fractions follow a long-tailed distribution: most tiles have low outlier fractions, while a minority show substantially elevated values. This pattern remains after normalizing by the number of spectra observed per tile
(Section~\ref{subsec:performance}), indicating that the variation is not driven solely by differences in the number of spectra inspected. The focal plane maps (Section~\ref{subsec:fplane}) show that these differences are spatially structured: elevated fractions concentrate near spectrograph boundaries.

This does not necessarily imply a systematic calibration failure between petals. 
Instead, it shows that the UMAP+FoF selection is sensitive to repeatable instrumental differences across the focal plane, including variations in throughput, noise, spectral resolution, and arm-dependent calibration residuals.
Some of these differences produce visibly problematic spectra, such as the arm-join discontinuities and sky-subtraction residuals identified during visual inspection (Section~\ref{subsec:pipeline}), while others may remain compatible with successful redshift measurement.

The quantitative assessment in Section~\ref{subsec:inspection} supports this interpretation. Of the 391 candidates reviewed, $66.8^{+4.6}_{-5.0}$\% show a visually identifiable anomaly, yet only $4.1^{+2.5}_{-1.7}$\% carry a non-zero \texttt{ZWARN} flag. 
This gap confirms that the embedding-based selection provides a complementary diagnostic to the pipeline's own quality bitmasks: it recovers a large population of problematic spectra that standard diagnostics miss. 
The anomaly fraction is notably higher in dark tiles ($79.7^{+8.3}_{-10.5}$\%) than in bright tiles ($63.5^{+5.3}_{-5.6}$\%), suggesting that the method's sensitivity varies with target population and observing conditions.

\subsection{Role of the method}

The pipeline functions as a scalable QA monitor rather than a classifier. 
It does not assign labels or correct spectra; it produces a ranked watch list that concentrates human review where it is most likely to be useful. 
By building embeddings per tile under fixed hyperparameters (Table~\ref{table:params}), the outlier fractions are directly comparable across tiles, programs, and data releases. 
This makes the method suitable for tracking changes in data quality as the pipeline evolves, for example, the kind of subtle processing fix between the \textit{Kibo} and \textit{Loa} productions described in Section~\ref{sec:desi}.

In practice, the dominant cost is not the UMAP+FoF computation, but the follow-up visual inspection. 
The manual inspection performed for this work took a few hours for the reviewed sample, since each candidate spectrum had to be opened, assessed, and assigned to a qualitative category. 
However, this cost should decrease in an operational setting. 
If the pipeline is run routinely after nightly reductions, the candidate list would be generated incrementally and only a small number of newly flagged spectra would require inspection at a time, rather than reviewing a large accumulated sample from an entire data release. 
The method is therefore most practical as a daily or near-real-time QA monitor: it reduces the inspection problem from the full set of observed spectra to a prioritized subset, while still leaving the final interpretation to human reviewers.

\subsection{Limitations}

The method has several limitations. 
It operates on coadded, calibrated spectra without any denoising or continuum normalization. 
This is intentional, since reduction artifacts are part of the signal we aim to identify, but it also means that per-arm systematics can dominate the embedding and push otherwise normal spectra into outlier regions. 

The results also depend on the UMAP and FoF configuration. 
The UMAP hyperparameters and distance metric, although fixed following \citet{Suarez_Art_2023}, shape the relative prominence of structures in the embedding in ways that are not always transparent; different choices of $N_n$ or $M_e$ could reveal different groupings or suppress the ones we observe. 
Similarly, FoF uses a single linking length across the entire embedding. 
In regions where the local density varies substantially, this fixed threshold creates a trade-off between isolating compact anomalous groups and over-fragmenting the dense core. 
As FoF is connectivity-based, it does not assign a continuous anomaly score, so candidates near the boundary between the dense background and sparse outskirts can depend on the adopted linking length.

Finally, tile-level outliers are local by construction. 
A DESI tile can contain a heterogeneous mix of target classes, redshifts, observing conditions, and spectral morphologies. 
As a result, a regular but rare spectrum may be flagged because it is uncommon within that tile, while coherent tile-wide reduction or calibration issues may be less visible, as they become part of the local reference distribution.

\subsection{Operational recommendations and future work}

The results suggest three concrete QA actions. 
First, tiles in the high-fraction tail of the outlier distribution (Fig.~\ref{fig:fracs}) should be prioritized for targeted review. Second, fibers near the edges of the 500-fiber petal/spectrograph blocks (\texttt{FIBERID}~$\approx 500\,k$) should receive dedicated QA checks, since they account for a disproportionate share of the candidates. 
Third, program-stratified outlier fractions (Fig.~\ref{fig:outliers_program})
should be reported routinely to separate astrophysical diversity from reduction-driven anomalies.

Looking ahead, several extensions would strengthen the framework. Analyzing individual exposures before coaddition would help isolate whether anomalies are present in a single exposure or accumulate across visits. 
Additionally, running the pipeline on successive data releases and comparing the resulting outlier maps would provide a direct, data-driven record of how pipeline changes affect data quality across the survey footprint.

Alternative distance metrics designed for spectroscopic data, such as metrics that weight emission-line regions differently from the continuum, could improve sensitivity to specific classes of artifacts. 
Score-based anomaly detectors, including Local Outlier Factor \citep{Breunig_2000a, Breunig_2000b} or Isolation Forest \citep{Liu_2008}, could also provide complementary rankings based on local density contrast or isolation rather than discrete connected components.

Another useful extension would be to explore higher-dimensional UMAP embeddings. 
Three-dimensional projections, for example, may retain additional neighborhood information while remaining visually inspectable. 
In this work, FoF is applied only to the two-dimensional UMAP projection; future analyses should test whether the resulting connected components are also close neighbors in the original high-dimensional flux space or in higher-dimensional embeddings.

Our aim is not to define a global anomaly taxonomy for all DESI spectra, but to develop an unsupervised quality-assessment tool that can be applied soon after nightly reductions and is naturally aligned with the observing workflow. 
Alternative embeddings trained on subsets defined by \texttt{redrock} class, redshift range, target class, observing program, or combinations of these quantities could provide complementary anomaly definitions. 
Such analyses would be valuable for distinguishing astrophysically rare spectra from reduction artifacts, and for identifying anomalies relative to more homogeneous parent populations. 
We leave these global or class-conditioned extensions to future work.

\section{Conclusions}
\label{sec:concl}

We have applied an unsupervised anomaly-detection framework to 58{,}291{,}334 coadded spectra from DESI DR2, the largest spectroscopic dataset analyzed with this type of method to date. 
The pipeline combines UMAP dimensionality reduction with Friends-of-Friends clustering, operates tile by tile under a fixed hyperparameter configuration (Table~\ref{table:params}), and preserves the standard DESI identifiers needed for direct retrieval and audit. 
Our main findings are as follows.

\begin{itemize}

  \item The pipeline cleanly separates a dense core of typical spectra from a compact set  of small components and singletons in each tile embedding. Across DR2, the mean outlier fraction is $\sim$1.96\% per tile.

  \item Candidate rates show a clear dependence on observing program. The mean outlier fraction increases from the Dark program ($\langle f_{\rm out}\rangle_{\rm Dark}=0.76\%$) to the Bright program ($\langle f_{\rm out}\rangle_{\rm Bright}=2.36\%$), reaching its highest value in the Backup program ($\langle f_{\rm out}\rangle_{\rm Backup}=7.31\%$). This pattern indicates that observing conditions, rather than tile sample size alone, are a dominant factor of the outlier fraction.

  \item Focal plane maps reveal spatially structured anomaly rates, with elevated counts  concentrated near spectrograph boundaries. The quasi-periodic modulation with  a $\sim$500-fiber period is consistent with the ten-camera architecture of the instrument.

  \item Visual inspection of 391 candidates across six representative tiles --- three  dark and three bright --- shows that $66.8^{+4.6}_{-5.0}$\% of the outliers exhibit identifiable spectral anomalies. 
  The anomaly fraction is higher in dark tiles ($79.7^{+8.3}_{-10.5}$\%) than in bright tiles ($63.5^{+5.3}_{-5.6}$\%), reflecting differences in target populations and observing conditions. 
  The three most common morphologies are a negative blue continuum, arm-join discontinuities, and spurious Z-arm emission, all consistent with known reduction and calibration effects.

  \item Only $4.1^{+2.5}_{-1.7}$\% of the inspected candidates carry a non-zero \texttt{ZWARN} flag from the standard pipeline. This gap with respect to the visually confirmed anomaly rate demonstrates that the embedding-based selection provides a complementary diagnostic to existing quality bitmasks and recovers a substantial population of problematic spectra that would otherwise go undetected.

  \item Extrapolating to the full Loa main-survey catalog, we estimate that approximately $218{,}000^{+32{,}000}_{-31{,}000}$ outliers show no identifiable reduction anomalies  and may correspond to genuine outliers in the context of DESI.
  This aspect will be explored in future publications. 
\end{itemize}

Taken together, these results show that the UMAP+FoF framework functions as a practical, survey-scale QA monitor: it concentrates human review on the spectra most likely to be problematic, produces fractions that are directly comparable across tiles and data releases, and provides a reproducible record of reduction performance.

Future work will focus on three extensions. 
First, applying the pipeline to successive DESI data releases will allow direct, data-driven tracking of how pipeline changes, such as the coaddition fix between \textit{Kibo} and \textit{Loa} described in Section~\ref{sec:desi}, propagate into the anomaly maps. 
Second, developing spectroscopically motivated metrics that may improve the sensitivity to specific classes of reduction artifacts.
Third, couple the pipeline to the existing daily QA processs that support the DESI survey operations.

The approach is directly applicable to other current and upcoming multi-object spectroscopic surveys, including WEAVE \citep{weave2024}, SDSS-V \citep{sdssv2020}, 4MOST \citep{4most2019}, the Subaru Prime Focus Spectrograph \citep{pfs2016}, and MOONS \citep{moons2020}. For next-generation facilities such as Spec-S5 \citep{specs5}, MUST \citep{must2024}, and WST \citep{wst2024}, which plan to observe over 100 million targets, automated and scalable quality assessment will be a basic operational requirement.

\section*{Data Availability}
The full list of outlier candidates from our analysis will be made public alongside  DESI Data Release 2 (see \url{https://data.desi.lbl.gov/doc/releases/}) as a Value-Added  Catalog. 
Each entry is identified by \texttt{TILEID}, \texttt{NIGHT}, \texttt{TARGETID},  and \texttt{FIBERID}. 
The data underlying the figures in this paper are available in the 
Zenodo repository at \url{https://doi.org/10.5281/zenodo.19373353}, with the exception  of Figures \ref{fig:jump_example}, \ref{fig:emission_example} and \ref{fig:negblue_example}, which were generated  using a spectrum visualizer internal to the DESI collaboration. 
All code to reproduce  the analysis and figures is available at 
\url{https://github.com/ValeriaTorresG/AssessingDesiData}.

\section*{Acknowledgments}
SP is supported by the International Gemini Observatory, a program of NSF NOIRLab, which is managed by the Association of Universities for Research in Astronomy (AURA) under a cooperative agreement with the U.S. National Science Foundation, on behalf of the Gemini partnership of Argentina, Brazil, Canada, Chile, the Republic of Korea, and the United States of America.

This material is based upon work supported by the U.S. Department of Energy (DOE), Office of Science, Office of High-Energy Physics, under Contract No. DE--AC02--05CH11231, and by the National Energy Research Scientific Computing Center, a DOE Office of Science User Facility under the same contract. Additional support for DESI was provided by the U.S. National Science Foundation (NSF), Division of Astronomical Sciences under Contract No. AST-0950945 to the NSF's National Optical-Infrared Astronomy Research Laboratory; the Science and Technology Facilities Council of the United Kingdom; the Gordon and Betty Moore Foundation; the Heising-Simons Foundation; the French Alternative Energies and Atomic Energy Commission (CEA); the Secretariat of Science, Humanities, Technology and Innovation (SECIHTI) of Mexico; the Ministry of Science, Innovation and Universities of Spain (MICIU/AEI/10.13039/501100011033), and by the DESI Member Institutions: \url{https://www.desi.lbl.gov/collaborating-institutions}. Any opinions, findings, and conclusions or recommendations expressed in this material are those of the author(s) and do not necessarily reflect the views of the U. S. National Science Foundation, the U. S. Department of Energy, or any of the listed funding agencies.

The authors are honored to be permitted to conduct scientific research on I'oligam Du'ag (Kitt Peak), a mountain with particular significance to the Tohono O'odham Nation.


\bibliographystyle{aasjournal}
\bibliography{oja_template}

@article{Clopper1934,
  author  = {Clopper, C. J. and Pearson, E. S.},
  title   = {The use of confidence or fiducial limits illustrated in the case of the binomial},
  journal = {Biometrika},
  year    = {1934},
  volume  = {26},
  number  = {4},
  pages   = {404--413},
  doi     = {10.1093/biomet/26.4.404}
}

@INPROCEEDINGS{umap-proceedings,
       author = {{Su{\'a}rez-P{\'e}rez}, John F. and {Forero-Romero}, Jaime},
        title = "{Assessing the Quality of Massive Spectroscopic Surveys with Unsupervised Machine Learning}",
    booktitle = {IAU Symposium},
         year = 2025,
       editor = {{McIver}, J. and {Mahabal}, A. and {Fluke}, C.},
       series = {IAU Symposium},
       volume = {19},
        month = aug,
        pages = {91-94},
          doi = {10.1017/S1743921322003568},
       adsurl = {https://ui.adsabs.harvard.edu/abs/2025IAUS..368...91S}
}

@ARTICLE{desiops,
       author = {{Schlafly}, Edward F. and {Kirkby}, David and {Schlegel}, David J. and {Myers}, Adam D. and {Raichoor}, Anand and {Dawson}, Kyle and {Aguilar}, Jessica and {Allende Prieto}, Carlos and {Bailey}, Stephen and {BenZvi}, Segev and {Bermejo-Climent}, Jose and {Brooks}, David and {de la Macorra}, Axel and {Dey}, Arjun and {Doel}, Peter and {Fanning}, Kevin and {Font-Ribera}, Andreu and {Forero-Romero}, Jaime E. and {Garc{\'\i}a-Bellido}, Juan and {Gontcho A Gontcho}, Satya and {Guy}, Julien and {Hahn}, ChangHoon and {Honscheid}, Klaus and {Ishak}, Mustapha and {Juneau}, St{\'e}phanie and {Kehoe}, Robert and {Kisner}, Theodore and {Kremin}, Anthony and {Landriau}, Martin and {Lang}, Dustin A. and {Lasker}, James and {Levi}, Michael E. and {Magneville}, Christophe and {Manser}, Christopher J. and {Martini}, Paul and {Meisner}, Aaron M. and {Miquel}, Ramon and {Moustakas}, John and {Newman}, Jeffrey A. and {Nie}, Jundan and {Palanque-Delabrouille}, Nathalie. and {Percival}, Will J. and {Poppett}, Claire and {Rockosi}, Constance and {Ross}, Ashley J. and {Rossi}, Graziano and {Tarl{\'e}}, Gregory and {Weaver}, Benjamin A. and {Y{\`e}che}, Christophe and {Zhou}, Rongpu and {DESI Collaboration}},
        title = "{Survey Operations for the Dark Energy Spectroscopic Instrument}",
      journal = {\aj},
         year = 2023,
        month = dec,
       volume = {166},
       number = {6},
          eid = {259},
        pages = {259},
          doi = {10.3847/1538-3881/ad0832},
archivePrefix = {arXiv},
       eprint = {2306.06309},
 primaryClass = {astro-ph.CO},
       adsurl = {https://ui.adsabs.harvard.edu/abs/2023AJ....166..259S}
}

@article{desi2025dr2bao,
  author       = {{DESI Collaboration}},
  title        = {{DESI DR2 Results II}: Measurements of Baryon Acoustic Oscillations
                  and Cosmological Constraints},
  journal      = {Phys. Rev. D},
  volume       = {112},
  pages        = {083515},
  year         = {2025},
  eprint       = {2503.14738},
  archivePrefix= {arXiv},
  primaryClass = {astro-ph.CO}
}

@article{desi2025lya,
  author       = {{DESI Collaboration}},
  title        = {{DESI DR2 Results I}: Baryon Acoustic Oscillations from the
                  {Lyman Alpha Forest}},
  journal      = {Phys. Rev. D},
  volume       = {112},
  pages        = {083514},
  year         = {2025},
  eprint       = {2503.14739},
  archivePrefix= {arXiv},
  primaryClass = {astro-ph.CO}
}

@article{desi2026dr1,
doi = {10.3847/1538-3881/ae4c43},
url = {https://doi.org/10.3847/1538-3881/ae4c43},
year = {2026},
month = {apr},
publisher = {The American Astronomical Society},
volume = {171},
number = {5},
pages = {285},
author = {{DESI Collaboration} and Abdul Karim, M. and Adame, A. G. and Aguado, D. and Aguilar, J. and Ahlen, S. and Alam, S. and Aldering, G. and Alexander, D. M. and Alfarsy, R. and Allen, L. and Allende Prieto, C. and Alves, O. and Anand, A. and Andrade, U. and Armengaud, E. and Avila, S. and Aviles, A. and Awan, H. and Bailey, S. and Baleato Lizancos, A. and Ballester, O. and Bault, A. and Bautista, J. and Bean, R. and Behera, J. and BenZvi, S. and Beraldo e Silva, L. and Bermejo-Climent, J. R. and Beutler, F. and Bianchi, D. and Blake, C. and Blum, R. and Bolton, A. S. and Bonici, M. and Brieden, S. and Brodzeller, A. and Brooks, D. and Buckley-Geer, E. and Burtin, E. and Byström, A. and Canning, R. and Carnero Rosell, A. and Carr, A. and Carrilho, P. and Casas, L. and Castander, F. J. and Cereskaite, R. and Cervantes-Cota, J. L. and Chaussidon, E. and Chaves-Montero, J. and Chen, S. and Chen, X. and Circosta, C. and Claybaugh, T. and Cole, S. and Cooper, A. P. and Cousinou, M.-C. and Cuceu, A. and Davis, T. M. and Dawson, K. S. and de Belsunce, R. and de la Cruz, R. and de la Macorra, A. and de Mattia, A. and Deiosso, N. and Della Costa, J. and Demina, R. and Demirbozan, U. and DeRose, J. and Dey, A. and Dey, B. and Ding, J. and Ding, Z. and Doel, P. and Douglass, K. and Dowicz, M. and Ebina, H. and Edelstein, J. and Eisenstein, D. J. and Elbers, W. and Emas, N. and Escoffier, S. and Fagrelius, P. and Fan, X. and Fanning, K. and Favole, G. and Fawcett, V. A. and Fernández-García, E. and Ferraro, S. and Findlay, N. and Font-Ribera, A. and Forero-Romero, J. E. and Forero-Sánchez, D. and Frenk, C. S. and Gänsicke, B. T. and Galbany, L. and García-Bellido, J. and Garcia-Quintero, C. and Garrison, L. H. and Gaztañaga, E. and Gil-Marín, H. and Gloudemans, A. and Gnedin, O. Y. and Gontcho A Gontcho, S. and Gonzalez, D. and Gonzalez-Morales, A. X. and Gonzalez-Perez, V. and Gordon, C. and Graur, O. and Green, D. and Gruen, D. and Gsponer, R. and Guandalin, C. and Gutierrez, G. and Guy, J. and Hahn, C. and Han, J. J. and Han, J. and He, S. and Herrera-Alcantar, H. K. and Heydenreich, S. and Honscheid, K. and Hou, J. and Howlett, C. and Huterer, D. and Iršič, V. and Ishak, M. and Jacques, A. and Jiang, L. and Jimenez, J. and Jing, Y. P. and Joachimi, B. and Joudaki, S. and Joyce, R. and Jullo, E. and Juneau, S. and Karaçaylı, N. G. and Karim, T. and Kehoe, R. and Kent, S. and Khederlarian, A. and Kirkby, D. and Kisner, T. and Kitaura, F.-S. and Kizhuprakkat, N. and Kong, H. and Koposov, S. E. and Kremin, A. and Krolewski, A. and Lahav, O. and Lai, Y. and Lamman, C. and Lan, T.-W. and Landriau, M. and Lang, D. and Lange, J. U. and Lasker, J. and Le Goff, J.M. and Le Guillou, L. and Leauthaud, A. and Levi, M. E. and Li, S. and Li, T. S. and Liu, W. and Lodha, K. and Lokken, M. and Luo, Y. and Luo, Y. and Magneville, C. and Manera, M. and Manser, C. J. and Margala, D. and Martini, P. and Maus, M. and McCullough, J. and McDonald, P. and Medina, G. E. and Medina-Varela, L. and Meisner, A. and Mena-Fernández, J. and Menegas, A. and Meneses-Rizo, J. and Mezcua, M. and Miquel, R. and Montero-Camacho, P. and Moon, J. and Moustakas, J. and Muñoz-Gutiérrez, A. and Mu noz-Santos, D. and Myers, A. D. and Myles, J. and Nadathur, S. and Najita, J. and Napolitano, L. and Newman, J. A. and Nikakhtar, F. and Nikutta, R. and Niz, G. and Noriega, H. E. and Nugent, P. and Padmanabhan, N. and Paillas, E. and Palanque-Delabrouille, N. and Palmese, A. and Pan, J. and Pan, Z. and Parkinson, D. and Peacock, J. A. and Ibanez, M. P. and Percival, W. J. and Pérez-Fernández, A. and Pérez-Ràfols, I. and Peterson, P. and Piat, J. and Pieri, M. M. and Pinon, M. and Poppett, C. and Porredon, A. and Prada, F. and Pucha, R. and Qin, F. and Rabinowitz, D. and Raichoor, A. and Ramírez-Pérez, C. and Ramirez-Solano, S. and Rashkovetskyi, M. and Ravoux, C. and Ried Guachalla, B. and Riley, A. H. and Rocher, A. and Rockosi, C. and Rohlf, J. and Rosado-Marín, A. J. and Ross, A. J. and Ross, C. and Rossi, G. and Ruggeri, R. and Ruhlmann-Kleider, V. and Sabiu, C. G. and Said, K. and Sailer, N. and Saintonge, A. and Salcedo Hernandez, Y. and Samushia, L. and Sanchez, E. and Sanders, N. and Sandford, N. and Satyavolu, S. and Saulder, C. and Saydjari, A. K. and Schlafly, E. F. and Schlegel, D. and Scholte, D. and Schubnell, M. and Semenaite, A. and Seo, H. and Shafieloo, A. and Sharples, R. and Silber, J. and Sinigaglia, F. and Siudek, M. and Slepian, Z. and Smith, A. and Soumagnac, M. and Sprayberry, D. and Suárez-Pérez, J. and Swanson, J. and Tan, T. and Tarlé, G. and Taylor, P. and Thomas, G. and Tojeiro, R. and Turner, R. J. and Turner, W. and Ureña-López, L. A. and Vaisakh, R. and Valluri, M. and Valogiannis, G. and Vargas-Magaña, M. and Verde, L. and Vielzeuf, P. and Walther, M. and Wang, B. and Wang, M. S. and Wang, W. and Weaver, B. A. and Weaverdyck, N. and Wechsler, R. H. and Weinberg, D. H. and White, M. and Whitford, A. and Wolfson, M. and Yang, J. and Yèche, C. and Youles, S. and Yu, J. and Yuan, S. and Zaborowski, E. A. and Zarrouk, P. and Zhang, H. and Zhao, C. and Zhao, R. and Zheng, Z. and Zhou, C. and Zhou, R. and Zhou, Y. and Zou, H. and Zou, S. and Zu, Y.},
title = {Data Release 1 of the Dark Energy Spectroscopic Instrument},
journal = {The Astronomical Journal}
}

@ARTICLE{wst2024,
       author = {{Bacon}, Roland and {Maineiri}, Vincenzo and {Randich}, Sofia and {Cimatti}, Andrea and {Kneib}, Jean-Paul and {Brinchmann}, Jarle and {Ellis}, Richard and {Tolstoi}, Eline and {Smiljanic}, Rodolfo and {Hill}, Vanessa and {Anderson}, Richard and {Sanchez Saez}, Paula and {Opitom}, Cyrielle and {Bryson}, Ian and {Dierickx}, Philippe and {Garilli}, Bianca and {Gonzalez}, Oscar and {de Jong}, Roelof and {Lee}, David and {Mieske}, Steffen and {Otarola}, Angel and {Schipani}, Pietro and {Travouillon}, Tony and {Vernet}, Joel and {Bryant}, Julia and {Casali}, Marc and {Colless}, Matthew and {Couch}, Warrick and {Driver}, Simon and {Fontana}, Adriano and {Lehnert}, Matthew and {Magrini}, Laura and {Montet}, Ben and {Pasquini}, Luca and {Roth}, Martin and {Sanchez-Janssen}, Ruben and {Steinmetz}, Matthias and {Tresse}, Laurence and {Yeche}, Christophe and {Ziegler}, Bodo},
        title = "{WST -- Widefield Spectroscopic Telescope: Motivation, science drivers and top-level requirements for a new dedicated facility}",
      journal = {arXiv e-prints},
         year = 2024,
        month = may,
          eid = {arXiv:2405.12518},
        pages = {arXiv:2405.12518},
          doi = {10.48550/arXiv.2405.12518},
archivePrefix = {arXiv},
       eprint = {2405.12518},
 primaryClass = {astro-ph.IM},
       adsurl = {https://ui.adsabs.harvard.edu/abs/2024arXiv240512518B}
}

@ARTICLE{specs5,
       author = {{Besuner}, Robert and {Dey}, Arjun and {Drlica-Wagner}, Alex and {Ebina}, Haruki and {Fernandez Moroni}, Guillermo and {Ferraro}, Simone and {Forero-Romero}, Jaime and {Honscheid}, Klaus and {Jelinsky}, Pat and {Lang}, Dustin and {Levi}, Michael and {Martini}, Paul and {Myers}, Adam and {Palanque-Delabrouille}, Nathalie and {Panda}, Swayamtrupta and {Poppett}, Claire and {Sailer}, Noah and {Schlegel}, David and {Shafieloo}, Arman and {Silber}, Joseph and {White}, Martin and {Abbott}, Timothy and {Allen}, Lori and {Avila}, Santiago and {Avil{\'e}s}, Roberto and {Bailey}, Stephen and {Bault}, Abby and {Bouri}, Mohamed and {Boutsia}, Konstantina and {Burtin}, Eienne and {Chierchie}, Fernando and {Coulton}, William and {Dawson}, Kyle and {Dey}, Biprateep and {Dor{\'e}}, Olivier and {Dunlop}, Patrick and {Eisenstein}, Daniel and {Emanuele}, Castorina and {Escoffier}, Stephanie and {Estrada}, Juan and {Fagrelius}, Parker and {Fanning}, Kevin and {Fanning}, Timothy and {Font-Ribera}, Andreu and {Frieman}, Joshua and {Galal}, Malak and {Gluscevic}, Vera and {Gontcho}, Satya Gontcho A and {Green}, Daniel and {Gutierrez}, Gaston and {Guy}, Julien and {Hashemi}, Kevan and {Heathcote}, Steve and {Holland}, Steve and {Hou}, Jiamin and {Huterer}, Dragan and {Irigoyen Gimenez}, Blas and {Ivanov}, Mikhail and {Joyce}, Richard and {Jullo}, Eric and {Juneau}, Stephanie and {Juramy}, Claire and {Karcher}, Armin and {Kent}, Stephen and {Kirkby}, David and {Kneib}, Jean-Paul and {Krause}, Elisabeth and {Krolewski}, Alex and {Lahav}, Ofer and {Lapi}, Agustin and {Leauthaud}, Alexie and {Lewandowski}, Matthew and {Li}, Ting and {Lin}, Kenneth and {Loverde}, Marilena and {MacBride}, Sean and {Magneville}, Christophe and {Marshall}, Jennifer and {McDonald}, Patrick and {Miller}, Timothy and {Moustakas}, John and {M{\"u}nchmeyer}, Moritz and {Najita}, Joan and {Newman}, Jeff and {Percival}, Will and {Philcox}, Oliver and {Pires}, Priscila and {Raichoor}, Anand and {Roach}, Brandon and {Rockosi}, Constance and {Rombach}, Maxime and {Ross}, Ashley and {Sanchez}, Eusebio and {Schmidt}, Luke and {Schubnell}, Michael and {Sebok}, Rebekah and {Seljak}, Uros and {Silverstein}, Eva and {Slepian}, Zachay and {Stone}, Chris and {Stupak}, Robert and {Tarl{\'e}}, Gregory and {Li}, Ting and {Tyas}, Luke and {Vargas-Maga{\~n}a}, Mariana and {Walker}, Alistair and {Wenner}, Nicholas and {Y{\`e}che}, Christophe and {Zhang}, Yuanyuan and {Zhou}, Rongpu},
        title = "{The Spectroscopic Stage-5 Experiment}",
      journal = {arXiv e-prints},
         year = 2025,
        month = mar,
          eid = {arXiv:2503.07923},
        pages = {arXiv:2503.07923},
          doi = {10.48550/arXiv.2503.07923},
archivePrefix = {arXiv},
       eprint = {2503.07923},
 primaryClass = {astro-ph.CO},
       adsurl = {https://ui.adsabs.harvard.edu/abs/2025arXiv250307923B}
}

@article{weave2024,
  title={wide-field, multiplexed, spectroscopic facility WEAVE: Survey design, overview, and simulated implementation},
  author={Dalton, G. and others},
  journal={Monthly Notices of the Royal Astronomical Society},
  volume={530},
  number={3},
  pages={2688--2730},
  year={2024},
  publisher={Oxford University Press}
}

@ARTICLE{sdssv2020,
       author = {{Kollmeier}, Juna A. and {Zasowski}, Gail and {Rix}, Hans-Walter and {Johns}, Matt and {Anderson}, Scott F. and {Drory}, Niv and {Johnson}, Jennifer A. and {Pogge}, Richard W. and {Bird}, Jonathan C. and {Blanc}, Guillermo A. and {Brownstein}, Joel R. and {Crane}, Jeffrey D. and {De Lee}, Nathan M. and {Klaene}, Mark A. and {Kreckel}, Kathryn and {MacDonald}, Nick and {Merloni}, Andrea and {Ness}, Melissa K. and {O'Brien}, Thomas and {Sanchez-Gallego}, Jose R. and {Sayres}, Conor C. and {Shen}, Yue and {Thakar}, Ani R. and {Tkachenko}, Andrew and {Aerts}, Conny and {Blanton}, Michael R. and {Eisenstein}, Daniel J. and {Holtzman}, Jon A. and {Maoz}, Dan and {Nandra}, Kirpal and {Rockosi}, Constance and {Weinberg}, David H. and {Bovy}, Jo and {Casey}, Andrew R. and {Chaname}, Julio and {Clerc}, Nicolas and {Conroy}, Charlie and {Eracleous}, Michael and {G{\"a}nsicke}, Boris T. and {Hekker}, Saskia and {Horne}, Keith and {Kauffmann}, Jens and {McQuinn}, Kristen B.~W. and {Pellegrini}, Eric W. and {Schinnerer}, Eva and {Schlafly}, Edward F. and {Schwope}, Axel D. and {Seibert}, Mark and {Teske}, Johanna K. and {van Saders}, Jennifer L.},
        title = "{SDSS-V: Pioneering Panoptic Spectroscopy}",
      journal = {arXiv e-prints},
         year = 2017,
        month = nov,
          eid = {arXiv:1711.03234},
        pages = {arXiv:1711.03234},
          doi = {10.48550/arXiv.1711.03234},
archivePrefix = {arXiv},
       eprint = {1711.03234},
 primaryClass = {astro-ph.GA},
       adsurl = {https://ui.adsabs.harvard.edu/abs/2017arXiv171103234K}
}

@article{4most2019,
  title={4MOST: Project overview and information for the First Call for Proposals},
  author={de Jong, R. S. and Agertz, O. and Berbel, A. A. and others},
  journal={The Messenger},
  volume={175},
  pages={3--11},
  year={2019}
}

@article{pfs2016,
  title={Prime Focus Spectrograph (PFS) for the Subaru telescope: overview, recent progress, and future perspectives},
  author={Takada, M. and others},
  journal={Proceedings of the SPIE},
  volume={9908},
  pages={99081A},
  year={2016}
}

@article{moons2020,
  title={MOONS: The New Multi-Object Spectrograph for the VLT},
  author={Cirasuolo, M. and others},
  journal={The Messenger},
  volume={180},
  pages={10--20},
  year={2020}
}

@ARTICLE{must2024,
       author = {{Zhao}, Cheng and {Huang}, Song and {He}, Mengfan and {Montero-Camacho}, Paulo and {Liu}, Yu and {Renard}, Pablo and {Tang}, Yunyi and {Verdier}, Aurelien and {Xu}, Wenshuo and {Yang}, Xiaorui and {Yu}, Jiaxi and {Zhang}, Yao and {Zhao}, Siyi and {Zhou}, Xingchen and {He}, Shengyu and {Kneib}, Jean-Paul and {Li}, Jiayi and {Li}, Zhuoyang and {Wang}, Wen-Ting and {Xianyu}, Zhong-Zhi and {Zhang}, Yidian and {Gsponer}, Rafaela and {Li}, Xiao-Dong and {Rocher}, Antoine and {Zou}, Siwei and {Tan}, Ting and {Huang}, Zhiqi and {Wang}, Zhuoxiao and {Li}, Pei and {Rombach}, Maxime and {Dong}, Chenxing and {Forero-Sanchez}, Daniel and {Ning}, Yuanhang and {Shan}, Huanyuan and {Wang}, Tao and {Li}, Yin and {Zhai}, Zhongxu and {Wang}, Yuting and {Zhao}, Gong-Bo and {Shi}, Yong and {Mao}, Shude and {Huang}, Lei and {Guo}, Liquan and {Cai}, Zheng},
        title = "{MUltiplexed Survey Telescope (MUST) Science White Paper I: Overview of Large-Scale Structure Cosmology in the Era of Stage-V Spectroscopic Surveys}",
      journal = {arXiv e-prints},
         year = 2024,
        month = nov,
          eid = {arXiv:2411.07970},
        pages = {arXiv:2411.07970},
          doi = {10.48550/arXiv.2411.07970},
archivePrefix = {arXiv},
       eprint = {2411.07970},
 primaryClass = {astro-ph.CO},
       adsurl = {https://ui.adsabs.harvard.edu/abs/2024arXiv241107970Z}
}

@article{edr_paper,
   title={The Early Data Release of the Dark Energy Spectroscopic Instrument},
   volume={168},
   ISSN={1538-3881},
   url={http://dx.doi.org/10.3847/1538-3881/ad3217},
   DOI={10.3847/1538-3881/ad3217},
   number={2},
   journal={The Astronomical Journal},
   publisher={American Astronomical Society},
   author={Adame, A. G. and Aguilar, J. and Ahlen, S. and Alam, S. and Aldering, G. and Alexander, D. M. and Alfarsy, R. and Allende Prieto, C. and Alvarez, M. and Alves, O. and Anand, A. and Andrade-Oliveira, F. and Armengaud, E. and Asorey, J. and Avila, S. and Aviles, A. and Bailey, S. and Balaguera-Antolínez, A. and Ballester, O. and Baltay, C. and Bault, A. and Bautista, J. and Behera, J. and Beltran, S. F. and BenZvi, S. and Beraldo e Silva, L. and Bermejo-Climent, J. R. and Berti, A. and Besuner, R. and Beutler, F. and Bianchi, D. and Blake, C. and Blum, R. and Bolton, A. S. and Brieden, S. and Brodzeller, A. and Brooks, D. and Brown, Z. and Buckley-Geer, E. and Burtin, E. and Cabayol-Garcia, L. and Cai, Z. and Canning, R. and Cardiel-Sas, L. and Carnero Rosell, A. and Castander, F. J. and Cervantes-Cota, J. L. and Chabanier, S. and Chaussidon, E. and Chaves-Montero, J. and Chen, S. and Chen, X. and Chuang, C. and Claybaugh, T. and Cole, S. and Cooper, A. P. and Cuceu, A. and Davis, T. M. and Dawson, K. and de Belsunce, R. and de la Cruz, R. and de la Macorra, A. and Della Costa, J. and de Mattia, A. and Demina, R. and Demirbozan, U. and DeRose, J. and Dey, A. and Dey, B. and Dhungana, G. and Ding, J. and Ding, Z. and Doel, P. and Doshi, R. and Douglass, K. and Edge, A. and Eftekharzadeh, S. and Eisenstein, D. J. and Elliott, A. and Ereza, J. and Escoffier, S. and Fagrelius, P. and Fan, X. and Fanning, K. and Fawcett, V. A. and Ferraro, S. and Flaugher, B. and Font-Ribera, A. and Forero-Romero, J. E. and Forero-Sánchez, D. and Frenk, C. S. and Gänsicke, B. T. and García, L. Á. and García-Bellido, J. and Garcia-Quintero, C. and Garrison, L. H. and Gil-Marín, H. and Golden-Marx, J. and Gontcho A Gontcho, S. and Gonzalez-Morales, A. X. and Gonzalez-Perez, V. and Gordon, C. and Graur, O. and Green, D. and Gruen, D. and Guy, J. and Hadzhiyska, B. and Hahn, C. and Han, J. J. and Hanif, M. M. S and Herrera-Alcantar, H. K. and Honscheid, K. and Hou, J. and Howlett, C. and Huterer, D. and Iršič, V. and Ishak, M. and Jacques, A. and Jana, A. and Jiang, L. and Jimenez, J. and Jing, Y. P. and Joudaki, S. and Joyce, R. and Jullo, E. and Juneau, S. and Karaçaylı, N. G. and Karim, T. and Kehoe, R. and Kent, S. and Khederlarian, A. and Kim, S. and Kirkby, D. and Kisner, T. and Kitaura, F. and Kizhuprakkat, N. and Kneib, J. and Koposov, S. E. and Kovács, A. and Kremin, A. and Krolewski, A. and L’Huillier, B. and Lahav, O. and Lambert, A. and Lamman, C. and Lan, T.-W. and Landriau, M. and Lang, D. and Lange, J. U. and Lasker, J. and Leauthaud, A. and Le Guillou, L. and Levi, M. E. and Li, T. S. and Linder, E. and Lyons, A. and Magneville, C. and Manera, M. and Manser, C. J. and Margala, D. and Martini, P. and McDonald, P. and Medina, G. E. and Medina-Varela, L. and Meisner, A. and Mena-Fernández, J. and Meneses-Rizo, J. and Mezcua, M. and Miquel, R. and Montero-Camacho, P. and Moon, J. and Moore, S. and Moustakas, J. and Mueller, E. and Mundet, J. and Muñoz-Gutiérrez, A. and Myers, A. D. and Nadathur, S. and Napolitano, L. and Neveux, R. and Newman, J. A. and Nie, J. and Nikutta, R. and Niz, G. and Norberg, P. and Noriega, H. E. and Paillas, E. and Palanque-Delabrouille, N. and Palmese, A. and Pan, Z. and Parkinson, D. and Penmetsa, S. and Percival, W. J. and Pérez-Fernández, A. and Pérez-Ràfols, I. and Pieri, M. and Poppett, C. and Porredon, A. and Pothier, S. and Prada, F. and Pucha, R. and Raichoor, A. and Ramírez-Pérez, C. and Ramirez-Solano, S. and Rashkovetskyi, M. and Ravoux, C. and Rocher, A. and Rockosi, C. and Ross, A. J. and Rossi, G. and Ruggeri, R. and Ruhlmann-Kleider, V. and Sabiu, C. G. and Said, K. and Saintonge, A. and Samushia, L. and Sanchez, E. and Saulder, C. and Schaan, E. and Schlafly, E. F. and Schlegel, D. and Scholte, D. and Schubnell, M. and Seo, H. and Shafieloo, A. and Sharples, R. and Sheu, W. and Silber, J. and Sinigaglia, F. and Siudek, M. and Slepian, Z. and Smith, A. and Soumagnac, M. T. and Sprayberry, D. and Stephey, L. and Suárez-Pérez, J. and Sun, Z. and Tan, T. and Tarlé, G. and Tojeiro, R. and Ureña-López, L. A. and Vaisakh, R. and Valcin, D. and Valdes, F. and Valluri, M. and Vargas-Magaña, M. and Variu, A. and Verde, L. and Walther, M. and Wang, B. and Wang, M. S. and Weaver, B. A. and Weaverdyck, N. and Wechsler, R. H. and White, M. and Xie, Y. and Yang, J. and Yèche, C. and Yu, J. and Yuan, S. and Zhang, H. and Zhang, Z. and Zhao, C. and Zheng, Z. and Zhou, R. and Zhou, Z. and Zou, H. and Zou, S. and Zu, Y.},
   year={2024},
   month=jul, pages={58} }

@article{Lan_2023,
   title={The DESI Survey Validation: Results from Visual Inspection of Bright Galaxies, Luminous Red Galaxies, and Emission-line Galaxies},
   volume={943},
   ISSN={1538-4357},
   url={http://dx.doi.org/10.3847/1538-4357/aca5fa},
   DOI={10.3847/1538-4357/aca5fa},
   number={1},
   journal={The Astrophysical Journal},
   publisher={American Astronomical Society},
   author={Lan, Ting-Wen and Tojeiro, R. and Armengaud, E. and Prochaska, J. Xavier and Davis, T. M. and Alexander, David M. and Raichoor, A. and Zhou, Rongpu and Yèche, Christophe and Balland, C. and BenZvi, S. and Berti, A. and Canning, R. and Carr, A. and Chittenden, H. and Cole, S. and Cousinou, M.-C. and Dawson, K. and Dey, Biprateep and Douglass, K. and Edge, A. and Escoffier, S. and Glanville, A. and A Gontcho, S. Gontcho and Guy, J. and Hahn, C. and Howlett, C. and Hwang, Ho Seong and Jiang, L. and Kovács, A. and Mezcua, M. and Moore, S. and Nadathur, S. and Oh, M. and Parkinson, D. and Rocher, A. and Ross, A. J. and Ruhlmann-Kleider, V. and Sabiu, C. G. and Said, K. and Saulder, C. and Sierra-Porta, D. and Weiner, B. and Yu, J. and Zarrouk, P. and Zhang, Y. and Zou, H. and Ahlen, S. and Bailey, S. and Brooks, D. and Cooper, A. P. and de la Macorra, A. and Dey, A. and Dhungana, G. and Doel, P. and Eftekharzadeh, S. and Fanning, K. and Font-Ribera, A. and Garrison, L. and Gaztañaga, E. and Kehoe, R. and Kisner, T. and Kremin, A. and Landriau, M. and Le Guillou, L. and Levi, Michael E. and Magneville, C. and Meisner, Aaron M. and Miquel, R. and Moustakas, J. and Myers, Adam D. and Newman, Jeffrey A. and Nie, J. D. and Palanque-Delabrouille, N. and Percival, W. J. and Poppett, C. and Prada, F. and Schubnell, M. and Tarlé, Gregory and Weaver, B. A. and Zhang, K. and Zhou, Zhimin},
   year={2023},
   month=jan, pages={68} }

@article{Alexander_2023,
   title={The DESI Survey Validation: Results from Visual Inspection of the Quasar Survey Spectra},
   volume={165},
   ISSN={1538-3881},
   url={http://dx.doi.org/10.3847/1538-3881/acacfc},
   DOI={10.3847/1538-3881/acacfc},
   number={3},
   journal={The Astronomical Journal},
   publisher={American Astronomical Society},
   author={Alexander, David M. and Davis, Tamara M. and Chaussidon, E. and Fawcett, V. A. and X. Gonzalez-Morales, Alma and Lan, Ting-Wen and Yèche, Christophe and Ahlen, S. and Aguilar, J. N. and Armengaud, E. and Bailey, S. and Brooks, D. and Cai, Z. and Canning, R. and Carr, A. and Chabanier, S. and Cousinou, Marie-Claude and Dawson, K. and de la Macorra, A. and Dey, A. and Dey, Biprateep and Dhungana, G. and Edge, A. C. and Eftekharzadeh, S. and Fanning, K. and Farr, James and Font-Ribera, A. and Garcia-Bellido, J. and Garrison, Lehman and Gaztañaga, E. and A Gontcho, Satya Gontcho and Gordon, C. and Medellin Gonzalez, Stefany Guadalupe and Guy, J. and Herrera-Alcantar, Hiram K. and Jiang, L. and Juneau, S. and Karaçaylı, N. G. and Kehoe, R. and Kisner, T. and Kovács, A. and Landriau, M. and Levi, Michael E. and Magneville, C. and Martini, P. and Meisner, Aaron M. and Mezcua, M. and Miquel, R. and Camacho, P. Montero and Moustakas, J. and Muñoz-Gutiérrez, Andrea and Myers, Adam D. and Nadathur, S. and Napolitano, L. and Nie, J. D. and Palanque-Delabrouille, N. and Pan, Z. and Percival, W. J. and Pérez-Ràfols, I. and Poppett, C. and Prada, F. and Ramírez-Pérez, César and Ravoux, C. and Rosario, D. J. and Schubnell, M. and Tarlé, Gregory and Walther, M. and Weiner, B. and Youles, S. and Zhou, Zhimin and Zou, H. and Zou, Siwei},
   year={2023},
   month=feb, pages={124} }

@misc{way_2012,
  author={Way, M. J. and Scargle, J. D. and Ali, K. and Srivastava, A. N.},
  title={Advances in Machine Learning and Data Mining for Astronomy},
  year={2012},
  publisher={Chapman and Hall/CRC},
  series={Data Mining and Knowledge Discovery Series},
  isbn={143984173X, 9781439841730}
}

@article{Nun_2014,
   title={SUPERVISED DETECTION OF ANOMALOUS LIGHT CURVES IN MASSIVE ASTRONOMICAL CATALOGS},
   volume={793},
   ISSN={1538-4357},
   url={http://dx.doi.org/10.1088/0004-637X/793/1/23},
   DOI={10.1088/0004-637x/793/1/23},
   number={1},
   journal={The Astrophysical Journal},
   publisher={American Astronomical Society},
   author={Nun, Isadora and Pichara, Karim and Protopapas, Pavlos and Kim, Dae-Won},
   year={2014},
   month=sep, pages={23}
}

@article{Baron_2016,
   title={The weirdest SDSS galaxies: results from an outlier detection algorithm},
   volume={465},
   ISSN={1365-2966},
   url={http://dx.doi.org/10.1093/mnras/stw3021},
   DOI={10.1093/mnras/stw3021},
   number={4},
   journal={Monthly Notices of the Royal Astronomical Society},
   publisher={Oxford University Press (OUP)},
   author={Baron, Dalya and Poznanski, Dovi},
   year={2016},
   month=nov, pages={4530–4555}
}

@article{Muthukrishna_2022,
   title={Real-time detection of anomalies in large-scale transient surveys},
   volume={517},
   ISSN={1365-2966},
   url={http://dx.doi.org/10.1093/mnras/stac2582},
   DOI={10.1093/mnras/stac2582},
   number={1},
   journal={Monthly Notices of the Royal Astronomical Society},
   publisher={Oxford University Press (OUP)},
   author={Muthukrishna, Daniel and Mandel, Kaisey S and Lochner, Michelle and Webb, Sara and Narayan, Gautham},
   year={2022},
   month=sep, pages={393–419}
}

@article{Sanchez_Saez_2021,
   title={Searching for Changing-state AGNs in Massive Data Sets. I. Applying Deep Learning and Anomaly-detection Techniques to Find AGNs with Anomalous Variability Behaviors},
   volume={162},
   ISSN={1538-3881},
   url={http://dx.doi.org/10.3847/1538-3881/ac1426},
   DOI={10.3847/1538-3881/ac1426},
   number={5},
   journal={The Astronomical Journal},
   publisher={American Astronomical Society},
   author={Sánchez-Sáez, P. and Lira, H. and Martí, L. and Sánchez-Pi, N. and Arredondo, J. and Bauer, F. E. and Bayo, A. and Cabrera-Vives, G. and Donoso-Oliva, C. and Estévez, P. A. and Eyheramendy, S. and Förster, F. and Hernández-García, L. and Arancibia, A. M. Muñoz and Pérez-Carrasco, M. and Sepúlveda, M. and Vergara, J. R.},
   year={2021},
   month=oct, pages={206}
}

@PHDTHESIS{Suarez_Art_2023,
    author={Suárez Pérez, John Fredy},
    title={Artificial Intelligence in astronomy: machine learning and deep learning approaches to DESI data},
    year={2023},
    school={Universidad de los Andes},
    url={https://hdl.handle.net/1992/68996},
}

@misc{Mcinnes_2020,
      title={UMAP: Uniform Manifold Approximation and Projection for Dimension Reduction}, 
      author={Leland McInnes and John Healy and James Melville},
      year={2020},
      eprint={1802.03426},
      archivePrefix={arXiv},
      primaryClass={stat.ML},
      url={https://arxiv.org/abs/1802.03426}, 
}

@article{Rosito_2023,
   title={Application of dimensionality reduction and clustering algorithms for the classification of kinematic morphologies of galaxies},
   volume={671},
   ISSN={1432-0746},
   url={http://dx.doi.org/10.1051/0004-6361/202244707},
   DOI={10.1051/0004-6361/202244707},
   journal={Astronomy \& Astrophysics},
   publisher={EDP Sciences},
   author={Rosito, M. S. and Bignone, L. A. and Tissera, P. B. and Pedrosa, S. E.},
   year={2023},
   month=mar, pages={A19}
}

@article{Cook_2024,
   title={Wide Area VISTA Extra-galactic Survey (WAVES): unsupervised star-galaxy separation on the WAVES-Wide photometric input catalogue using UMAP and <scp>hdbscan</scp>},
   volume={535},
   ISSN={1365-2966},
   url={http://dx.doi.org/10.1093/mnras/stae2389},
   DOI={10.1093/mnras/stae2389},
   number={3},
   journal={Monthly Notices of the Royal Astronomical Society},
   publisher={Oxford University Press (OUP)},
   author={Cook, Todd L and Bandi, Behnood and Philipsborn, Sam and Loveday, Jon and Bellstedt, Sabine and Driver, Simon P and Robotham, Aaron S G and Bilicki, Maciej and Kaur, Gursharanjit and Tempel, Elmo and Baldry, Ivan and Gruen, Daniel and Longhetti, Marcella and Iovino, Angela and Holwerda, Benne W and Demarco, Ricardo},
   year={2024},
   month=oct, pages={2129–2148}
}

@ARTICLE{HuchraGeller_1982,
  author  = {Huchra, J. P. and Geller, M. J.},
  title   = {Groups of Galaxies. I. Nearby Groups},
  journal = {Astrophysical Journal},
  year    = {1982},
  month   = jun,
  volume  = {257},
  pages   = {423--437},
  doi     = {10.1086/160000},
  bibcode = {1982ApJ...257..423H}
}

@misc{desicollaboration2016desiexperimentiiinstrument,
      title={The DESI Experiment Part II: Instrument Design}, 
      author={DESI Collaboration and Amir Aghamousa and Jessica Aguilar and Steve Ahlen and Shadab Alam and Lori E. Allen and Carlos Allende Prieto and James Annis and Stephen Bailey and Christophe Balland and Otger Ballester and Charles Baltay and Lucas Beaufore and Chris Bebek and Timothy C. Beers and Eric F. Bell and José Luis Bernal and Robert Besuner and Florian Beutler and Chris Blake and Hannes Bleuler and Michael Blomqvist and Robert Blum and Adam S. Bolton and Cesar Briceno and David Brooks and Joel R. Brownstein and Elizabeth Buckley-Geer and Angela Burden and Etienne Burtin and Nicolas G. Busca and Robert N. Cahn and Yan-Chuan Cai and Laia Cardiel-Sas and Raymond G. Carlberg and Pierre-Henri Carton and Ricard Casas and Francisco J. Castander and Jorge L. Cervantes-Cota and Todd M. Claybaugh and Madeline Close and Carl T. Coker and Shaun Cole and Johan Comparat and Andrew P. Cooper and M. -C. Cousinou and Martin Crocce and Jean-Gabriel Cuby and Daniel P. Cunningham and Tamara M. Davis and Kyle S. Dawson and Axel de la Macorra and Juan De Vicente and Timothée Delubac and Mark Derwent and Arjun Dey and Govinda Dhungana and Zhejie Ding and Peter Doel and Yutong T. Duan and Anne Ealet and Jerry Edelstein and Sarah Eftekharzadeh and Daniel J. Eisenstein and Ann Elliott and Stéphanie Escoffier and Matthew Evatt and Parker Fagrelius and Xiaohui Fan and Kevin Fanning and Arya Farahi and Jay Farihi and Ginevra Favole and Yu Feng and Enrique Fernandez and Joseph R. Findlay and Douglas P. Finkbeiner and Michael J. Fitzpatrick and Brenna Flaugher and Samuel Flender and Andreu Font-Ribera and Jaime E. Forero-Romero and Pablo Fosalba and Carlos S. Frenk and Michele Fumagalli and Boris T. Gaensicke and Giuseppe Gallo and Juan Garcia-Bellido and Enrique Gaztanaga and Nicola Pietro Gentile Fusillo and Terry Gerard and Irena Gershkovich and Tommaso Giannantonio and Denis Gillet and Guillermo Gonzalez-de-Rivera and Violeta Gonzalez-Perez and Shelby Gott and Or Graur and Gaston Gutierrez and Julien Guy and Salman Habib and Henry Heetderks and Ian Heetderks and Katrin Heitmann and Wojciech A. Hellwing and David A. Herrera and Shirley Ho and Stephen Holland and Klaus Honscheid and Eric Huff and Timothy A. Hutchinson and Dragan Huterer and Ho Seong Hwang and Joseph Maria Illa Laguna and Yuzo Ishikawa and Dianna Jacobs and Niall Jeffrey and Patrick Jelinsky and Elise Jennings and Linhua Jiang and Jorge Jimenez and Jennifer Johnson and Richard Joyce and Eric Jullo and Stéphanie Juneau and Sami Kama and Armin Karcher and Sonia Karkar and Robert Kehoe and Noble Kennamer and Stephen Kent and Martin Kilbinger and Alex G. Kim and David Kirkby and Theodore Kisner and Ellie Kitanidis and Jean-Paul Kneib and Sergey Koposov and Eve Kovacs and Kazuya Koyama and Anthony Kremin and Richard Kron and Luzius Kronig and Andrea Kueter-Young and Cedric G. Lacey and Robin Lafever and Ofer Lahav and Andrew Lambert and Michael Lampton and Martin Landriau and Dustin Lang and Tod R. Lauer and Jean-Marc Le Goff and Laurent Le Guillou and Auguste Le Van Suu and Jae Hyeon Lee and Su-Jeong Lee and Daniela Leitner and Michael Lesser and Michael E. Levi and Benjamin L'Huillier and Baojiu Li and Ming Liang and Huan Lin and Eric Linder and Sarah R. Loebman and Zarija Lukić and Jun Ma and Niall MacCrann and Christophe Magneville and Laleh Makarem and Marc Manera and Christopher J. Manser and Robert Marshall and Paul Martini and Richard Massey and Thomas Matheson and Jeremy McCauley and Patrick McDonald and Ian D. McGreer and Aaron Meisner and Nigel Metcalfe and Timothy N. Miller and Ramon Miquel and John Moustakas and Adam Myers and Milind Naik and Jeffrey A. Newman and Robert C. Nichol and Andrina Nicola and Luiz Nicolati da Costa and Jundan Nie and Gustavo Niz and Peder Norberg and Brian Nord and Dara Norman and Peter Nugent and Thomas O'Brien and Minji Oh and Knut A. G. Olsen and Cristobal Padilla and Hamsa Padmanabhan and Nikhil Padmanabhan and Nathalie Palanque-Delabrouille and Antonella Palmese and Daniel Pappalardo and Isabelle Pâris and Changbom Park and Anna Patej and John A. Peacock and Hiranya V. Peiris and Xiyan Peng and Will J. Percival and Sandrine Perruchot and Matthew M. Pieri and Richard Pogge and Jennifer E. Pollack and Claire Poppett and Francisco Prada and Abhishek Prakash and Ronald G. Probst and David Rabinowitz and Anand Raichoor and Chang Hee Ree and Alexandre Refregier and Xavier Regal and Beth Reid and Kevin Reil and Mehdi Rezaie and Constance M. Rockosi and Natalie Roe and Samuel Ronayette and Aaron Roodman and Ashley J. Ross and Nicholas P. Ross and Graziano Rossi and Eduardo Rozo and Vanina Ruhlmann-Kleider and Eli S. Rykoff and Cristiano Sabiu and Lado Samushia and Eusebio Sanchez and Javier Sanchez and David J. Schlegel and Michael Schneider and Michael Schubnell and Aurélia Secroun and Uros Seljak and Hee-Jong Seo and Santiago Serrano and Arman Shafieloo and Huanyuan Shan and Ray Sharples and Michael J. Sholl and William V. Shourt and Joseph H. Silber and David R. Silva and Martin M. Sirk and Anze Slosar and Alex Smith and George F. Smoot and Debopam Som and Yong-Seon Song and David Sprayberry and Ryan Staten and Andy Stefanik and Gregory Tarle and Suk Sien Tie and Jeremy L. Tinker and Rita Tojeiro and Francisco Valdes and Octavio Valenzuela and Monica Valluri and Mariana Vargas-Magana and Licia Verde and Alistair R. Walker and Jiali Wang and Yuting Wang and Benjamin A. Weaver and Curtis Weaverdyck and Risa H. Wechsler and David H. Weinberg and Martin White and Qian Yang and Christophe Yeche and Tianmeng Zhang and Gong-Bo Zhao and Yi Zheng and Xu Zhou and Zhimin Zhou and Yaling Zhu and Hu Zou and Ying Zu},
      year={2016},
      eprint={1611.00037},
      archivePrefix={arXiv},
      primaryClass={astro-ph.IM},
      url={https://arxiv.org/abs/1611.00037}, 
}

@article{Guy_2023,
   title={The Spectroscopic Data Processing Pipeline for the Dark Energy Spectroscopic Instrument},
   volume={165},
   ISSN={1538-3881},
   url={http://dx.doi.org/10.3847/1538-3881/acb212},
   DOI={10.3847/1538-3881/acb212},
   number={4},
   journal={The Astronomical Journal},
   publisher={American Astronomical Society},
   author={Guy, J. and Bailey, S. and Kremin, A. and Alam, Shadab and Alexander, D. M. and Allende Prieto, C. and BenZvi, S. and Bolton, A. S. and Brooks, D. and Chaussidon, E. and Cooper, A. P. and Dawson, K. and de la Macorra, A. and Dey, A. and Dey, Biprateep and Dhungana, G. and Eisenstein, D. J. and Font-Ribera, A. and Forero-Romero, J. E. and Gaztañaga, E. and Gontcho A Gontcho, S. and Green, D. and Honscheid, K. and Ishak, M. and Kehoe, R. and Kirkby, D. and Kisner, T. and Koposov, Sergey E. and Lan, Ting-Wen and Landriau, M. and Le Guillou, L. and Levi, Michael E. and Magneville, C. and Manser, Christopher J. and Martini, P. and Meisner, Aaron M. and Miquel, R. and Moustakas, J. and Myers, Adam D. and Newman, Jeffrey A. and Nie, Jundan and Palanque-Delabrouille, N. and Percival, W. J. and Poppett, C. and Prada, F. and Raichoor, A. and Ravoux, C. and Ross, A. J. and Schlafly, E. F. and Schlegel, D. and Schubnell, M. and Sharples, Ray M. and Tarlé, Gregory and Weaver, B. A. and Yéche, Christophe and Zhou, Rongpu and Zhou, Zhimin and Zou, H.},
   year={2023},
   month=mar, pages={144} }

@article{Dey_2019,
   title={Overview of the DESI Legacy Imaging Surveys},
   volume={157},
   ISSN={1538-3881},
   url={http://dx.doi.org/10.3847/1538-3881/ab089d},
   DOI={10.3847/1538-3881/ab089d},
   number={5},
   journal={The Astronomical Journal},
   publisher={American Astronomical Society},
   author={Dey, Arjun and Schlegel, David J. and Lang, Dustin and Blum, Robert and Burleigh, Kaylan and Fan, Xiaohui and Findlay, Joseph R. and Finkbeiner, Doug and Herrera, David and Juneau, Stéphanie and Landriau, Martin and Levi, Michael and McGreer, Ian and Meisner, Aaron and Myers, Adam D. and Moustakas, John and Nugent, Peter and Patej, Anna and Schlafly, Edward F. and Walker, Alistair R. and Valdes, Francisco and Weaver, Benjamin A. and Yèche, Christophe and Zou, Hu and Zhou, Xu and Abareshi, Behzad and Abbott, T. M. C. and Abolfathi, Bela and Aguilera, C. and Alam, Shadab and Allen, Lori and Alvarez, A. and Annis, James and Ansarinejad, Behzad and Aubert, Marie and Beechert, Jacqueline and Bell, Eric F. and BenZvi, Segev Y. and Beutler, Florian and Bielby, Richard M. and Bolton, Adam S. and Briceño, César and Buckley-Geer, Elizabeth J. and Butler, Karen and Calamida, Annalisa and Carlberg, Raymond G. and Carter, Paul and Casas, Ricard and Castander, Francisco J. and Choi, Yumi and Comparat, Johan and Cukanovaite, Elena and Delubac, Timothée and DeVries, Kaitlin and Dey, Sharmila and Dhungana, Govinda and Dickinson, Mark and Ding, Zhejie and Donaldson, John B. and Duan, Yutong and Duckworth, Christopher J. and Eftekharzadeh, Sarah and Eisenstein, Daniel J. and Etourneau, Thomas and Fagrelius, Parker A. and Farihi, Jay and Fitzpatrick, Mike and Font-Ribera, Andreu and Fulmer, Leah and Gänsicke, Boris T. and Gaztanaga, Enrique and George, Koshy and Gerdes, David W. and A Gontcho, Satya Gontcho and Gorgoni, Claudio and Green, Gregory and Guy, Julien and Harmer, Diane and Hernandez, M. and Honscheid, Klaus and Huang, Lijuan (Wendy) and James, David J. and Jannuzi, Buell T. and Jiang, Linhua and Joyce, Richard and Karcher, Armin and Karkar, Sonia and Kehoe, Robert and Kneib, Jean-Paul and Kueter-Young, Andrea and Lan, Ting-Wen and Lauer, Tod R. and Guillou, Laurent Le and Van Suu, Auguste Le and Lee, Jae Hyeon and Lesser, Michael and Levasseur, Laurence Perreault and Li, Ting S. and Mann, Justin L. and Marshall, Robert and Martínez-Vázquez, C. E. and Martini, Paul and du Mas des Bourboux, Hélion and McManus, Sean and Meier, Tobias Gabriel and Ménard, Brice and Metcalfe, Nigel and Muñoz-Gutiérrez, Andrea and Najita, Joan and Napier, Kevin and Narayan, Gautham and Newman, Jeffrey A. and Nie, Jundan and Nord, Brian and Norman, Dara J. and Olsen, Knut A. G. and Paat, Anthony and Palanque-Delabrouille, Nathalie and Peng, Xiyan and Poppett, Claire L. and Poremba, Megan R. and Prakash, Abhishek and Rabinowitz, David and Raichoor, Anand and Rezaie, Mehdi and Robertson, A. N. and Roe, Natalie A. and Ross, Ashley J. and Ross, Nicholas P. and Rudnick, Gregory and Gaines, Sasha and Saha, Abhijit and Sánchez, F. Javier and Savary, Elodie and Schweiker, Heidi and Scott, Adam and Seo, Hee-Jong and Shan, Huanyuan and Silva, David R. and Slepian, Zachary and Soto, Christian and Sprayberry, David and Staten, Ryan and Stillman, Coley M. and Stupak, Robert J. and Summers, David L. and Tie, Suk Sien and Tirado, H. and Vargas-Magaña, Mariana and Vivas, A. Katherina and Wechsler, Risa H. and Williams, Doug and Yang, Jinyi and Yang, Qian and Yapici, Tolga and Zaritsky, Dennis and Zenteno, A. and Zhang, Kai and Zhang, Tianmeng and Zhou, Rongpu and Zhou, Zhimin},
   year={2019},
   month=apr, pages={168}
}

@article{Myers_2023,
   title={The Target-selection Pipeline for the Dark Energy Spectroscopic Instrument},
   volume={165},
   ISSN={1538-3881},
   url={http://dx.doi.org/10.3847/1538-3881/aca5f9},
   DOI={10.3847/1538-3881/aca5f9},
   number={2},
   journal={The Astronomical Journal},
   publisher={American Astronomical Society},
   author={Myers, Adam D. and Moustakas, John and Bailey, Stephen and Weaver, Benjamin A. and Cooper, Andrew P. and Forero-Romero, Jaime E. and Abolfathi, Bela and Alexander, David M. and Brooks, David and Chaussidon, Edmond and Chuang, Chia-Hsun and Dawson, Kyle and Dey, Arjun and Dey, Biprateep and Dhungana, Govinda and Doel, Peter and Fanning, Kevin and Gaztañaga, Enrique and A Gontcho, Satya Gontcho and Gonzalez-Morales, Alma X. and Hahn, ChangHoon and Herrera-Alcantar, Hiram K. and Honscheid, Klaus and Ishak, Mustapha and Karim, Tanveer and Kirkby, David and Kisner, Theodore and Koposov, Sergey E. and Kremin, Anthony and Lan, Ting-Wen and Landriau, Martin and Lang, Dustin and Levi, Michael E. and Magneville, Christophe and Napolitano, Lucas and Martini, Paul and Meisner, Aaron and Newman, Jeffrey A. and Palanque-Delabrouille, Nathalie and Percival, Will and Poppett, Claire and Prada, Francisco and Raichoor, Anand and Ross, Ashley J. and Schlafly, Edward F. and Schlegel, David and Schubnell, Michael and Tan, Ting and Tarle, Gregory and Wilson, Michael J. and Yèche, Christophe and Zhou, Rongpu and Zhou, Zhimin and Zou, Hu},
   year={2023},
   month=jan, pages={50}
}

@misc{desicollaboration2016desiexperimentisciencetargeting,
      title={The DESI Experiment Part I: Science,Targeting, and Survey Design}, 
      author={{DESI Collaboration} and Amir Aghamousa and Jessica Aguilar and Steve Ahlen and Shadab Alam and Lori E. Allen and Carlos Allende Prieto and James Annis and Stephen Bailey and Christophe Balland and Otger Ballester and Charles Baltay and Lucas Beaufore and Chris Bebek and Timothy C. Beers and Eric F. Bell and José Luis Bernal and Robert Besuner and Florian Beutler and Chris Blake and Hannes Bleuler and Michael Blomqvist and Robert Blum and Adam S. Bolton and Cesar Briceno and David Brooks and Joel R. Brownstein and Elizabeth Buckley-Geer and Angela Burden and Etienne Burtin and Nicolas G. Busca and Robert N. Cahn and Yan-Chuan Cai and Laia Cardiel-Sas and Raymond G. Carlberg and Pierre-Henri Carton and Ricard Casas and Francisco J. Castander and Jorge L. Cervantes-Cota and Todd M. Claybaugh and Madeline Close and Carl T. Coker and Shaun Cole and Johan Comparat and Andrew P. Cooper and M. -C. Cousinou and Martin Crocce and Jean-Gabriel Cuby and Daniel P. Cunningham and Tamara M. Davis and Kyle S. Dawson and Axel de la Macorra and Juan De Vicente and Timothée Delubac and Mark Derwent and Arjun Dey and Govinda Dhungana and Zhejie Ding and Peter Doel and Yutong T. Duan and Anne Ealet and Jerry Edelstein and Sarah Eftekharzadeh and Daniel J. Eisenstein and Ann Elliott and Stéphanie Escoffier and Matthew Evatt and Parker Fagrelius and Xiaohui Fan and Kevin Fanning and Arya Farahi and Jay Farihi and Ginevra Favole and Yu Feng and Enrique Fernandez and Joseph R. Findlay and Douglas P. Finkbeiner and Michael J. Fitzpatrick and Brenna Flaugher and Samuel Flender and Andreu Font-Ribera and Jaime E. Forero-Romero and Pablo Fosalba and Carlos S. Frenk and Michele Fumagalli and Boris T. Gaensicke and Giuseppe Gallo and Juan Garcia-Bellido and Enrique Gaztanaga and Nicola Pietro Gentile Fusillo and Terry Gerard and Irena Gershkovich and Tommaso Giannantonio and Denis Gillet and Guillermo Gonzalez-de-Rivera and Violeta Gonzalez-Perez and Shelby Gott and Or Graur and Gaston Gutierrez and Julien Guy and Salman Habib and Henry Heetderks and Ian Heetderks and Katrin Heitmann and Wojciech A. Hellwing and David A. Herrera and Shirley Ho and Stephen Holland and Klaus Honscheid and Eric Huff and Timothy A. Hutchinson and Dragan Huterer and Ho Seong Hwang and Joseph Maria Illa Laguna and Yuzo Ishikawa and Dianna Jacobs and Niall Jeffrey and Patrick Jelinsky and Elise Jennings and Linhua Jiang and Jorge Jimenez and Jennifer Johnson and Richard Joyce and Eric Jullo and Stéphanie Juneau and Sami Kama and Armin Karcher and Sonia Karkar and Robert Kehoe and Noble Kennamer and Stephen Kent and Martin Kilbinger and Alex G. Kim and David Kirkby and Theodore Kisner and Ellie Kitanidis and Jean-Paul Kneib and Sergey Koposov and Eve Kovacs and Kazuya Koyama and Anthony Kremin and Richard Kron and Luzius Kronig and Andrea Kueter-Young and Cedric G. Lacey and Robin Lafever and Ofer Lahav and Andrew Lambert and Michael Lampton and Martin Landriau and Dustin Lang and Tod R. Lauer and Jean-Marc Le Goff and Laurent Le Guillou and Auguste Le Van Suu and Jae Hyeon Lee and Su-Jeong Lee and Daniela Leitner and Michael Lesser and Michael E. Levi and Benjamin L'Huillier and Baojiu Li and Ming Liang and Huan Lin and Eric Linder and Sarah R. Loebman and Zarija Lukić and Jun Ma and Niall MacCrann and Christophe Magneville and Laleh Makarem and Marc Manera and Christopher J. Manser and Robert Marshall and Paul Martini and Richard Massey and Thomas Matheson and Jeremy McCauley and Patrick McDonald and Ian D. McGreer and Aaron Meisner and Nigel Metcalfe and Timothy N. Miller and Ramon Miquel and John Moustakas and Adam Myers and Milind Naik and Jeffrey A. Newman and Robert C. Nichol and Andrina Nicola and Luiz Nicolati da Costa and Jundan Nie and Gustavo Niz and Peder Norberg and Brian Nord and Dara Norman and Peter Nugent and Thomas O'Brien and Minji Oh and Knut A. G. Olsen and Cristobal Padilla and Hamsa Padmanabhan and Nikhil Padmanabhan and Nathalie Palanque-Delabrouille and Antonella Palmese and Daniel Pappalardo and Isabelle Pâris and Changbom Park and Anna Patej and John A. Peacock and Hiranya V. Peiris and Xiyan Peng and Will J. Percival and Sandrine Perruchot and Matthew M. Pieri and Richard Pogge and Jennifer E. Pollack and Claire Poppett and Francisco Prada and Abhishek Prakash and Ronald G. Probst and David Rabinowitz and Anand Raichoor and Chang Hee Ree and Alexandre Refregier and Xavier Regal and Beth Reid and Kevin Reil and Mehdi Rezaie and Constance M. Rockosi and Natalie Roe and Samuel Ronayette and Aaron Roodman and Ashley J. Ross and Nicholas P. Ross and Graziano Rossi and Eduardo Rozo and Vanina Ruhlmann-Kleider and Eli S. Rykoff and Cristiano Sabiu and Lado Samushia and Eusebio Sanchez and Javier Sanchez and David J. Schlegel and Michael Schneider and Michael Schubnell and Aurélia Secroun and Uros Seljak and Hee-Jong Seo and Santiago Serrano and Arman Shafieloo and Huanyuan Shan and Ray Sharples and Michael J. Sholl and William V. Shourt and Joseph H. Silber and David R. Silva and Martin M. Sirk and Anze Slosar and Alex Smith and George F. Smoot and Debopam Som and Yong-Seon Song and David Sprayberry and Ryan Staten and Andy Stefanik and Gregory Tarle and Suk Sien Tie and Jeremy L. Tinker and Rita Tojeiro and Francisco Valdes and Octavio Valenzuela and Monica Valluri and Mariana Vargas-Magana and Licia Verde and Alistair R. Walker and Jiali Wang and Yuting Wang and Benjamin A. Weaver and Curtis Weaverdyck and Risa H. Wechsler and David H. Weinberg and Martin White and Qian Yang and Christophe Yeche and Tianmeng Zhang and Gong-Bo Zhao and Yi Zheng and Xu Zhou and Zhimin Zhou and Yaling Zhu and Hu Zou and Ying Zu},
      year={2016},
      eprint={1611.00036},
      archivePrefix={arXiv},
      primaryClass={astro-ph.IM},
      url={https://arxiv.org/abs/1611.00036}, 
}

@article{Raichoor_2023,
   title={Target Selection and Validation of DESI Emission Line Galaxies},
   volume={165},
   ISSN={1538-3881},
   url={http://dx.doi.org/10.3847/1538-3881/acb213},
   DOI={10.3847/1538-3881/acb213},
   number={3},
   journal={The Astronomical Journal},
   publisher={American Astronomical Society},
   author={Raichoor, A. and Moustakas, J. and Newman, Jeffrey A. and Karim, T. and Ahlen, S. and Alam, Shadab and Bailey, S. and Brooks, D. and Dawson, K. and de la Macorra, A. and de Mattia, A. and Dey, A. and Dey, Biprateep and Dhungana, G. and Eftekharzadeh, S. and Eisenstein, D. J. and Fanning, K. and Font-Ribera, A. and García-Bellido, J. and Gaztañaga, E. and A Gontcho, S. Gontcho and Guy, J. and Honscheid, K. and Ishak, M. and Kehoe, R. and Kisner, T. and Kremin, Anthony and Lan, Ting-Wen and Landriau, M. and Le Guillou, L. and Levi, Michael E. and Magneville, C. and Manera, M. and Martini, P. and Meisner, Aaron M. and Myers, Adam D. and Nie, Jundan and Palanque-Delabrouille, N. and Percival, W. J. and Poppett, C. and Prada, F. and Ross, A. J. and Ruhlmann-Kleider, V. and Sabiu, C. G. and Schlafly, E. F. and Schlegel, D. and Tarlé, Gregory and Weaver, B. A. and Yèche, Christophe and Zhou, Rongpu and Zhou, Zhimin and Zou, H.},
   year={2023},
   month=feb, pages={126}
}

@article{Chaussidon_2023,
   title={Target Selection and Validation of DESI Quasars},
   volume={944},
   ISSN={1538-4357},
   url={http://dx.doi.org/10.3847/1538-4357/acb3c2},
   DOI={10.3847/1538-4357/acb3c2},
   number={1},
   journal={The Astrophysical Journal},
   publisher={American Astronomical Society},
   author={Chaussidon, Edmond and Yèche, Christophe and Palanque-Delabrouille, Nathalie and Alexander, David M. and Yang, Jinyi and Ahlen, Steven and Bailey, Stephen and Brooks, David and Cai, Zheng and Chabanier, Solène and Davis, Tamara M. and Dawson, Kyle and de laMacorra, Axel and Dey, Arjun and Dey, Biprateep and Eftekharzadeh, Sarah and Eisenstein, Daniel J. and Fanning, Kevin and Font-Ribera, Andreu and Gaztañaga, Enrique and A Gontcho, Satya Gontcho and Gonzalez-Morales, Alma X. and Guy, Julien and Herrera-Alcantar, Hiram K. and Honscheid, Klaus and Ishak, Mustapha and Jiang, Linhua and Juneau, Stephanie and Kehoe, Robert and Kisner, Theodore and Kovács, Andras and Kremin, Anthony and Lan, Ting-Wen and Landriau, Martin and Le Guillou, Laurent and Levi, Michael E. and Magneville, Christophe and Martini, Paul and Meisner, Aaron M. and Moustakas, John and Muñoz-Gutiérrez, Andrea and Myers, Adam D. and Newman, Jeffrey A. and Nie, Jundan and Percival, Will J. and Poppett, Claire and Prada, Francisco and Raichoor, Anand and Ravoux, Corentin and Ross, Ashley J. and Schlafly, Edward and Schlegel, David and Tan, Ting and Tarlé, Gregory and Zhou, Rongpu and Zhou, Zhimin and Zou, Hu},
   year={2023},
   month=feb, pages={107}
}

@article{Hahn_2023,
   title={The DESI Bright Galaxy Survey: Final Target Selection, Design, and Validation},
   volume={165},
   ISSN={1538-3881},
   url={http://dx.doi.org/10.3847/1538-3881/accff8},
   DOI={10.3847/1538-3881/accff8},
   number={6},
   journal={The Astronomical Journal},
   publisher={American Astronomical Society},
   author={Hahn, ChangHoon and Wilson, Michael J. and Ruiz-Macias, Omar and Cole, Shaun and Weinberg, David H. and Moustakas, John and Kremin, Anthony and Tinker, Jeremy L. and Smith, Alex and Wechsler, Risa H. and Ahlen, Steven and Alam, Shadab and Bailey, Stephen and Brooks, David and Cooper, Andrew P. and Davis, Tamara M. and Dawson, Kyle and Dey, Arjun and Dey, Biprateep and Eftekharzadeh, Sarah and Eisenstein, Daniel J. and Fanning, Kevin and Forero-Romero, Jaime E. and Frenk, Carlos S. and Gaztañaga, Enrique and A Gontcho, Satya Gontcho and Guy, Julien and Honscheid, Klaus and Ishak, Mustapha and Juneau, Stéphanie and Kehoe, Robert and Kisner, Theodore and Lan, Ting-Wen and Landriau, Martin and Le Guillou, Laurent and Levi, Michael E. and Magneville, Christophe and Martini, Paul and Meisner, Aaron and Myers, Adam D. and Nie, Jundan and Norberg, Peder and Palanque-Delabrouille, Nathalie and Percival, Will J. and Poppett, Claire and Prada, Francisco and Raichoor, Anand and Ross, Ashley J. and Gaines, Sasha and Saulder, Christoph and Schlafly, Eddie and Schlegel, David and Sierra-Porta, David and Tarle, Gregory and Weaver, Benjamin A. and Yèche, Christophe and Zarrouk, Pauline and Zhou, Rongpu and Zhou, Zhimin and Zou, Hu},
   year={2023},
   month=may, pages={253}
}

@article{Cooper_2023,
   title={Overview of the DESI Milky Way Survey},
   volume={947},
   ISSN={1538-4357},
   url={http://dx.doi.org/10.3847/1538-4357/acb3c0},
   DOI={10.3847/1538-4357/acb3c0},
   number={1},
   journal={The Astrophysical Journal},
   publisher={American Astronomical Society},
   author={Cooper, Andrew P. and Koposov, Sergey E. and Allende Prieto, Carlos and Manser, Christopher J. and Kizhuprakkat, Namitha and Myers, Adam D. and Dey, Arjun and Gänsicke, Boris T. and Li, Ting S. and Rockosi, Constance and Valluri, Monica and Najita, Joan and Deason, Alis and Raichoor, Anand and Wang, M.-Y. and Ting, Y.-S. and Kim, Bokyoung and Carrillo, Andreia and Wang, Wenting and Beraldo e Silva, Leandro and Han, Jiwon Jesse and Ding, Jiani and Sánchez-Conde, Miguel and Aguilar, Jessica N. and Ahlen, Steven and Bailey, Stephen and Belokurov, Vasily and Brooks, David and Cunha, Katia and Dawson, Kyle and de la Macorra, Axel and Doel, Peter and Eisenstein, Daniel J. and Fagrelius, Parker and Fanning, Kevin and Font-Ribera, Andreu and Forero-Romero, Jaime E. and Gaztañaga, Enrique and A Gontcho, Satya Gontcho and Guy, Julien and Honscheid, Klaus and Kehoe, Robert and Kisner, Theodore and Kremin, Anthony and Landriau, Martin and Levi, Michael E. and Martini, Paul and Meisner, Aaron M. and Miquel, Ramon and Moustakas, John and Nie, Jundan J. D. and Palanque-Delabrouille, Nathalie and Percival, Will J. and Poppett, Claire and Prada, Francisco and Rehemtulla, Nabeel and Schlafly, Edward and Schlegel, David and Schubnell, Michael and Sharples, Ray M. and Tarlé, Gregory and Wechsler, Risa H. and Weinberg, David H. and Zhou, Zhimin and Zou, Hu},
   year={2023},
   month=apr, pages={37}
}

@article{Ruiz-Macias_2021,
    author = {Ruiz-Macias, Omar and Zarrouk, Pauline and Cole, Shaun and Baugh, Carlton M and Norberg, Peder and Lucey, John and Dey, Arjun and Eisenstein, Daniel J and Doel, Peter and Gaztañaga, Enrique and Hahn, ChangHoon and Kehoe, Robert and Kitanidis, Ellie and Landriau, Martin and Lang, Dustin and Moustakas, John and Myers, Adam D and Prada, Francisco and Schubnell, Michael and Weinberg, David H and Wilson, M J},
    title = "{Characterizing the target selection pipeline for the Dark Energy Spectroscopic Instrument Bright Galaxy Survey}",
    journal = {Monthly Notices of the Royal Astronomical Society},
    volume = {502},
    number = {3},
    pages = {4328-4349},
    year = {2021},
    month = {02},
    issn = {0035-8711},
    doi = {10.1093/mnras/stab292},
    url = {https://doi.org/10.1093/mnras/stab292},
    eprint = {https://academic.oup.com/mnras/article-pdf/502/3/4328/39112755/stab292.pdf},
}

@article{Zhou_2023,
   title={Target Selection and Validation of DESI Luminous Red Galaxies},
   volume={165},
   ISSN={1538-3881},
   url={http://dx.doi.org/10.3847/1538-3881/aca5fb},
   DOI={10.3847/1538-3881/aca5fb},
   number={2},
   journal={The Astronomical Journal},
   publisher={American Astronomical Society},
   author={Zhou, Rongpu and Dey, Biprateep and Newman, Jeffrey A. and Eisenstein, Daniel J. and Dawson, K. and Bailey, S. and Berti, A. and Guy, J. and Lan, Ting-Wen and Zou, H. and Aguilar, J. and Ahlen, S. and Alam, Shadab and Brooks, D. and de la Macorra, A. and Dey, A. and Dhungana, G. and Fanning, K. and Font-Ribera, A. and Gontcho, S. Gontcho A. and Honscheid, K. and Ishak, Mustapha and Kisner, T. and Kovács, A. and Kremin, A. and Landriau, M. and Levi, Michael E. and Magneville, C. and Manera, Marc and Martini, P. and Meisner, Aaron M. and Miquel, R. and Moustakas, J. and Myers, Adam D. and Nie, Jundan and Palanque-Delabrouille, N. and Percival, W. J. and Poppett, C. and Prada, F. and Raichoor, A. and Ross, A. J. and Schlafly, E. and Schlegel, D. and Schubnell, M. and Tarlé, Gregory and Weaver, B. A. and Wechsler, R. H. and Yéche, Christophe and Zhou, Zhimin},
   year={2023},
   month=jan, pages={58}
}

@article{Raichoor_2020,
   title={Preliminary Target Selection for the DESI Emission Line Galaxy (ELG) Sample},
   volume={4},
   ISSN={2515-5172},
   url={http://dx.doi.org/10.3847/2515-5172/abc078},
   DOI={10.3847/2515-5172/abc078},
   number={10},
   journal={Research Notes of the AAS},
   publisher={American Astronomical Society},
   author={Raichoor, Anand and Eisenstein, Daniel J. and Karim, Tanveer and Newman, Jeffrey A. and Moustakas, John and Brooks, David D. and Dawson, Kyle S. and Dey, Arjun and Duan, Yutong and Eftekharzadeh, Sarah and Gaztañaga, Enrique and Kehoe, Robert and Landriau, Martin and Lang, Dustin and Lee, Jae H. and Levi, Michael E. and Meisner, Aaron M. and Myers, Adam D. and Palanque-Delabrouille, Nathalie and Poppett, Claire and Prada, Francisco and Ross, Ashley J. and Schlegel, David J. and Schubnell, Michael and Staten, Ryan and Tarlé, Gregory and Tojeiro, Rita and Yèche, Christophe and Zhou, Rongpu},
   year={2020},
   month=oct, pages={180}
}

@article{DESI_Collaboration_2022,
   title={Overview of the Instrumentation for the Dark Energy Spectroscopic Instrument},
   volume={164},
   ISSN={1538-3881},
   url={http://dx.doi.org/10.3847/1538-3881/ac882b},
   DOI={10.3847/1538-3881/ac882b},
   number={5},
   journal={The Astronomical Journal},
   publisher={American Astronomical Society},
   author={{DESI Collaboration} and Abareshi, B. and Aguilar, J. and Ahlen, S. and Alam, Shadab and Alexander, David M. and Alfarsy, R. and Allen, L. and Allende Prieto, C. and Alves, O. and Ameel, J. and Armengaud, E. and Asorey, J. and Aviles, Alejandro and Bailey, S. and Balaguera-Antolínez, A. and Ballester, O. and Baltay, C. and Bault, A. and Beltran, S. F. and Benavides, B. and BenZvi, S. and Berti, A. and Besuner, R. and Beutler, Florian and Bianchi, D. and Blake, C. and Blanc, P. and Blum, R. and Bolton, A. and Bose, S. and Bramall, D. and Brieden, S. and Brodzeller, A. and Brooks, D. and Brownewell, C. and Buckley-Geer, E. and Cahn, R. N. and Cai, Z. and Canning, R. and Capasso, R. and Carnero Rosell, A. and Carton, P. and Casas, R. and Castander, F. J. and Cervantes-Cota, J. L. and Chabanier, S. and Chaussidon, E. and Chuang, C. and Circosta, C. and Cole, S. and Cooper, A. P. and da Costa, L. and Cousinou, M.-C. and Cuceu, A. and Davis, T. M. and Dawson, K. and de la Cruz-Noriega, R. and de la Macorra, A. and de Mattia, A. and Della Costa, J. and Demmer, P. and Derwent, M. and Dey, A. and Dey, B. and Dhungana, G. and Ding, Z. and Dobson, C. and Doel, P. and Donald-McCann, J. and Donaldson, J. and Douglass, K. and Duan, Y. and Dunlop, P. and Edelstein, J. and Eftekharzadeh, S. and Eisenstein, D. J. and Enriquez-Vargas, M. and Escoffier, S. and Evatt, M. and Fagrelius, P. and Fan, X. and Fanning, K. and Fawcett, V. A. and Ferraro, S. and Ereza, J. and Flaugher, B. and Font-Ribera, A. and Forero-Romero, J. E. and Frenk, C. S. and Fromenteau, S. and Gänsicke, B. T. and Garcia-Quintero, C. and Garrison, L. and Gaztañaga, E. and Gerardi, F. and Gil-Marín, H. and Gontcho A Gontcho, S. and Gonzalez-Morales, Alma X. and Gonzalez-de-Rivera, G. and Gonzalez-Perez, V. and Gordon, C. and Graur, O. and Green, D. and Grove, C. and Gruen, D. and Gutierrez, G. and Guy, J. and Hahn, C. and Harris, S. and Herrera, D. and Herrera-Alcantar, Hiram K. and Honscheid, K. and Howlett, C. and Huterer, D. and Iršič, V. and Ishak, M. and Jelinsky, P. and Jiang, L. and Jimenez, J. and Jing, Y. P. and Joyce, R. and Jullo, E. and Juneau, S. and Karaçaylı, N. G. and Karamanis, M. and Karcher, A. and Karim, T. and Kehoe, R. and Kent, S. and Kirkby, D. and Kisner, T. and Kitaura, F. and Koposov, S. E. and Kovács, A. and Kremin, A. and Krolewski, Alex and L’Huillier, B. and Lahav, O. and Lambert, A. and Lamman, C. and Lan, Ting-Wen and Landriau, M. and Lane, S. and Lang, D. and Lange, J. U. and Lasker, J. and Le Guillou, L. and Leauthaud, A. and Le Van Suu, A. and Levi, Michael E. and Li, T. S. and Magneville, C. and Manera, M. and Manser, Christopher J. and Marshall, B. and Martini, Paul and McCollam, W. and McDonald, P. and Meisner, Aaron M. and Mena-Fernández, J. and Meneses-Rizo, J. and Mezcua, M. and Miller, T. and Miquel, R. and Montero-Camacho, P. and Moon, J. and Moustakas, J. and Mueller, E. and Muñoz-Gutiérrez, Andrea and Myers, Adam D. and Nadathur, S. and Najita, J. and Napolitano, L. and Neilsen, E. and Newman, Jeffrey A. and Nie, J. D. and Ning, Y. and Niz, G. and Norberg, P. and Noriega, Hernán E. and O’Brien, T. and Obuljen, A. and Palanque-Delabrouille, N. and Palmese, A. and Zhiwei, P. and Pappalardo, D. and PENG, X. and Percival, W. J. and Perruchot, S. and Pogge, R. and Poppett, C. and Porredon, A. and Prada, F. and Prochaska, J. and Pucha, R. and Pérez-Fernández, A. and Pérez-Ràfols, I. and Rabinowitz, D. and Raichoor, A. and Ramirez-Solano, S. and Ramírez-Pérez, César and Ravoux, C. and Reil, K. and Rezaie, M. and Rocher, A. and Rockosi, C. and Roe, N. A. and Roodman, A. and Ross, A. J. and Rossi, G. and Ruggeri, R. and Ruhlmann-Kleider, V. and Sabiu, C. G. and Gaines, S. and Said, K. and Saintonge, A. and Salas Catonga, Javier and Samushia, L. and Sanchez, E. and Saulder, C. and Schaan, E. and Schlafly, E. and Schlegel, D. and Schmoll, J. and Scholte, D. and Schubnell, M. and Secroun, A. and Seo, H. and Serrano, S. and Sharples, Ray M. and Sholl, Michael J. and Silber, Joseph Harry and Silva, D. R. and Sirk, M. and Siudek, M. and Smith, A. and Sprayberry, D. and Staten, R. and Stupak, B. and Tan, T. and Tarlé, Gregory and Tie, Suk Sien and Tojeiro, R. and Ureña-López, L. A. and Valdes, F. and Valenzuela, O. and Valluri, M. and Vargas-Magaña, M. and Verde, L. and Walther, M. and Wang, B. and Wang, M. S. and Weaver, B. A. and Weaverdyck, C. and Wechsler, R. and Wilson, Michael J. and Yang, J. and Yu, Y. and Yuan, S. and Yèche, Christophe and Zhang, H. and Zhang, K. and Zhao, Cheng and Zhou, Rongpu and Zhou, Zhimin and Zou, H. and Zou, J. and Zou, S. and Zu, Y.},
   year={2022},
   month=oct, pages={207}
}

@article{Ross_2016,
   title={The clustering of galaxies in the completed SDSS-III Baryon Oscillation Spectroscopic Survey: observational systematics and baryon acoustic oscillations in the correlation function},
   volume={464},
   ISSN={1365-2966},
   url={http://dx.doi.org/10.1093/mnras/stw2372},
   DOI={10.1093/mnras/stw2372},
   number={1},
   journal={Monthly Notices of the Royal Astronomical Society},
   publisher={Oxford University Press (OUP)},
   author={Ross, Ashley J. and Beutler, Florian and Chuang, Chia-Hsun and Pellejero-Ibanez, Marcos and Seo, Hee-Jong and Vargas-Magaña, Mariana and Cuesta, Antonio J. and Percival, Will J. and Burden, Angela and Sánchez, Ariel G. and Grieb, Jan Niklas and Reid, Beth and Brownstein, Joel R. and Dawson, Kyle S. and Eisenstein, Daniel J. and Ho, Shirley and Kitaura, Francisco-Shu and Nichol, Robert C. and Olmstead, Matthew D. and Prada, Francisco and Rodríguez-Torres, Sergio A. and Saito, Shun and Salazar-Albornoz, Salvador and Schneider, Donald P. and Thomas, Daniel and Tinker, Jeremy and Tojeiro, Rita and Wang, Yuting and White, Martin and Zhao, Gong-bo},
   year={2016},
   month=sep, pages={1168–1191}
}

@article{Bault_2025,
   title={Impact of systematic redshift errors on the cross-correlation of the Lyman-α forest with quasars at small scales using DESI Early Data},
   volume={2025},
   ISSN={1475-7516},
   url={http://dx.doi.org/10.1088/1475-7516/2025/01/130},
   DOI={10.1088/1475-7516/2025/01/130},
   number={01},
   journal={Journal of Cosmology and Astroparticle Physics},
   publisher={IOP Publishing},
   author={Bault, Abby and Kirkby, David and Guy, Julien and Brodzeller, Allyson and Aguilar, J. and Ahlen, S. and Bailey, S. and Brooks, D. and Cabayol-Garcia, L. and Chaves-Montero, J. and Claybaugh, T. and Cuceu, A. and Dawson, K. and de la Cruz, R. and de la Macorra, A. and Dey, A. and Doel, P. and Filbert, S. and Font-Ribera, A. and Forero-Romero, J.E. and Gaztañaga, E. and Gontcho, S.Gontcho A. and Gordon, C. and Herrera-Alcantar, H.K. and Honscheid, K. and Iršič, V. and Karaçaylı, N.G. and Kehoe, R. and Kisner, T. and Kremin, A. and Lambert, A. and Landriau, M. and Le Guillou, L. and Levi, M.E. and Manera, M. and Martini, P. and Meisner, A. and Miquel, R. and Montero-Camacho, P. and Moustakas, J. and Muñoz-Gutiérrez, A. and Nie, J. and Niz, G. and Palanque-Delabrouille, N. and Percival, W.J. and Pérez-Ràfols, I. and Poppett, C. and Prada, F. and Ramírez-Pérez, C. and Ravoux, C. and Rezaie, M. and Rossi, G. and Sanchez, E. and Schlafly, E.F. and Schlegel, D. and Schubnell, M. and Silber, J. and Tan, T. and Tarlé, G. and Walther, M. and Weaver, B.A. and Zhou, Z.},
   year={2025},
   month=jan, pages={130} 
}

@misc{miller2023opticalcorrectordarkenergy,
      title={The Optical Corrector for the Dark Energy Spectroscopic Instrument}, 
      author={Timothy N. Miller and Peter Doel and Gaston Gutierrez and Robert Besuner and David Brooks and Giuseppe Gallo and Henry Heetderks and Patrick Jelinsky and Stephen M. Kent and Michael Lampton and Michael Levi and Ming Liang and Aaron Meisner and Michael J. Sholl and Joseph Harry Silber and David Sprayberry and Jessica Nicole Aguilar and Axel de la Macorra and Daniel Eisenstein and Kevin Fanning and Andreu Font-Ribera and Enrique Gaztanaga and Satya Gontcho A Gontcho and Klaus Honscheid and Jorge Jimenez and Dick Joyce and Robert Kehoe and Theodore Kisner and Anthony Kremin and Martin Landriau and Laurent Le Guillou and Christophe Magneville and Paul Martini and Ramon Miquel and John Moustakas and Jundan Nie and Will Percival and Claire Poppett and Francisco Prada and Graziano Rossi and David Schlegel and Michael Schubnell and Hee-Jong Seo and Ray Sharples and Gregory Tarle and Mariana Vargas-Magana and Zhimin Zhou},
      year={2023},
      eprint={2306.06310},
      archivePrefix={arXiv},
      primaryClass={astro-ph.IM},
      url={https://arxiv.org/abs/2306.06310}, 
}

@article{Poppett_2024,
  author  = {Poppett, Claire and Tyas, Luke and Aguilar, J. and others},
  title   = {Overview of the Fiber System for the Dark Energy Spectroscopic Instrument},
  journal = {The Astronomical Journal},
  volume  = {168},
  number  = {6},
  pages   = {245},
  year    = {2024},
  doi     = {10.3847/1538-3881/ad76a4}
}

@article{Silber_2022,
   title={The Robotic Multiobject Focal Plane System of the Dark Energy Spectroscopic Instrument (DESI)},
   volume={165},
   ISSN={1538-3881},
   url={http://dx.doi.org/10.3847/1538-3881/ac9ab1},
   DOI={10.3847/1538-3881/ac9ab1},
   number={1},
   journal={The Astronomical Journal},
   publisher={American Astronomical Society},
   author={Silber, Joseph Harry and Fagrelius, Parker and Fanning, Kevin and Schubnell, Michael and Aguilar, Jessica Nicole and Ahlen, Steven and Ameel, Jon and Ballester, Otger and Baltay, Charles and Bebek, Chris and Beard, Dominic Benton and Besuner, Robert and Cardiel-Sas, Laia and Casas, Ricard and Castander, Francisco Javier and Claybaugh, Todd and Dobson, Carl and Duan, Yutong and Dunlop, Patrick and Edelstein, Jerry and Emmet, William T. and Elliott, Ann and Evatt, Matthew and Gershkovich, Irena and Guy, Julien and Harris, Stu and Heetderks, Henry and Heetderks, Ian and Honscheid, Klaus and Illa, Jose Maria and Jelinsky, Patrick and Jelinsky, Sharon R. and Jimenez, Jorge and Karcher, Armin and Kent, Stephen and Kirkby, David and Kneib, Jean-Paul and Lambert, Andrew and Lampton, Mike and Leitner, Daniela and Levi, Michael and McCauley, Jeremy and Meisner, Aaron and Miller, Timothy N. and Miquel, Ramon and Mundet, Juliá and Poppett, Claire and Rabinowitz, David and Reil, Kevin and Roman, David and Schlegel, David and Serrano, Santiago and Van Shourt, William and Sprayberry, David and Tarlé, Gregory and Tie, Suk Sien and Weaverdyck, Curtis and Zhang, Kai and Azzaro, Marco and Bailey, Stephen and Becerril, Santiago and Blackwell, Tami and Bouri, Mohamed and Brooks, David and Buckley-Geer, Elizabeth and Castro, Jose Peñate and Derwent, Mark and Dey, Arjun and Dhungana, Govinda and Doel, Peter and Eisenstein, Daniel J. and Fahim, Nasib and Garcia-Bellido, Juan and Gaztañaga, Enrique and A Gontcho, Satya Gontcho and Gutierrez, Gaston and Hörler, Philipp and Kehoe, Robert and Kisner, Theodore and Kremin, Anthony and Kronig, Luzius and Landriau, Martin and Le Guillou, Laurent and Martini, Paul and Moustakas, John and Palanque-Delabrouille, Nathalie and Peng, Xiyan and Percival, Will and Prada, Francisco and Prieto, Carlos Allende and de Rivera, Guillermo Gonzalez and Sanchez, Eusebio and Sanchez, Justo and Sharples, Ray and Soares-Santos, Marcelle and Schlafly, Edward and Weaver, Benjamin Alan and Zhou, Zhimin and Zhu, Yaling and Zou, Hu},
   year={2022},
   month=dec, pages={9} }

@article{Adame_2025,
   title={DESI 2024 VII: cosmological constraints from the full-shape modeling of clustering measurements},
   volume={2025},
   ISSN={1475-7516},
   url={http://dx.doi.org/10.1088/1475-7516/2025/07/028},
   DOI={10.1088/1475-7516/2025/07/028},
   number={07},
   journal={Journal of Cosmology and Astroparticle Physics},
   publisher={IOP Publishing},
   author={Adame, A.G. and Aguilar, J. and Ahlen, S. and Alam, S. and Alexander, D.M. and Allende Prieto, C. and Alvarez, M. and Alves, O. and Anand, A. and Andrade, U. and Armengaud, E. and Avila, S. and Aviles, A. and Awan, H. and Bahr-Kalus, B. and Bailey, S. and Baltay, C. and Bault, A. and Behera, J. and BenZvi, S. and Beutler, F. and Bianchi, D. and Blake, C. and Blum, R. and Bonici, M. and Brieden, S. and Brodzeller, A. and Brooks, D. and Buckley-Geer, E. and Burtin, E. and Calderon, R. and Canning, R. and Carnero Rosell, A. and Cereskaite, R. and Cervantes-Cota, J.L. and Chabanier, S. and Chaussidon, E. and Chaves-Montero, J. and Chebat, D. and Chen, S. and Chen, X. and Claybaugh, T. and Cole, S. and Cuceu, A. and Davis, T.M. and Dawson, K. and de la Macorra, A. and de Mattia, A. and Deiosso, N. and Dey, A. and Dey, B. and Ding, Z. and Doel, P. and Edelstein, J. and Eftekharzadeh, S. and Eisenstein, D.J. and Elbers, W. and Elliott, A. and Fagrelius, P. and Fanning, K. and Ferraro, S. and Ereza, J. and Findlay, N. and Flaugher, B. and Font-Ribera, A. and Forero-Sánchez, D. and Forero-Romero, J.E. and Frenk, C.S. and Garcia-Quintero, C. and Garrison, L.H. and Gaztañaga, E. and Gil-Marín, H. and Gontcho, S.Gontcho A. and Gonzalez-Morales, A.X. and Gonzalez-Perez, V. and Gordon, C. and Green, D. and Gruen, D. and Gsponer, R. and Gutierrez, G. and Guy, J. and Hadzhiyska, B. and Hahn, C. and Hanif, M.M.S. and Herrera-Alcantar, H.K. and Honscheid, K. and Howlett, C. and Huterer, D. and Iršič, V. and Ishak, M. and Joyce, R. and Juneau, S. and Karaçaylı, N.G. and Kehoe, R. and Kent, S. and Kirkby, D. and Kong, H. and Koposov, S.E. and Kremin, A. and Krolewski, A. and Lahav, O. and Lai, Y. and Lan, T.-W. and Landriau, M. and Lang, D. and Lasker, J. and Le Goff, J.M. and Le Guillou, L. and Leauthaud, A. and Levi, M.E. and Li, T.S. and Lodha, K. and Magneville, C. and Manera, M. and Margala, D. and Martini, P. and Matthewson, W. and Maus, M. and McDonald, P. and Medina-Varela, L. and Meisner, A. and Mena-Fernández, J. and Miquel, R. and Moon, J. and Moore, S. and Moustakas, J. and Mudur, N. and Mueller, E. and Muñoz-Gutiérrez, A. and Myers, A.D. and Nadathur, S. and Napolitano, L. and Neveux, R. and Newman, J.A. and Nguyen, N.M. and Nie, J. and Niz, G. and Noriega, H.E. and Padmanabhan, N. and Paillas, E. and Palanque-Delabrouille, N. and Pan, J. and Penmetsa, S. and Percival, W.J. and Pieri, M.M. and Pinon, M. and Poppett, C. and Porredon, A. and Prada, F. and Pérez-Fernández, A. and Pérez-Ràfols, I. and Rabinowitz, D. and Raichoor, A. and Ramírez-Pérez, C. and Ramirez-Solano, S. and Rashkovetskyi, M. and Ravoux, C. and Rezaie, M. and Rich, J. and Rocher, A. and Rockosi, C. and Roe, N.A. and Rosado-Marin, A. and Ross, A.J. and Rossi, G. and Ruggeri, R. and Ruhlmann-Kleider, V. and Samushia, L. and Sanchez, E. and Saulder, C. and Schlafly, E.F. and Schlegel, D. and Schubnell, M. and Seo, H. and Shafieloo, A. and Sharples, R. and Silber, J. and Slosar, A. and Smith, A. and Sprayberry, D. and Tan, T. and Tarlé, G. and Taylor, P. and Trusov, S. and Vaisakh, R. and Valcin, D. and Valdes, F. and Valogiannis, G. and Vargas-Magaña, M. and Verde, L. and Walther, M. and Wang, B. and Wang, M.S. and Weaver, B.A. and Weaverdyck, N. and Wechsler, R.H. and Weinberg, D.H. and White, M. and Wilson, M.J. and Yi, L. and Yu, J. and Yu, Y. and Yuan, S. and Yèche, C. and Zaborowski, E.A. and Zarrouk, P. and Zhang, H. and Zhao, C. and Zhao, R. and Zhou, R. and Zhuang, T. and Zou, H.},
   year={2025},
   month=jul, pages={028} }

@article{Nicolaou_2026,
    author = {Nicolaou, C and Nathan, R P and Lahav, O and Palmese, A and Saintonge, A and Aguilar, J and Ahlen, S and Prieto, C Allende and Bailey, S and BenZvi, S and Bianchi, D and Brodzeller, A and Brooks, D and Claybaugh, T and de la Macorra, A and Della Costa, J and Dey, Arjun and Doel, P and Forero-Romero, J E and Gaztañaga, E and Gontcho A Gontcho, S and Gutierrez, G and Honscheid, K and Howlett, C and Ishak, M and Kehoe, R and Kirkby, D and Kisner, T and Kremin, A and Lambert, A and Landriau, M and Le Guillou, L and Meisner, A and Miquel, R and Moustakas, J and Nadathur, S and Prada, F and Pérez-Ràfols, I and Rossi, G and Sanchez, E and Schubnell, M and Siudek, M and Sprayberry, D and Tarlé, G and Weaver, B A and Zou, H},
    title = {Identifying Anomalous DESI Galaxy Spectra with a Variational Autoencoder},
    journal = {Monthly Notices of the Royal Astronomical Society},
    volume = {547},
    number = {2},
    pages = {stag010},
    year = {2026},
    month = {04},
    issn = {0035-8711},
    doi = {10.1093/mnras/stag010},
    url = {https://doi.org/10.1093/mnras/stag010},
    eprint = {https://academic.oup.com/mnras/article-pdf/547/2/stag010/66292288/stag010.pdf},
}

@ARTICLE{liang2023outlierdetectiondesibright,
       author = {{Liang}, Yan and {Melchior}, Peter and {Hahn}, ChangHoon and {Shen}, Jeff and {Goulding}, Andy and {Ward}, Charlotte},
        title = "{Outlier Detection in the DESI Bright Galaxy Survey}",
      journal = {\apjl},
         year = 2023,
        month = oct,
       volume = {956},
       number = {1},
          eid = {L6},
        pages = {L6},
          doi = {10.3847/2041-8213/acfa03},
archivePrefix = {arXiv},
       eprint = {2307.07664},
 primaryClass = {astro-ph.GA},
       adsurl = {https://ui.adsabs.harvard.edu/abs/2023ApJ...956L...6L}
}

@article{Yip_2004a,
   title={Distributions of Galaxy Spectral Types in the Sloan Digital Sky Survey},
   volume={128},
   ISSN={1538-3881},
   url={http://dx.doi.org/10.1086/422429},
   DOI={10.1086/422429},
   number={2},
   journal={The Astronomical Journal},
   publisher={American Astronomical Society},
   author={Yip, C. W. and Connolly, A. J. and Szalay, A. S. and Budavári, T. and SubbaRao, M. and Frieman, J. A. and Nichol, R. C. and Hopkins, A. M. and York, D. G. and Okamura, S. and Brinkmann, J. and Csabai, I. and Thakar, A. R. and Fukugita, M. and Ivezić, Ž.},
   year={2004},
   month=Aug, pages={585–609} }

@article{Portillo_2020,
   title={Dimensionality Reduction of SDSS Spectra with Variational Autoencoders},
   volume={160},
   ISSN={1538-3881},
   url={http://dx.doi.org/10.3847/1538-3881/ab9644},
   DOI={10.3847/1538-3881/ab9644},
   number={1},
   journal={The Astronomical Journal},
   publisher={American Astronomical Society},
   author={Portillo, Stephen K. N. and Parejko, John K. and Vergara, Jorge R. and Connolly, Andrew J.},
   year={2020},
   month=June, pages={45} }

@article{Yip_2004b,
   title={Spectral Classification of Quasars in the Sloan Digital Sky Survey: Eigenspectra, Redshift, and Luminosity Effects},
   volume={128},
   ISSN={1538-3881},
   url={http://dx.doi.org/10.1086/425626},
   DOI={10.1086/425626},
   number={6},
   journal={The Astronomical Journal},
   publisher={American Astronomical Society},
   author={Yip, C. W. and Connolly, A. J. and Vanden Berk, D. E. and Ma, Z. and Frieman, J. A. and SubbaRao, M. and Szalay, A. S. and Richards, G. T. and Hall, P. B. and Schneider, D. P. and Hopkins, A. M. and Trump, J. and Brinkmann, J.},
   year={2004},
   month=Dec, pages={2603–2630} }

@article{Ortiz_2025,
   title={Interpreting the detection of anomalies in SDSS spectra},
   volume={703},
   ISSN={1432-0746},
   url={http://dx.doi.org/10.1051/0004-6361/202556339},
   DOI={10.1051/0004-6361/202556339},
   journal={Astronomy \& Astrophysics},
   publisher={EDP Sciences},
   author={Ortiz, E. and Boquien, M.},
   year={2025},
   month=Nov, pages={A242} }

@article{McGurk_2010,
   title = {Principal Component Analysis of Sloan Digital Sky Survey Stellar Spectra},
   volume = {139},
   issn = {1538-3881},
   url = {http://dx.doi.org/10.1088/0004-6256/139/3/1261},
   doi = {10.1088/0004-6256/139/3/1261},
   number = {3},
   journal = {The Astronomical Journal},
   publisher = {American Astronomical Society},
   author = {McGurk, Rosalie C. and Kimball, Amy E. and Ivezi{\'c}, {\v Z}eljko},
   year = {2010},
   month = feb,
   pages = {1261--1268}
}

@article{Sharbaf_2023,
    author = {Sharbaf, Zahra and Ferreras, Ignacio and Lahav, Ofer},
    title = {What drives the variance of galaxy spectra?},
    journal = {Monthly Notices of the Royal Astronomical Society},
    volume = {526},
    number = {1},
    pages = {585-599},
    year = {2023},
    month = {11},
    issn = {0035-8711},
    doi = {10.1093/mnras/stad2668},
    url = {https://doi.org/10.1093/mnras/stad2668},
    eprint = {https://academic.oup.com/mnras/article-pdf/526/1/585/51770721/stad2668.pdf},
}

@article{vanderMaaten_2008,
  author  = {van der Maaten, Laurens and Hinton, Geoffrey},
  title   = {Visualizing Data using t-SNE},
  journal = {Journal of Machine Learning Research},
  volume  = {9},
  pages   = {2579--2605},
  year    = {2008}
}

@article{Breunig_2000a,
author = {Breunig, Markus M. and Kriegel, Hans-Peter and Ng, Raymond T. and Sander, J\"{o}rg},
title = {LOF: identifying density-based local outliers},
year = {2000},
issue_date = {June 2000},
publisher = {Association for Computing Machinery},
address = {New York, NY, USA},
volume = {29},
number = {2},
issn = {0163-5808},
url = {https://doi.org/10.1145/335191.335388},
doi = {10.1145/335191.335388},
journal = {SIGMOD Rec.},
month = may,
pages = {93–104},
numpages = {12}
}

@inproceedings{Breunig_2000b,
author = {Breunig, Markus M. and Kriegel, Hans-Peter and Ng, Raymond T. and Sander, J\"{o}rg},
title = {LOF: identifying density-based local outliers},
year = {2000},
isbn = {1581132174},
publisher = {Association for Computing Machinery},
address = {New York, NY, USA},
url = {https://doi.org/10.1145/342009.335388},
doi = {10.1145/342009.335388},
booktitle = {Proceedings of the 2000 ACM SIGMOD International Conference on Management of Data},
pages = {93–104},
numpages = {12},
location = {Dallas, Texas, USA},
series = {SIGMOD '00}
}

@INPROCEEDINGS{Liu_2008,
  author={Liu, Fei Tony and Ting, Kai Ming and Zhou, Zhi-Hua},
  booktitle={2008 Eighth IEEE International Conference on Data Mining}, 
  title={Isolation Forest}, 
  year={2008},
  volume={},
  number={},
  pages={413-422},
  doi={10.1109/ICDM.2008.17}}

@ARTICLE{Krolewski_2025,
       author = {{Krolewski}, A. and {Yu}, J. and {Ross}, A.~J. and {Penmetsa}, S. and {Percival}, W.~J. and {Zhou}, R. and {Wilson}, M.~J. and {Hou}, J. and {Aguilar}, J. and {Ahlen}, S. and {Brooks}, D. and {Chaussidon}, E. and {Claybaugh}, T. and {de la Macorra}, A. and {Dey}, Biprateep and {Forero-Romero}, J.~E. and {Gontcho}, S. Gontcho A. and {Guy}, J. and {Honscheid}, K. and {Juneau}, S. and {Kirkby}, D. and {Kisner}, T. and {Kremin}, A. and {Lambert}, A. and {Le Guillou}, L. and {Levi}, M.~E. and {Martini}, P. and {Meisner}, A. and {Miquel}, R. and {Moustakas}, J. and {Myers}, A.~D. and {Newman}, J.~A. and {Niz}, G. and {Palanque-Delabrouille}, N. and {Rossi}, G. and {Sanchez}, E. and {Schlafly}, E.~F. and {Schlegel}, D. and {Schubnell}, M. and {Seo}, H. and {Sprayberry}, D. and {Tarl{\'e}}, G. and {Weaver}, B.~A. and {Zhao}, C.},
        title = "{Impact and mitigation of spectroscopic systematics on DESI DR1 clustering measurements}",
      journal = {\jcap},
         year = 2025,
        month = jan,
       volume = {2025},
       number = {1},
          eid = {147},
        pages = {147},
          doi = {10.1088/1475-7516/2025/01/147},
archivePrefix = {arXiv},
       eprint = {2405.17208},
 primaryClass = {astro-ph.CO},
       adsurl = {https://ui.adsabs.harvard.edu/abs/2025JCAP...01..147K}
}

@ARTICLE{Levi_2013,
       author = {{Levi}, Michael and {Bebek}, Chris and {Beers}, Timothy and
                 {Blum}, Robert and {Cahn}, Robert and {Eisenstein}, Daniel and
                 {Flaugher}, Brenna and {Honscheid}, Klaus and {Kron}, Richard and
                 {Lahav}, Ofer and {McDonald}, Patrick and {Roe}, Natalie and
                 {Schlegel}, David and {representing the DESI collaboration}},
        title = "{The DESI Experiment, a whitepaper for Snowmass 2013}",
      journal = {arXiv e-prints},
         year = 2013,
        month = aug,
          eid = {arXiv:1308.0847},
        pages = {arXiv:1308.0847},
          doi = {10.48550/arXiv.1308.0847},
archivePrefix = {arXiv},
       eprint = {1308.0847},
 primaryClass = {astro-ph.CO},
       adsurl = {https://ui.adsabs.harvard.edu/abs/2013arXiv1308.0847L}
}

\end{document}